\documentclass[aps,nofootinbib,showkeys,noshowpacs,preprintnumbers,amsmath,amssymb]{revtex4}
\pdfoutput=1

\def\be{\begin{equation}}
\def\ee{\end{equation}}
\def\ba{\begin{eqnarray}}
\def\ea{\end{eqnarray}}
\def\bea{\begin{eqnarray}}
\def\eea{\end{eqnarray}}
\def\bes{\begin{subequations}}
\def\ees{\end{subequations}}

\newcommand{\sm}{{\sigma_{\rm m}}}
\newcommand{\sg}{{\sigma}}
\newcommand{\tkap}{{\widetilde \kappa}}
\newcommand{\tk}{{\widetilde k}}
\newcommand{\tK}{{\widetilde K}}

\newcommand{\ta}{{\widetilde a}}
\newcommand{\tal}{{\widetilde \alpha}}

\newcommand{\tb}{{\widetilde b}}
\newcommand{\td}{{\widetilde d}}
\newcommand{\tr}{{\widetilde r}}

\newcommand{\tg}{{\widetilde {\gamma}}}
\newcommand{\bg}{{\overline {\gamma}}}
\newcommand{\MSbar}{\overline{\rm MS}}  

\usepackage{graphics}
\usepackage{graphicx}
\usepackage{dcolumn}
\usepackage{bm}
\usepackage{epsfig}
\usepackage{graphicx,color}
\usepackage{multirow}

\begin{document}


\title{Determination of perturbative QCD coupling from ALEPH $\tau$ decay data using pinched Borel-Laplace and Finite Energy Sum Rules}

\author{C\'esar Ayala}
\email{c.ayala86@gmail.com}
\author{Gorazd Cveti\v{c}}
\email{gorazd.cvetic@gmail.com}
\author{Diego Teca}
\email{diegotecawellmann@gmail.com}

\affiliation{Department of Physics, Universidad T{\'e}cnica Federico Santa Mar{\'\i}a (UTFSM), Casilla 110-V, Valpara{\'\i}so, Chile}

\date{\today}

\begin{abstract}
  We present a determination of the perturbative QCD (pQCD) coupling using the V+A channel ALEPH $\tau$-decay data. The determination involves the double-pinched Borel-Laplace Sum Rules and Finite Energy Sum Rules. The theoretical basis is the Operator Product Expansion (OPE) of the V+A channel Adler function in which the higher order terms of the leading-twist part originate from a model based on the known structure of the leading renormalons of this quantity. The applied evaluation methods are contour-improved perturbation theory (CIPT), fixed-order perturbation theory (FOPT), and Principal Value of the Borel resummation (PV). All the methods involve truncations in the order of the coupling. In contrast to the truncated CIPT method, the truncated FOPT and PV methods account correctly for the suppression of various renormalon contributions of the Adler function in the mentioned sum rules. The extracted value of the ${\overline {\rm MS}}$ coupling is $\alpha_s(m_{\tau}^2) = 0.3116 \pm 0.0073$ [$\alpha_s(M_Z^2)=0.1176 \pm 0.0010$] for the average of the FOPT and PV methods, which we regard as our main result. On the other hand, if we include in the average also the CIPT method, the resulting values are significantly higher, $\alpha_s(m_{\tau}^2) = 0.3194 \pm 0.0167$ [$\alpha_s(M_Z^2)=0.1186 \pm 0.0021$].

\end{abstract}

\maketitle


\section{Introduction}
\label{sec:Int}

The physics of semihadronic $\tau$ lepton decays is an important area of QCD, because it describes QCD at relatively low momenta $Q \lesssim m_{\tau} \sim 1$ GeV and, at the same time, has high precision experimental results. The latter are principally from ALEPH Collaboration \cite{ALEPH2,DDHMZ,ALEPHfin,ALEPHwww}, where the spectral function $\omega(\sigma)$ was measured with high precision.\footnote{$\omega(\sigma) \propto {\rm Im} \Pi(-\sigma - i \epsilon)$, where $\Pi(Q^2)$ is the polarisation function of the quark current correlator. The related Adler function ${\cal D}(Q^2)$ is proportional to $d \Pi(Q^2)/d \ln Q^2$. We will use  the notation $Q^2 \equiv - q^2 = - (q^0)^2 + {\vec q}^2$.} The extraction of the value of the running QCD coupling $\alpha_s(Q^2)$ at such low momenta $Q^2 \approx m^2_{\tau}$ represents a test of QCD, especially when comparing, via renormalisation group equation (RGE) evolution, with the extraction of the running coupling from experiments at higher energies $Q^2 \gg m^2_{\tau}$ where the perturbative methods of evaluation work very well \cite{DBTrev,alpha2019,PDG2020}.

The theoretical framework for the calculation of the QCD corrections $r_{\tau}$ to the $\tau$ decay ratio $R_{\tau} \propto \Gamma(\tau \to \nu_{\tau} {\rm hadrons})$,\footnote{The QCD V+A quantity $r_{\tau}=r_{\tau}^{(D=0)} + \delta r_{\tau}(m_{u,d} \not= 0) + \sum_{D \geq 4} r_{\tau}^{(D)}$ appears in the semihadronic strangeless $V+A$ $\tau$-decay ratio $R_{\tau}$ via the relation $R_{\tau}= 3 |V_{ud}|^2 S_{\rm EW} ( 1 + \delta'_{\rm EW} + r_{\tau})$, where $S_{\rm EW}=1.0198 \pm 0.0006$ \cite{SEW} and $\delta'_{\rm EW}=0.0010 \pm 0.0010$ \cite{dpEW} are electroweak corrections, $V_{ud}$ is the CKM matrix element, and $\delta r_{\tau}(m_{u,d} \not= 0) \approx - 8 \pi^2 f_{\pi}^2 m_{\pi}^2/m_{\tau}^4 \approx -0.0026$ (where $f_{\pi}=0.1305$ GeV), cf.~\cite{NP88,B88B89,BNP92,GCTK01}.}
  and of other related sum rules, is well-established \cite{NP88,B88B89,BNP92}. The perturbative part of the related quark current correlator is known up to ${\cal O}(\alpha_s^4)$ \cite{BCK}. The nonperturbative corrections to $r_{\tau}$ are also well understood and were shown to be small \cite{BNP92,DP92}.

Nonetheless, the extraction of $\alpha_s$ from the $\tau$-decay data shows a significant ambiguity which has to do with the way the (re)summations and subsequent truncations are performed in the perturbative part of the decay width ratio and of other related sum rules. These evaluations involve integration of functions containing the QCD running coupling parameter $a(Q^2) \equiv \alpha_s(Q^2)/\pi$ along the circle in the complex $Q^2$-plane with the radius $|Q^2|=\sigma_{\rm max}$ ($\sim m^2_{\tau}$). The integration is usually performed by Taylor-expanding the integrand around $Q^2=\sigma_{\rm max} >0$ (fixed order [FI]) or RGE-evolving the integrand along the countour $Q^2=\sigma_{\rm max} e^{i \phi}$ (contour improved [CIPT]) \cite{CI1,CI2,CIAPT}. Since these two main methods involve truncations in powers of $a$, they give different results. The CIPT method in general gives significantly higher value of the extracted $\alpha_s$ than the FOPT method, cf.~\cite{BCK,Davetal,Pich}. In the work of \cite{Bo2015}, which is concentrated on the analysis of the QCD duality violation effects in $\tau$-decays \cite{Cata}, a similar discrepancy is obtained, although the values of $\alpha_s$ are in general lower than those obtained from the FOPT and CIPT approaches of Refs.~\cite{BCK,Davetal,Pich}.  

This problem of FOPT vs CIPT was addressed in the works \cite{BJ,BJ2}. There it was argued, on the basis of the large-$\beta_0$ (LB) approximation and on numerical evidence, that the truncated FOPT method accounts for certain renormalon cancellations in $r_{\tau}^{(D=0)} \equiv r_{\tau}(m_{\tau}^2)^{(D=0)}$ and in related Finite Energy Sum Rules (FESRs), and that the truncated CIPT method does not account for such cancellations.\footnote{In Ref.~\cite{BoiOl} this argument was extended beyond LB when a modified Borel transform in a specific renormalisation scheme is used.}. Such arguments necessarily involve an extension of the perturbative part of the Adler function, $d(Q^2)_{(D=0)}$, beyond $\sim \alpha_s^4$ so as to account for the theoretically expected renormalon structure. The resulting FOPT and CIPT (truncated) evaluations of $r_{\tau}^{(D=0)}$ and of other related FESRs were then compared with the evaluation of these quantities when the Adler function is calculated as the inverse Borel transformation (Borel sum) and the renormalon ambiguity in the Borel sum is fixed by the Principal Value (PV) prescription.

In this work, we perform a QCD analysis of various sum rules related with the strangeless semihadronic $\tau$-decays, following in part the work of Ref.~\cite{Pich}. In order to discern the role of the renormalon singularities, we use an extension of the Adler function $d(Q^2)_{(D=0)}$ beyond the  order $\alpha_s^4$, based on the renormalon-motivated construction of Ref.~\cite{renmod}. We apply, in the theoretical Operator Product Expansion (OPE) of various FESRs and of Borel-Laplace sum rules, the (truncated) FOPT and CIPT methods and the Borel sum (PV) method, and then we extract the corresponding values of $\alpha_s$ from the ALEPH experimental data.  All the sum rules are double-pinched, and the V+A channel of the ALEPH data was used; we believe that these two aspects suppress sufficiently the duality violating effects, cf.~\cite{Pich} (cf.~also \cite{BNP92,DP92,Chib,Malt,DomSch,Cir,GonzAl,Dom,RSan}). We further argue (beyond the LB approximation) that in the considered sum rules important renormalon contributions of the Adler function get cancelled in the truncated FOPT approach, and that such a cancellation is in general not expected in the truncated CIPT approach. The Borel sum approach, on the other hand, is expected to sum correctly in the sum rules the main renormalon contributions of the Adler functions. The extracted values of $\alpha_s$ appear to be consistent with these considerations; namely, they turn out to be similar in the truncated FOPT approach and in the Borel sum approach, and they are consistently higher in the truncated CIPT approach. For these reasons, our main numerical results for $\alpha_s$ are obtained from the combination of the FOPT and Borel sum results.\footnote{Our Borel sum of the Adler extension $d(Q^2)_{(D=0)}$ consists, in addition, of a correction part in a form of a (truncated) perturbation series, cf.~Eq.~(\ref{deld}) in Sec.~\ref{ssubs:PV}.} 

The paper is organised in the following way. In Sec.~\ref{sec:SRgen} we recapitulate the main elements of the QCD sum rules in the context of the semihadronic $\tau$ decays, and their relation to the Adler function. In Sec.~\ref{sec:AdlRM} we resume the main aspects of the renormalon-motivated extension of the Adler function in the $\MSbar$ scheme, as constructed in Ref.~\cite{renmod}. In Sec.~\ref{sec:extr} we present the specific sum rules (Sec.~\ref{subs:weight}) to be considered in the numerical analysis, and the methods of evaluation of the Adler function extension: FOPT, CIPT and Borel sum (PV) (Sec.~\ref{subs:evalm}). In the related Appendix we show how certain renormalon contributions of the Adler function get cancelled in the various considered sum rules (FESRs and Borel-Laplace sum rule), at any loop level (i.e., beyond the LB approximation) and in any renormalisation scheme. In Sec.~\ref{sec:res} we then present the numerical results for the extracted values of the coupling $\alpha_s$ and of the low-dimension condesates. In Sec.~\ref{sec:summ} we make conclusions, summarise our results and make a brief comparison with the results of other works.

\section{Sum rules and Adler function}
\label{sec:SRgen}

The Adler function ${\cal D}(Q^2)$ is a logarithmic derivative of the quark current polarisation function $\Pi(Q^2)$
\be
{\cal D}(Q^2) \equiv - 2 \pi^2 \frac{d \Pi(Q^2)}{d \ln Q^2} ,
\label{Ddef}
\ee   
where $\Pi(Q^2)$ stands for the total (V+A)-channel polarisation function
\be
\Pi(Q^2) = \Pi_{\rm V}^{(1)}(Q^2) + \Pi_{\rm A}^{(1)}(Q^2) + \Pi_{\rm A}^{(0)}(Q^2).
\label{Pidef}
\ee
These functions appear in the quark current-current correlator
\be
\Pi_{\rm{J}, \mu\nu}(q) =  i \int  d^4 x \; e^{i q \cdot x} 
\langle T J_{\mu}(x) J_{\nu}(0)^{\dagger} \rangle
=  (q_{\mu} q_{\nu} - g_{\mu \nu} q^2) \Pi_{\rm J}^{(1)}(Q^2)
+ q_{\mu} q_{\nu} \Pi_{\rm J}^{(0)}(Q^2),
\label{PiJ}
\ee
where $q^2 \equiv -Q^2$ is the square of the momentum transfer. Further, $J_{\mu}$ are up-down quark currents, $J_{\mu} = {\bar u} \gamma_{\mu} d$ and ${\bar u} \gamma_{\mu} \gamma_5 d$ for $J=V, A$, respectively. In the $V+A$ sum (\ref{Pidef}), the contribution $\Pi_{\rm V}^{(0)}(Q^2)$ is neglected since ${\rm Im} \Pi_{\rm V}^{(0)}(-\sigma + i \epsilon) \propto (m_d-m_u)^2$. Further, in our analysis we will not include corrections ${\cal O}(m^2_{u,d})$ and  ${\cal O}(m^4_{u,d})$ for being numerically negligible.

\begin{figure}[htb] 
\centering\includegraphics[width=70mm]{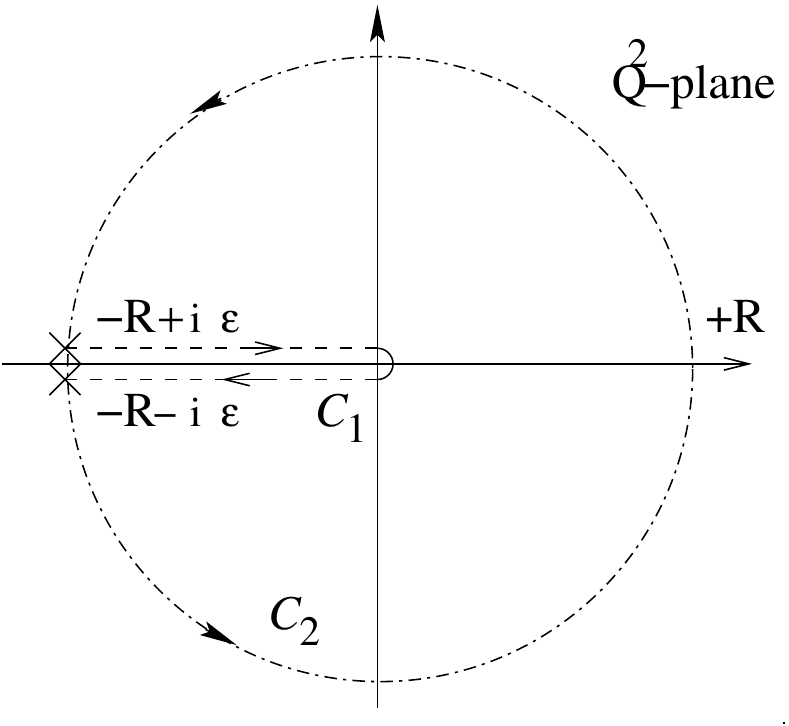}
\caption{\footnotesize The closed contour $C_1+C_2$ for integration of $g(Q^2) \Pi(Q^2)$, where the contour radius is $R=\sigma_{\rm max}$ ($\equiv \sm$) ($\leq m_{\tau}^2$).}
\label{Figcont}
 \end{figure}

\begin{figure}[htb] 
  \centering\includegraphics[width=110mm]{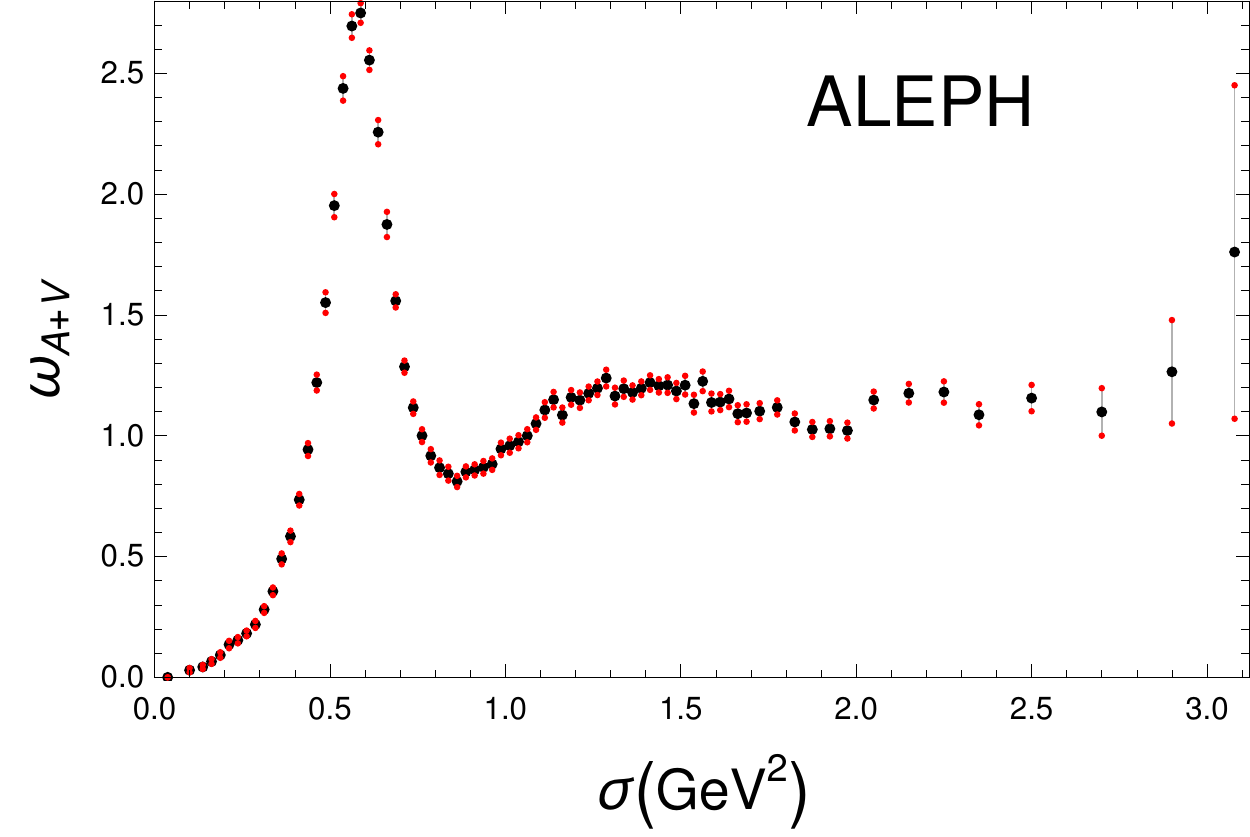}
\vspace{-0.2cm}
\caption{\footnotesize  The spectral function $\omega(\sigma)$ for the (V+A)-channel, measured by ALEPH Collaboration \cite{ALEPH2,DDHMZ,ALEPHfin,ALEPHwww}, without the pion peak. The pion peak contribution $2 \pi^2 f^2_{\pi} \delta(\sigma - m^2_{\pi})$ has to be added (where we took $f_{\pi}=0.1305$ GeV). In the sum rules we will take $\sm =2.80 \ {\rm GeV}^2$ in order to exclude the last two bins with large uncertainties.}
\label{FigOmega}
\end{figure}

The polarisation function has a theoretical expression in the form of OPE \cite{SVZ}
\be
\Pi_{\rm (th)}(Q^2; \mu^2) = - \frac{1}{2 \pi^2} \ln \left( \frac{Q^2}{\mu^2} \right) + \Pi(Q^2)_{(D=0)} + \sum_{k \geq 2} \frac{ \langle O_{2 k} \rangle_{V+A}}{(Q^2)^k} \left( 1 + {\cal O}(a) \right),
\label{PiOPE}
\ee
where $\mu^2$ is the squared renormalisation scale, and $\langle O_{2 k} \rangle_{V+A}$ are vacuum expectation values (condensates) of dimension $D=2 k$ ($\geq 4$). The ${\cal O}(a)$ terms in the Wilson coefficients turn out to be negligible \cite{GCCV2012}.\footnote{In Ref.~\cite{GCCV2012} nonpinched Borel-Laplace sum rules were applied.}
Using the OPE (\ref{PiOPE}), the corresponding Adler function is obtained using the relation (\ref{Ddef})
\be
{\cal D}_{\rm (th)}(Q^2) \equiv  - 2 \pi^2 \frac{d \Pi_{\rm (th)}(Q^2)}{d \ln Q^2} = d(Q^2)_{(D=0)} + 1 + 2 \pi^2 \sum_{k \geq 2} \frac{k \langle O_{2 k} \rangle_{V+A}}{ (Q^2)^k}.
\label{DOPE}
\ee
According to the general principles of Quantum Field Theory, the polarisation function $\Pi(Q^2; \mu^2)$, and thus the Adler function ${\cal D}(Q^2)$, are holomorphic (analytic) functions of $Q^2$ in the complex $Q^2$-plane with the exception of the negative semiaxis, $Q^2 \in \mathbb{C} \backslash (-\infty, -m_{\pi}^2)$. The associated QCD sum rules are obtained then in the following way. If $g(Q^2)$ is any holomorphic function in the complex $Q^2$-plane, then the integration of the integrand $g(Q^2) \Pi(Q^2)$ along the closed path $C_1+C_2$ presented in Fig.~\ref{Figcont}
gives zero by Cauchy theorem
\be
\oint_{C_1+C_2} d Q^2 g(Q^2) \Pi(Q^2) =  0 ,
\label{Cauchy}
\ee
which then leads to the $g$-function associated QCD sum rule
\be
\int_0^{\sm} d \sigma g(-\sigma) \omega_{\rm (exp)}(\sigma)  =
-i \pi  \oint_{|Q^2|=\sm}
d Q^2 g(Q^2) \Pi_{\rm (th)}(Q^2) .
\label{sr1} \ee
Here, the integration on the right-hand side is in the counter-clockwise direction in the complex $Q^2$-plane, and we denoted with  $\omega(\sigma)$ the spectral function of $\Pi(Q^2)$ (along the cut)
\be
\omega(\sigma) \equiv 2 \pi \; {\rm Im} \ \Pi(Q^2=-\sigma - i \epsilon) \ ,
\label{om1}
\ee
which was measured by OPAL \cite{OPAL,PerisPC1} and ALEPH Collaborations \cite{ALEPH2,DDHMZ,ALEPHfin,ALEPHwww} in strangeless semihadronic $\tau$ decays. We will use the ALEPH data as they have less experimental uncertainty; these data are presented in Fig.~\ref{FigOmega}.
In the sum rule (\ref{sr1}) the theoretical polarisation function (\ref{PiOPE}) can be replaced by the Adler function (\ref{DOPE}) by application of integration by parts
\be
\int_0^{\sm} d \sigma g(-\sigma) \omega_{\rm (exp)}(\sigma)  =
-\frac{i}{2 \pi}   \oint_{|Q^2|=\sm}
\frac{d Q^2}{Q^2} {\cal D}_{\rm (th)}(Q^2) G(Q^2) ,
\label{sr2}
\ee
where, as in Eq.~(\ref{sr1}), the integration on the right-hand side is in the counter-clockwise direction in the complex $Q^2$-plane, ${\cal D}_{\rm (th)}(Q^2)$ is given by the OPE expansion (\ref{DOPE}), and 
$G(Q^2)$ is 
\be
G(Q^2)= \int_{-\sm}^{Q^2} d Q^{'2} g(Q^{'2}).
\label{GQ2}
\ee
The Adler function ${\cal D}(Q^2)$ as a (quasi)observable is a spacelike quantity, i.e., it is holomorphic in the complex $Q^2$-plane with the exception of the negative semiaxis. On the other hand, the quantities (\ref{sr2}) are timelike observables, they are functions of the squared energy $\sm >0$ ($=-Q^2$). In the case of the sum rules (\ref{sr2}), the timelike squared energy $\sm$ is in an intermediate range $\sm \sim m_{\tau}^2 \sim 1 \ {\rm GeV}^2$ (we have here $\sm=2.8 \ {\rm GeV}^2$). There exist several other timelike quantities in form of integrals of ${\cal D}(Q^2)$ that are phenomenologically important \cite{Nesterenko:2016pmx}, among them: (a) the production ratio for $e^+ e^- \to$ hadrons, $R(s)$ \cite{AKR,ANR}, where the squared energy $|Q^2|=s>0$ is in principle not constrained; (b) the leading order hadronic vacuum polarisation contribution to the anomalous magnetic moment of $\mu$ lepton,  $(g_{\mu}/2-1)^{\rm had(1)}$ \cite{amurev,amuZoltan}, where the dominant momenta $Q^2$ of ${\cal D}(Q^2)$ are in the deep IR regime $Q^2 \sim m^2_{\mu}$ ($\sim 0.01 \ {\rm GeV}^2$) \cite{NestJPG42,amuO}.

\section{Adler function: renormalon-motivated extension}
\label{sec:AdlRM}

In the sum rule (\ref{sr2}), the theoretical expression for the Adler function is the OPE Eq.~(\ref{DOPE}), where the leading-twist ($D=0$) QCD part is given by the perturbation expansion (pt)
\be
d(Q^2)_{(D=0), {\rm pt}}= d_0 a(Q^2) + d_1(\kappa) \; a(\kappa Q^2)^2 + \ldots + d_n(\kappa) \; a(\kappa Q^2)^{n+1} + \ldots,
\label{dpt}
\ee
where $a(\mu^2) \equiv \alpha_s(\mu^2)/\pi$ is the pQCD coupling, $d_0=1$ in our normalisation, and we will work in the $\MSbar$ renormalisation scheme. Here, $\mu^2=\kappa Q^2$ is the renormalisation scale ($0 < \kappa \sim 1$ is the renormalisation scale parameter), and the $\kappa$-dependence of the coupling is determined by the (five-loop) $\MSbar$ RGE \cite{5lMSbarbeta}
\be
\frac{d a(\kappa Q^2)}{d \ln \kappa} = - \beta_0 a(\kappa Q^2)^2 -\beta_1 a(\kappa Q^2)^3 - \sum_{j=2}^4 {\beta}_j a(\kappa Q^2)^{j+2},
\label{RGE}
\ee
where $\beta_0 = (11 - 2 N_f/3)/4$ ($=9/4$ for $N_f=3$) and $\beta_1=(1/16)(102 - 38 N_f/3)$ are universal (i.e., scheme independent) in mass independent schemes, and ${\beta}_j$ ($j \geq 2$) depend on the renormalisation scheme. We will always use the five-loop $\MSbar$ RGE when varying the renormalisation scale, independent of the truncation index. The first four terms in the expansion (\ref{dpt}) (i.e., the coefficients $d_j$, $j=0,1,2,3$) are exactly known \cite{d1,d2,BCK}. In Ref.~\cite{renmod}, a renormalon-motivated extension of this expansion to all orders was constructed. It was based on the following considerations. The perturbation expansion (\ref{dpt}) in powers of $a$ can be reorganised in another expansion
\be
d(Q^2)_{(D=0), {\rm lpt}}= {\td}_0 a(Q^2) + {\td}_1(\kappa) \; {\ta}_2(\kappa Q^2) + \ldots + {\td}_n(\kappa) \; {\ta}_{n+1}(\kappa Q^2) + \ldots,
\label{dlpt} \ee
where ${\ta}_{n+1}(Q^{'2})$ are logarithmic derivatives
\be
\ta_{n+1}(Q^{'2}) \equiv \frac{(-1)^n}{n! \beta_0^n} \left( \frac{d}{d \ln Q^{'2}} \right)^n a(Q^{'2}) \qquad (n=0,1,2,\ldots),
\label{tan} \ee
which can be expressed in powers of $a$ (by using the RGE)
\be
\ta_{n+1}(Q^{'2}) = a(Q^{'2})^{n+1} + \sum_{m \geq 1} k_m(n+1) \; a(Q^{'2})^{n+1+m}.
\label{tanan}
\ee
These relations can be inverted and have the form
\be a(Q^{'2})^{n+1} =  \ta_{n+1}(Q^{'2}) + \sum_{m \geq 1} \tk_m(n+1) \; \ta_{n+1+m}(Q^{'2}).
\label{antan} \ee
The relations (\ref{tanan}) imply linear relations between the expansion coefficients $d_n$ and $\td_k$
\be
d_n(\kappa) = \td_n(\kappa) + \sum_{s=1}^{n-1} k_s(n+1-s) \; \td_{n-s}(\kappa) \quad
(n=0,1,2, \ldots),
\label{dntdk} \ee
where the coefficients $k_s(n+1-s)$ are ($\kappa$-independent) combinations of the beta-function related coefficients $c_j \equiv \beta_j/\beta_0$ \cite{renmod}. We note that $d_0 = \td_0$ ($=1$ in our normalisation).

If we formally replace in the expansion (\ref{dlpt}) the logarithmic derivatives by the corresponding powers, ${\ta}_{n+1}(Q^{'2}) \mapsto a(Q^{'2})^{n+1}$, we obtain another associated quantity ${\td}(Q^2;\kappa)_{(D=0)}$
\be
{\td}(Q^2;\kappa)_{(D=0), {\rm pt}} = {\td}_0 a(Q^2) + {\td}_1(\kappa) \; a(\kappa Q^2)^2 + \ldots + {\td}_n(\kappa) \; a(\kappa Q^2)^{n+1} + \ldots,
\label{tdpt} \ee
which agrees with $d(Q^2)_{(D=0), {\rm pt}}$ Eq.~(\ref{dpt}) only at the one-loop level, and is $\kappa$-independent only at the one-loop level [$a^{(1 \ell)}(Q^{'2})^{n+1} = {\ta}^{(1 \ell)}_{n+1}(Q^{'2})$]. It turns out that the exact $\kappa$-dependence of the coefficients ${\td}_n(\kappa)$ has the one-loop-type form
\be
\frac{d}{d \ln \kappa} {\td}_n(\kappa) = n \beta_0 {\td}_{n-1}(\kappa).
\label{tdnkap} \ee
As a consequence, the Borel transform ${\cal B}[\td]$ of the power expansion (\ref{tdpt})
\be
{\cal B}[\td](u; \kappa) \equiv {\td}_0 + \frac{{\td}_1(\kappa)}{1! \beta_0} u + \ldots + \frac{{\td}_n(\kappa)}{n! \beta_0^n} u^n + \ldots
\label{Btdexp} \ee
has the simple one-loop-type (or: large-$\beta_0$-type) $\kappa$-dependence
\be
{\cal B}[\td](u; \kappa) = \kappa^u {\cal B}[\td](u).
\label{Btdkappa} \ee
This would suggest that this Borel transformation has the renormalon structure of the form of the the large-$\beta_0$ Borel transform of the Adler function , i.e., in terms of single or multiple poles (and not noninteger-multiplicity poles)\be
{\cal B}[\td](u; \kappa) \sim \frac{1}{(2-u)}, \frac{1}{(3-u)^2},  \frac{1}{(3-u)^1}, \frac{1}{(1+u)^2}, \; {\rm etc.}
\label{Btdstr} \ee
The ansatz as made in Ref.~\cite{renmod} includes these leading renormalon singularities, as well as the ``zero''-multiplicity $u=2$ infrared renormalon singularity $\sim \ln(1 - u/2)$:
\bea
{\cal B}[\td](u) & = & \exp \left( \tK u \right) \pi {\Big \{}
\td_{2,1}^{\rm IR} \left[ \frac{1}{(2-u)} + \tal (-1) \ln \left( 1 - \frac{u}{2} \right) \right] + \frac{ \td_{3,2}^{\rm IR} }{(3 - u)^2} + \frac{ \td_{3,1}^{\rm IR} }{(3 - u)} + \frac{ \td_{1,2}^{\rm UV} }{(1 + u)^2} {\Big \}},
\label{Btd5P}
\eea
where $\kappa=1$ and the values of the parameters, for the $\MSbar$ scheme case, are
\bea
\tK &=& 0.5160; \qquad \td_{2,1}^{\rm IR}=1.10826; \qquad {\widetilde \alpha}=-0.255;
\nonumber\\
\td_{3,2}^{\rm IR}&=&-0.481538; \qquad \td_{3,1}^{\rm IR}=-0.511642; \qquad
\td_{1,2}^{\rm UV}=-0.0117704.
\label{paramrenmod}
\eea
We refer to Ref.~\cite{renmod} for details on how these parameter values were obtained. We point out that the model, by construction, reproduces the first four exactly known expansion coefficients (${\td}_0=1$, ${\td}_j$, $j=1,2,3$). Further, the next unknown expansion coefficient (at $\kappa=1$, in $\MSbar$) is predicted to be $d_4=338.19$ ($\td_4=37.77$); this prediction comes from consideration of this approach in the lattice-related MiniMOM scheme\footnote{The MiniMOM $\beta$-function has been evaluated up to four-loops \cite{MiniMOM,BoucaudMM,CheRet}. In the work \cite{AKGCR} it has been shown that the MiniMOM scheme, in the Landau gauge, respects the $\beta$-function factorisation property of the conformal symmetry breaking contribution to the generalised Crewther relation.}
where the number of adjustable parameters was one less (i.e., without the term $\propto \td_{3,1}^{\rm IR}$).\footnote{The effective charge (ECH) method \cite{ECH} gives the estimate $(d_4)_{\rm ECH}=275$ \cite{KatStar,BCK}. The estimate of Ref.~\cite{BJ} is $0 < d_4 < 642$ [their preferred value is: $d_4=283$]. Recent estimates based on Pad\'e approximants give  $d_4=277 \pm 51$ \cite{Boitoetal}, and on conformal mappings in the Borel plane give $d_4=287 \pm 40$ \cite{Caprini2019}. In Refs.~\cite{Pich2,Pich} the estimate $d_4=275 \pm 400$ was used. We will use for the uncertainty of $d_4$ the range $d_4=338.19 \pm 338.19$.}

The value of the parameter ${\widetilde \alpha}$ was determined on the basis of the knowledge of the subleading Wilson coefficient ${\hat c}_1^{D=4}$ of the $D=4$ condensate. On the other hand, the values of the other five parameters in the ansatz (\ref{Btd5P}) were fixed by the knowledge of the first five perturbation coefficients $d_j$ ($j=0,1,\ldots,4;$ where $d_4=338.19$).

The expression (\ref{Btd5P}) generates the coefficients $\td_k$ [cf.~Eq.~(\ref{Btdexp})]. Then, the coefficients $d_n$ are obtained via the relations (\ref{dntdk}). As shown in Ref.~\cite{renmod}, the coefficients $d_n$ obtained in this way then lead to the following Borel transform of the quantity $d(Q^2)_{D=0}$ Eq.~(\ref{dpt})
\bea
{\cal B}[d](u;\kappa=e^{-\tK}) & = & \pi {\Bigg \{} \frac{d_{2,1}^{\rm IR}}{(2 - u)^{\tg_2}} \left[ 1 + \frac{(b_1^{(4)}+{\cal C}_{1,1}^{(4)})}{\beta_0 (\tg_2-1)} (2-u) + \frac{(b_2^{(4)}+{\cal C}_{1,1}^{(4)} b_1^{(4)} +{\cal C}_{2,1}^{(4)})}{ \beta_0^2 (\tg_2-1)(\tg_2-2)} (2-u)^2 + \ldots \right]
\nonumber\\ &&
+ \frac{d_{2,1}^{\rm IR} \alpha}{(2 - u)^{\tg_2-1}}  \left[ 1 + \frac{(b_1^{(4)}+{\cal C}_{1,0}^{(4)})}{\beta_0 (\tg_2-2)} (2-u) + \ldots \right]
\nonumber\\ &&
+ \frac{d_{3,2}^{\rm IR}}{(3 - u)^{\tg_3+1}} \left[ 1 + \frac{(b_1^{(6)}+{\cal C}_{1,2}^{(6)})}{\beta_0 \tg_3} (3-u) + \frac{(b_2^{(6)}+{\cal C}_{1,2}^{(6)} b_1^{(6)} +{\cal C}_{2,2}^{(6)})}{ \beta_0^2 \tg_3 (\tg_3-1)} (3 - u)^2 + \ldots \right]
\nonumber\\ &&
+ \frac{d_{3,1}^{\rm IR}}{(3 - u)^{\tg_3}} \left[ 1 + \frac{(b_1^{(6)}+{\cal C}_{1,1}^{(6)})}{\beta_0 (\tg_3-1)} (3-u) + \ldots \right]
\nonumber\\ &&
+ \frac{d_{1,2}^{\rm UV}}{(1+ u)^{\bg_1+1}} \left[ 1 + \frac{(b_1^{(-2)}+{\cal C}_{1,2}^{(-2)})}{(-\beta_0) \bg_1} (1+u) + \frac{(b_2^{(-2)}+{\cal C}_{1,2}^{(-2)} b_1^{(-2)} +{\cal C}_{2,2}^{(-2)})}{ (-\beta_0)^2 \bg_1 (\bg_1-1)} (1+ u)^2 + \ldots \right] {\Bigg \}},
\label{Bd5P}
\eea
where the main beyond-one-loop effects are contained in the coefficients $\tg_p$ and $\bg_p$
\be
\tg_p \equiv 1 + p \frac{c_1}{\beta_0}, \quad \bg_p \equiv  1 - p \frac{c_1}{\beta_0}, \quad (p=1,2,3),
\label{tgbg}
\ee
the coefficients $b_j^{(D)}$ are
\bes
\label{bjDs}
\bea
b_1^{(D)} &=& \frac{D}{2 \beta_0} ( c_1^2 - c_2),
\label{b1D}
\\
b_2^{(D)} &=& \frac{1}{2} (b_1^{(D)})^2 - \frac{D}{4 \beta_0} ( c_1^3 - 2 c_1 c_2 + c_3).
\label{b2D}
\eea
\ees
We recall that $c_j \equiv \beta_j/\beta_0$.
The numerical values of the coefficients ${\cal C}^{(D)}_{j,k}$ and the residue ratios $d^{\rm X}_{p,k}/\td^{\rm X}_{p,k}$ are given here in Table \ref{tabCd} (cf.~Ref.~\cite{renmod}).
\begin{table}
  \caption{The coefficients ${\cal C}^{(D)}_{j,k}$ and the residue ratios $d^{\rm X}_{p,k}/\td^{\rm X}_{p,k}$ in the $\MSbar$ scheme (${\bar c}_j=0$ for $j \geq 5$), with $N_f=3$. Note that $D=2 p$ for IR renormalons, and $D=-2 p$ for UV renormalons.}
\label{tabCd}
\begin{ruledtabular}
\begin{tabular}{l|rrr}
  type &  ${\cal C}^{(D)}_{1,k}$ & ${\cal C}^{(D)}_{2,k}$ & $d_{p,k}^{\rm X}/\td_{p,k}^{\rm X}$
\\
\hline
X=IR, $p=2, {\rm SP}(k=1)$ &  $(-0.03 \pm 0.02)$ & $(+1.7 \pm 0.3)$ & $(+1.7995 \pm 0.0001)$ 
\\
X=IR, $p=2, {\rm SL}(k=0)$ &  $(+7.7 \pm 0.4)$ & $\dots$ & $(+1.155 \pm 0.005)$  
\\
X-=IR, $p=2, {\rm SSL}(k=-1)$ & $(+17. \pm 1.)$ & $\ldots$ & $(2.00 \pm 0.002)$
\\
\hline
X=IR, $p=3, {\rm DP}(k=2)$ & $(-7.2 \pm 1.2)$ & $(+20. \pm 9.)$ & $(+29.7 \pm 0.8)$ 
\\
X=IR, $p=3, {\rm SP}(k=1)$ & $(-0.07 \pm 0.06)$ & $(+3.0 \pm 0.8)$ & $(+9.03 \pm 0.01)$ 
\\
\hline
X=UV, $p=1, {\rm DP}(k=2)$ & $(-10.1 \pm 2.1)$ & $(-83. \pm 8.)$ & $(+1.056 \pm 0.014)$
\\
X=UV, $p=1, {\rm SP}(k=1)$ & $(0.0 \pm 0.0)$ & $(+0.5. \pm 0.1)$ & $(+5.0098 \pm 0.0001)$
\end{tabular}
\end{ruledtabular}
\end{table}
Further, the coefficient $\alpha$ is
\be
\alpha =\frac{ ({\hat c}_1^{(4)} - {\cal C}_{1,1}^{(4)}) }{\beta_0 (\tg_2 -1)},
\label{al}
\ee
where ${\hat c}_1^{(4)}=(7/6)-c_1$ ($=-11/18$ when $N_f=3$) is the known subleading Wilson coefficient of the $D=4$ condensate of the V+A channel Adler function [${\widetilde \alpha} = \alpha (d_{2,1}^{\rm IR}/\td_{2,1}^{\rm IR}) (d_{2,0}^{\rm IR}/\td_{2,0}^{\rm IR})^{-1}$]. The result (\ref{Bd5P}) was obtained from the expression (\ref{Btd5P}) to a large precision, by generating first the coefficients $d_n$ from the coefficients $\td_k$ via the relations (\ref{dntdk}) and going up to high $n$ ($n_{\rm max}=70$). It is interesting that the form of the expression (\ref{Bd5P}) is also expected by the arguments of the theory of renormalons.

We can interpret the transition from the coefficients $\td_k$ to the coefficients $d_n$ [the relation (\ref{dntdk})], or equivalently, the transition from the Borel transform Eq.~(\ref{Btd5P}) of the quantity $\td(Q^2;\kappa)_{D=0}$ to the Borel transform Eq.~(\ref{Bd5P}) of the quantity $d(Q^2)_{D=0}$, as a procedure of ``dressing'' with the beyond-one-loop effects. Nonetheless, we point out that the coefficients $\td_k$ contain all the information about the quantity $d(Q^2)_{D=0}$ (to all loop levels), because they are in one-to-one correspondence with the coefficients $d_n$, cf.~Eq.~(\ref{dntdk}). This in spite of the fact that the Borel transform Eq.~(\ref{Btd5P}) of the quantity $\td(Q^2;\kappa)_{D=0}$ behaves under the variation of the renormalisation scale parameter $\kappa$ as if it were the Borel transform of the  quantity $d(Q^2)_{D=0}$ in the one-loop approximation, cf.~Eq.~(\ref{Btdkappa}).

\begin{table}
  \caption{The $\MSbar$ coefficients $\td_n$ and $d_n$ (with $\kappa=1$) of the considered renormalon-motivated Adler function extension: the coefficients $\td_n$ are generated by the Borel transform Eq.~(\ref{Btd5P}) of the extended auxiliary quantity $\td(Q^2;\kappa)$ Eq.~(\ref{tdpt}), and the coefficients $d_n$ are then generated by the relations (\ref{dntdk}). The values of the first four coefficients ($n=0,1,2,3$) coincide with the exactly known values. See the text for details.}
\label{tabtdndn}
\begin{ruledtabular}
\begin{tabular}{r|rr|rrr}
 $n$ & $\td_n$ & $d_n$ & $\td_n/((n+1)! (-\beta_0)^n)$  & $d_n/J(n)^{(0)}$ &  $d_n/J(n)^{(1)}$
\\
\hline
0       &  1       & 1         &  1  & 1.09217  & 0.0657431 \\
1       &  1.63982 & 1.63982   &  -0.364405  & -0.657905   & -0.177413 \\
2       &  3.45578 & 6.37101 &  0.11377  &  0.514073   & 0.207056 \\
3       &  26.3849 & 49.0757 &  -0.0965156  & -0.548292  & -0.27132 \\
4       & 37.7719 & 338.19 &  0.0122817   & 0.398891  & 0.224305 \\
5       &  1732.04 & 3799.99 & -0.0417171 & -0.382355  & -0.234725 \\
6       & -9949.19 & 29672.9 & -0.0152147 & 0.213687  & 0.139878 \\
7       & 322129. & 465315. & -0.0273673 & -0.206564 &  -0.142019 \\
8       & $-5.1117 \times 10^{6}$ & $3.21051 \times 10^{6}$ & -0.0214458  & 0.0771547 &  0.0551451 \\
9       & $1.28702 \times 10^{8}$ & $8.8993 \times 10^{7}$ & -0.0239983 & -0.103207  & -0.0761233 \\
10      & $-3.00623 \times 10^{9}$ & $1.7999 \times 10^{8}$ & -0.0226486 & 0.0090865  & 0.00687883 \\
11      & $8.29374 \times 10^{10}$ & $2.86115 \times 10^{10}$ & -0.0231423 & -0.0572673  & -0.044314 \\
12      & $-2.38986 \times 10^{12}$ & $-2.42769 \times 10^{11}$ & -0.0227982 & -0.0176875  &-0.0139451 \\
\hline
20      & $-1.27389 \times 10^{25}$ & $-1.50938 \times 10^{24}$ & -0.0225496  & -0.0297333 & -0.025585 \\
\hline
30      & $-6.7647 \times 10^{42}$ & $-5.72077 \times 10^{41}$ & -0.0223744 & -0.0286077 & -0.0258083 \\
\hline
40      & $-9.11451 \times 10^{61}$ & $-6.06045 \times 10^{60}$ & -0.0222845 & -0.0279225 & -0.0258185 \\
\hline
50      & $-1.40187 \times 10^{82}$ & $-7.74379 \times 10^{80}$ & -0.0222299 & -0.0274841 & -0.0258003 \\
\hline
60      & $-1.52292 \times 10^{103}$ & $-5.72077 \times 10^{41}$ &-0.0221932 & -0.0271776 & -0.0257749 \\
\hline
70      & $-8.47507 \times 10^{124}$ & $-3.54627 \times 10^{123}$ & -0.0221668 & -0.0269503 & -0.0257486 
\end{tabular}
\end{ruledtabular}
\end{table}
In Table \ref{tabtdndn} we present the values of some of the coefficients $\td_n$ and $d_n$ (for $\kappa=1$). In the Table we include the ratios $\td_n/((n+1)! (-\beta_0)^n)$ and $d_n/J(n)^{\rm (X)}$, where $J(n)^{\rm (X)}$ (X=0, 1) describe the leading ($\sim 1$) or next-to-leading (up to $\sim 1/n$) asymptotic behaviour factor of $d_n$ as follows from the expression containing the UV ($u=-1$) renormalon contribution [i.e., the contribution to $d_n$ from the term containing $d_{1,2}^{\rm UV}$ in Eq.~(\ref{Bd5P})]
\bes
\label{Jn}
\bea
J(n)^{(0)} &=&  \Gamma(\bg_1+1+n) (-\beta_0)^n,
  \label{JnLO}
  \\
  J(n)^{(1)} &=&  \Gamma(\bg_1+1+n) (-\beta_0)^n \left[1 + \left( b_1^{(-2)} + {\cal C}^{(-2)}_{1,2} \right) \frac{1}{(-\beta_0)} \frac{1}{(\bg_1 + n)} \right].
\label{JnNLO}
\eea
\ees
We see from the Table that the two ratios $\td_n/((n+1)! (-\beta_0)^n)$ and  $d_n/J(n)^{(1)}$ converge to specific values at large $n$ (approximately to $-0.0221$ and $-0.0257$, respectively), which confirms that the $p=1$ UV renormalon contribution is really the dominant contribution to these coefficients at large $n$. The ratio involving $d_n$ in the last column converges even faster if we included the terms ${\cal O}(1/n^2)$ in the asymptotic form (i.e., $d_n/J^{(2)}(n)$).

We point out that the starting point for the construction of the higher order coefficients $d_n$ ($n \geq 4$) of an extended Adler function in \cite{renmod} (and here) was not the Borel transform of the ($D=0$) Adler function, ${\cal B}[d](u)$, but a renormalon-motivated ansatz for the Borel transform  of the auxiliary quantity $\td$,  ${\cal B}[\td](u)$, which has a particularly simple strucure of poles with integer multiplicity, Eq.~(\ref{Btd5P}). On the other hand, the works \cite{BJ,BJ2} construct and use a renormalon-motivated  ansatz for the Borel transform ${\cal B}[d](u)$ (which has poles of noninteger multiplicity) in order to generate the higher order coefficients $d_n$. The authors of \cite{CF1,CF2,AACF,Caprini2019,Caprini2020} generated the higher order coefficients $d_n$ by a combination of a renormalon-motivated ansatz for ${\cal B}[d](u)$ and application of an optimal conformal mapping in the Borel plane. For a classical review on renormalons, we refer to \cite{ren}, and for some recent developments on the subject of renormalons we refer to \cite{Maiezza,Cavalc,Pineda1,Pineda2}.

\section{Methods of evaluation of Adler function, extraction of $\alpha_s$}
\label{sec:extr}

In the previous Section we described the renormalon-motivated extension of the known truncated perturbation series for the Adler function $d(Q^2)_{(D=0)}$. We will use various methods of evaluation of this function in the sum rule approach described in Sec.~\ref{sec:SRgen}, and will use various weight functions $g(Q^2)$ in the sum rules Eq.~(\ref{Cauchy}). In the analysis, we will use the ALEPH experimental data, and will extract the corresponding values of the ($\MSbar$) QCD coupling $\alpha_s(m_{\tau}^2)$.

\subsection{Weight functions for sum rules}
\label{subs:weight}

In order to suppress significantly the duality violation effects, most of the chosen weight functions $g(Q^2)$ will have double zero (double pinch) at the Minkowskian end $Q^2=-\sigma_{\rm max}$ ($\equiv -\sm$) where the OPE expansion is not expected to work \cite{Pich} (cf.~also \cite{BNP92,DP92,Chib,Malt,DomSch,Cir,GonzAl,Dom,RSan}).

We will consider the FESRs with moments $a^{(2,n)}$ associated with the following weight functions $g^{(2,n)}$ ($n=0,1,2,\ldots$):
\bes
\label{a2n}
\bea
g^{(2,n)}(Q^2) & = & \left( \frac{n+3}{n+1} \right) \frac{1}{\sm} \left( 1 + \frac{Q^2}{\sm} \right)^2 \sum_{k=0}^n (k+1) (-1)^k \left( \frac{Q^2}{\sm} \right)^k
\nonumber\\
& = & \left( \frac{n+3}{n+1} \right) \frac{1}{\sm} \left[ 1 - (n+2) \left(- \frac{Q^2}{\sm} \right)^{n+1} + (n+1)  \left(- \frac{Q^2}{\sm} \right)^{n+2} \right] \quad \Rightarrow
\label{g2n} \\
G^{(2,n)}(Q^2) & = &   \left( \frac{n+3}{n+1} \right) \frac{Q^2}{\sm} \left[ 1 - \left( - \frac{Q^2}{\sm} \right)^{n+1} \right] + \left[ 1 - \left( - \frac{Q^2}{\sm} \right)^{n+3} \right], 
\label{G2n} \\
a^{(2,n)}_{\rm exp}(\sm) & = & \int_0^{\sm} d \sg \; g^{(2,n)}(-\sg) \omega_{\rm (exp)}(\sg) - 1,
\label{a2nexp} \\
a^{(2,n)}_{\rm th}(\sm) & = &  \frac{1}{2 \pi} \int_{-\pi}^{+\pi} d \phi \;
G^{(2,n)} \left(\sm e^{i \phi} \right) d \left( \sm e^{i \phi} \right)_{(D=0)}
\nonumber\\
&& + \left( \frac{n+3}{n+1} \right) 2 \pi^2 (-1)^n \left\{ (n+2) \frac{\langle O_{2 n + 4} \rangle}{\sm^{n+2}} +  (n+1) \frac{\langle O_{2 n + 6} \rangle}{\sm^{n+3}} \right\}.
\label{a2nth}
\eea
\ees
We recall that we assume that $\langle O_D \rangle$ are $Q^2$-independent.
The weight function $G^{(2,n)}(Q^2)$ is related with $g^{(2,n)}(Q^2)$ via the relation (\ref{GQ2}), and the theoretical expression (\ref{a2nth}) represents the right-hand side of the sum rule (\ref{sr2}) (minus unity) where for the entire Adler function ${\cal D}_{\rm (th)}(Q^2)$ the OPE Eq.~(\ref{DOPE}) was used, and $Q^2 \equiv \sm \exp(i \phi)$ ($-\pi \leq \phi < +\pi$) on the countour. The coefficient $(n+3)/(n+1)$ appearing in the weight functions was used so that the unity in the OPE expansion (\ref{DOPE}) of the Adler function gives exactly the unity in the contour integration on the right-hand side  of the sum rule (\ref{sr2}).

We will use the sum rules with the above moments for $n=5 \pm 1$, by assuming that the contributions of the high dimension condensates in Eq.~(\ref{a2nth}) are negligible. This will allow us to extract the values of $\alpha_s(m_{\tau}^2)$ without consideration of these condensates.

In addition to these sum rules, we will consider the sum rules with the (double-pinched) Borel-Laplace transforms $B(M^2)$ (where $M$ is a complex scale parameter)
\bes
\label{BL}
\bea
g_{M^2}(Q^2) &=&  \left( 1 + \frac{Q^2}{\sm} \right)^2  \frac{1}{M^2} \exp \left( \frac{Q^2}{M^2} \right) \quad \Rightarrow
\label{gM2} \\
G_{M^2}(Q^2) & = & \left\{  \left[  \left( 1 + \frac{Q^2}{\sm} \right)^2 - 2 \frac{M^2}{\sm}  \left( 1 + \frac{Q^2}{\sm} \right) + 2 \left( \frac{M^2}{\sm} \right)^2 \right] \exp \left( \frac{Q^2}{M^2} \right) - 2  \left( \frac{M^2}{\sm} \right)^2 \exp \left( - \frac{\sm}{M^2} \right) \right\},
\label{GM2} \\
B_{\rm exp}(M^2;\sm) & = & \int_0^{\sm} d \sg \; g_{M^2}(-\sg) \omega_{\rm (exp)}(\sg)
= \frac{1}{M^2} \int_0^{\sm} d \sg \; \left( 1 - \frac{\sg}{\sm} \right)^2  \exp \left( -\frac{\sg}{M^2} \right) \omega_{\rm (exp)}(\sg),
\label{Bexp}
\\
B_{\rm th}(M^2; \sm) &=&   \frac{1}{2 \pi} \int_{-\pi}^{+\pi} d \phi \;
  G_{M^2} \left(\sm e^{i \phi} \right) {\cal D}_{\rm (th)} \left( \sm e^{i \phi} \right)
  \nonumber\\ &&
\!\!\!\!\!\!\!\!\!\!\!\!\!\!\!\!\!\!\!\!\!\!\!\!\!\!
  =  \left[ \left( 1 - 2 \frac{M^2}{\sm} \right) + 2 \left( \frac{M^2}{\sm} \right)^2 \left(1 - \exp \left( - \frac{\sm}{M^2} \right) \right) \right]  
  \nonumber\\ &&
\!\!\!\!\!\!\!\!\!\!\!\!\!\!\!\!\!\!\!\!\!\!\!\!\!\!\!  
+ \frac{1}{2 \pi}  \int_{-\pi}^{+\pi} d \phi \left\{ \left[ \left(1 + e^{i \phi} \right)^2 - 2 \frac{M^2}{\sm}  \left(1 + e^{i \phi} \right) +  2 \left( \frac{M^2}{\sm} \right)^2 \right] \exp \left( \frac{\sm}{M^2} e^{i \phi} \right) -  2 \left( \frac{M^2}{\sm} \right)^2 \exp \left( - \frac{\sm}{M^2} \right) \right\} d \left( \sm e^{i \phi} \right)_{(D=0)}
  \nonumber\\ &&
\!\!\!\!\!  + \sum_{k \geq 2} B_{\rm th}(M^2; \sm)_{(D=2 k)},
\label{Bth}
\eea
\ees
where in Eq.~(\ref{Bth}) the dimension $D=2 k$ condensates contribute to the Borel-Laplace
\be
B_{\rm th}(M^2; \sm)_{(D=2 k)} = \frac{2 \pi^2}{(k-1)!} \frac{ \langle O_{2 k} \rangle}{(M^2)^k}
\left[ 1 + 2 (k-1) \frac{M^2}{\sm} + (k-1) (k-2) \left(\frac{M^2}{\sm}\right)^2 \right].
\label{Bth2k}
\ee
In our application of these (double-pinched) Borel-Laplace sum rules, we will use only the real part of the Borel-Laplace, ${\rm Re} B(M^2; \sm)$, because in this way the $D=4$ condensate contribution dominates over the $D=6$ condensate contribution when $M^2$ varies along the ray $M^2 = |M^2| \exp(i \pi/6)$, and $D=6$ dominates over $D=4$ when $M^2=|M^2| \exp(i \pi/4)$.\footnote{When the Borel-Laplace is not pinched, then ${\rm Re}  B(M^2; \sm)$ is completely independent of $\langle O_6 \rangle$ for  $M^2 = |M^2| \exp(i \pi/6)$, and  completely independent of $\langle O_4 \rangle$ for  $M^2 = |M^2| \exp(i \pi/4)$, cf.~Ref.~\cite{3dAQCD}.}
Further, while including the (small) $D=8$ contribution, we will neglect higher dimension contributions
\bea
\sum_{k=2}^4 B_{\rm th}(M^2; \sm)_{(D=2 k)} & = & 2 \pi^2 {\bigg [} \frac{1}{M^2} \frac{1}{\sm} \left( 2 \langle O_4 \rangle + \frac{ \langle O_6 \rangle}{\sm} \right) + \frac{1}{(M^2)^2} \left( \langle O_4 \rangle + 2 \frac{ \langle O_6 \rangle}{\sm} + \frac{ \langle O_8 \rangle}{\sm^2} \right)
\nonumber\\
&& + \frac{\sm}{2 (M^2)^3} \left( \frac{ \langle O_6 \rangle}{\sm}  +2 \frac{ \langle O_8 \rangle}{\sm^2} \right) + \frac{\sm^2}{6 (M^2)^4} \frac{\langle O_8 \rangle}{\sm^2} {\bigg ]}.
\label{BthD468} \eea
In general, while smaller scales $|M^2|$ tend to minimize the duality violations they make the (higher) condensate contributions larger \cite{Pich}. Further, larger values of $|M^2|$ lead to large experimental uncertainties [$\omega_{\rm (exp)}(\sg)$ has larger uncertainties at large $\sg$]. We consider as a reasonable range
\be
0.9 \ {\rm GeV}^2 \leq |M^2| \leq 1.5 \ {\rm GeV}^2,
\label{rangeM2} \ee
and we will use this range in our fits. Further, for $\Psi \equiv {\rm Arg}(M^2)$ we will use the rays with $\Psi=0, \pi/6$ and $\pi/4$ [we recall that, along the last two rays, specific condensate contributions are suppressed in ${\rm Re} B_{\rm th}(M^2; \sm)$].

\subsection{Evaluation methods for the Adler function $d(Q^2)_{(D=0)}$}
\label{subs:evalm}

\subsubsection{Fixed-order (FOPT)}
\label{ssubs:FO}

The basis of the known fixed-order (FOPT) approach is the application of the Taylor expansion to the $D=0$ (leading-twist) part $d(Q^2)_{(D=0)}$ of the Adler function ${\cal D}_{\rm (th)}(Q^2)$ Eq.~(\ref{DOPE}) on the contour $Q^2=\sm \exp(i \phi)$ on the right-hand side of the sum rule (\ref{sr2})
\be
d(\sm e^{i \phi})_{(D=0)} =  d(\sm)_{(D=0)} + i \phi \frac{d}{d \ln Q^2} d(Q^2)_{(D=0)} {\big |}_{Q^2=\sm} + \ldots \frac{1}{k!} ( i \phi)^k \left( \frac{d}{d \ln Q^2} \right)^k d(Q^2)_{(D=0)} {\big |}_{Q^2=\sm} + \ldots
\label{dTay1} \ee
When appying this Taylor expansion to the expansion (\ref{dlpt}) of  $d(Q^2)_{(D=0)}$ in logarithmic detivatives ${\ta}_{n+1}(\kappa Q^2)$, and using the identity
\be
\left(  \frac{d}{d \ln Q^2} \right)^k  \ta_{n+1}(\kappa Q^2) = (-\beta_0)^k \frac{(n+k)!}{n!} {\ta}_{n+k+1}(\kappa Q^2),
\label{dkta} \ee
which is a direct consequence of the definition (\ref{tan}), we obtain
\be
d(\sm e^{i \phi})_{(D=0)} =  a(\kappa \sm) + {\td}_1(\phi;\kappa) \ta_2(\kappa \sm) + \ldots
+  {\td}_n(\phi;\kappa) \ta_{n+1}(\kappa \sm) + \ldots,
\label{dTay2}
\ee
where the $\phi$-dependent expansion coefficients ${\td}_n(\phi;\kappa)$ are the following combination of the coefficients ${\td}_n(\kappa)$ of the expansion (\ref{dlpt}) of $d(Q^2)_{(D=0)}$:
\be
{\td}_n(\phi;\kappa) = \sum_{k=0}^{n}  \binom{n}{k} (- i \phi \beta_0)^k {\td}_{n-k}(\kappa), \qquad ({\td}_0(\kappa)= 1).
\label{tdnphi} \ee
We note that ${\td}_n(\phi=0;\kappa) = {\td}_n(\kappa)$.
When inserting the expansion (\ref{dTay2}) in the right-hand side of the $D=0$ part of the sum rule (\ref{sr2}), where $Q^2=\sm \exp(i \phi)$  ($-\pi \leq \phi < \pi$), we obtain the expression
\bea
\left( -\frac{i}{2 \pi}   \oint_{|Q^2|=\sm}
\frac{d Q^2}{Q^2} d(Q^2)_{(D=0)} G(Q^2) \right)^{(\widetilde{\rm FO},[N])} & = & {\tr}_0^{(G)} a(\kappa \sm) + {\tr}_1^{(G)}(\kappa) {\ta}_2(\kappa \sm) + \ldots + {\tr}_{N-1}^{(G)}(\kappa) {\ta}_{N}(\kappa \sm) ,
\label{sr2tFO} \eea
where the expansion coefficients ${\tr}_n^{(G)}(\kappa)$ are
\be
 {\tr}_n^{(G)}(\kappa) =  
\sum_{\ell=0}^{n}  \binom{n}{n - \ell} \beta_0^{n-\ell} {\cal K}_{n - \ell}^{(G)}(\sm) {\td}_{\ell}(\kappa),
\label{trn} \ee
where ${\td}_0(\kappa)=1$ and the coefficients ${\cal K}_{n - \ell}^{(G)}(\sm)$ are the following contour integrals:
\be
{\cal K}_{k}^{(G)}(\sm) = \frac{1}{2 \pi} \int_{-\pi}^{\pi} d \phi (- i \phi)^k G(\sm e^{i \phi}).
\label{calKk} \ee
The expansion (\ref{sr2tFO}) represents the FOPT expansion in logarithmic derivatives $\ta_{n+1}(\kappa Q^2)$ at $Q^2=\sm$, which we denoted as  $(\widetilde{{\rm FOPT}})$. This expansion involves in practice a truncation, say at ${\ta}_N(\kappa \sm)$, which we will denote with the superscript $(\widetilde{{\rm FO}}, [N])$. This expansion can be reorganised in terms of powers $a(\kappa \sm)^k$ using the relations (\ref{tanan}), and then truncating at the power $a(\kappa \sm)^N$; this represent the usual FOPT approach, and we will denote it with the superscript (FO), and its truncated version with $({\rm FO}; [N])$
\bea
\left( -\frac{i}{2 \pi}   \oint_{|Q^2|=\sm}
\frac{d Q^2}{Q^2} d(Q^2)_{(D=0)} G(Q^2) \right)^{({\rm FO}, [N])} & = &
     r_0^{(G)} a(\kappa \sm) + r_1^{(G)}(\kappa) a(\kappa \sm)^2 + \ldots + r_{N-1}^{(G)}(\kappa) a(\kappa \sm)^{N},
  \label{sr2FO} \eea
 where 
\be
r_n(\kappa) = \tr_n + \sum_{s=1}^{n-1} k_s(n+1-s) \; \tr_{n-s}(\kappa) \quad
(n=0, 1,2, \ldots),
\label{rntrk} \ee
in complete analogy with the relations (\ref{dntdk}).
We point out that, while the approaches $(\widetilde{{\rm FOPT}})$ and FOPT give in principle equal results, in practice it is not so due to the truncation. Namely, both types of series are divergent due to the renormalon-dominated growth of ${\tr}_n$ and $r_n$ coefficients when $n$ increases; a truncation is needed (at $\ta_{N}$ and $a^N$, respectively), which then gives somewhat different results.

\subsubsection{Contour-improved (CIPT)}
\label{ssubs:CI}

The contour-improved method is represented by the direct integration along the contour of the integrand $d(Q^2)_{(D=0)} G(Q^2)$ on the right-hand side of the sum rule (\ref{sr2}) where $d(Q^2)_{(D=0)}$ has the form of the perturbation series (\ref{dpt}) truncated at a specific power $a(\kappa Q^2)^N$
\bea
\left( -\frac{i}{2 \pi}   \oint_{|Q^2|=\sm}
\frac{d Q^2}{Q^2} d(Q^2)_{(D=0)} G(Q^2) \right)^{({\rm CI},[N])} & = &
\frac{1}{2 \pi} \int_{-\pi}^{\pi} d \phi \; d(\sm e^{i \phi}; \kappa)^{[N]}_{(D=0); {\rm pt}}  G(\sm  e^{i \phi}),
\label{sr2CI} \eea
where $d^{[N]}_{(D=0); \rm pt}$ is the truncated series
\be
d(Q^2; \kappa)^{[N]}_{(D=0), {\rm pt}}= a(Q^2) + d_1(\kappa) \; a(\kappa Q^2)^2 + \ldots + d_n(\kappa) \; a(\kappa Q^2)^{N}.
\label{dNpt} \ee
Here, the renormalisation parameter $\kappa$-dependence appears because of truncation. 

\subsubsection{Principal-Value (PV)}
\label{ssubs:PV}

In this approach, the Adler function is evaluated as the Principal Value (PV) of the inverse Borel transformation
\be
d(Q^2)_{(D=0),{\rm PV}} = \frac{1}{\beta_0} \frac{1}{2} \left( \int_{{\cal C}_{+}} + \int_{{\cal C}_{-}} \right) d u \exp \left[ - \frac{u}{\beta_0 a(\kappa Q^2)} \right] {\cal B}[d](u; \kappa),
\label{PV1}
\ee
where the paths ${\cal C}_{\pm}$ go from $u=0$ to $u=+\infty$ within the upper and lower half of the complex $u$-plane; application of the Cauchy theorem shows that the details of these two paths are irrelevant, because the Borel transform $ {\cal B}[d](u; \kappa)$ has singularities only along the real axis. For example, we can choose for ${\cal C}_{+}$ the path going as straight line from $u=0$ to $u=i \varepsilon$ (for any $\varepsilon >0$) and then parallel to the real axis from $u=+i \varepsilon$ to $+\infty + i \varepsilon$. Another example is the path from $u=0$ along a ray $u=|u| \exp(i \phi_0)$ ($|u|$ from zero to $+\infty$) where $\phi_0$ is a fixed angle, $0 < \phi_0 < \pi/2$. We note that in the integration (\ref{PV1}) in the sum rules, the value of $Q^2$ is in general complex nonreal [$Q^2 = \sm \exp(i \phi)$].

In applying the integration (\ref{PV1}), we use for the Borel transform ${\cal B}[d](u; \kappa)$ the expression (\ref{Bd5P}), but now at a given general $\kappa$-parameter value, and truncated. In practice, this truncation requires to include a polynomial correction form 
\be
{\delta d}(Q^2)(Q^2; \kappa)^{[N]}_{(D=0)} = {\delta d}_0(\kappa) a(\kappa Q^2) + \ldots + {\delta d}_{N-1}(\kappa) a(\kappa Q^2)^N,
\label{deld}
\ee
so that the full expansion coefficients $d_n(\kappa)$ of the Adler function are restored. The leading part of the renormalon growth of the coefficients $d_n(\kappa)$ is contained in the PV of the inverse Borel integral of the truncated singular transform ${\cal B}[d](u; \kappa)$. The latter transform, at a general $\kappa$, is obtained in the following way. The starting point is the Borel transform of the auxiliary quantity ${\td}(Q^2; \kappa)_{(D=0)}$, cf.~Eqs.~(\ref{Btd5P}) and (\ref{Btdkappa}), which can be written for the general case of the renormalisation scale parameter $\kappa$ as
\bea
{\cal B}[\td](u; \kappa)_{\rm sing} & = & \pi {\Bigg \{}
\left[ \frac{\td_{2,1}^{\rm IR}(\tkap)}{(2-u)} +  \td_{2,0}^{\rm IR}(\tkap) (-1) \ln \left( 1 - \frac{u}{2} \right) + \td_{2,-1}^{\rm IR}(\tkap) (2-u) \ln \left( 1 - \frac{u}{2} \right) \right]
\nonumber\\ &&
+ \left[ \frac{ \td_{3,2}^{\rm IR}(\tkap) }{(3 - u)^2}  + \frac{ \td_{3,1}^{\rm IR}(\tkap) }{(3 - u)}  \right]
+ \left[ \frac{ \td_{1,2}^{\rm UV}(\tkap) }{(1 + u)^2} + \frac{ \td_{1,1}^{\rm UV}(\tkap) }{(1 + u)} \right] {\Bigg \}},
\label{Btd5Pkap}
\eea
where
\bes
\label{tdkap}
\bea
\tkap & \equiv & \kappa \exp( \tK ) \; (\approx 1.68 \times \kappa) ,
\label{tkap}
\\
\td_{2,1}^{\rm IR}(\tkap) & = & \tkap^2 \td_{2,1}^{\rm IR}, \quad
\td_{2,0}^{\rm IR}(\tkap) = \tal \td_{2,1}^{\rm IR}(\tkap) , \quad
\td_{2,-1}^{\rm IR}(\tkap) = (\ln \tkap)  \tal \td_{2,1}^{\rm IR}(\tkap),
\label{td2jIR} \\
\td_{3,2}^{\rm IR}(\tkap) & = & \tkap^3 \td_{3,2}^{\rm IR}, \quad
\td_{3,1}^{\rm IR}(\tkap) = \tkap^3 \left( \td_{3,1}^{\rm IR} - (\ln \tkap) \td_{3,2}^{\rm IR} \right),
\label{td3jIR} \\
\td_{1,2}^{\rm UV}(\tkap) & = & \frac{1}{\tkap} \td_{1,2}^{\rm UV}, \quad
\td_{1,1}^{\rm UV}(\tkap) = (\ln \tkap) \td_{1,2}^{\rm UV}(\tkap).
\label{td1jUV}
\eea
\ees
These relations are obtained by using the $\kappa$-dependence Eq.~(\ref{Btdkappa}) and the expression Eq.~(\ref{Btd5P}), performing the corresponding expansions of $\exp(u (\ln \kappa + \tK))$ around $u=2, 3, -1$, and ignoring the terms $\sim (2 -u)^2 \ln(1 - u/2)$, $(2-u)^0$, $(3-u)^0$, $(1+u)^0$. This truncation means that we include in the expression (\ref{Btd5Pkap}) only the singular contributions. We note that for $\tkap=1$ [$\kappa=\exp(-\tK)$] the values of the residues $\td_{p,k}^X(\tkap)$ reduce to the values $\td_{p,k}^X$ given in Eq.~(\ref{paramrenmod}). 

Then, according to the data of Table \ref{tabCd}, the corresponding truncated Borel transform of the Adler function is [in analogy with Eq.~(\ref{Bd5P})]
\bea
\frac{1}{\pi}{\cal B}[d](u;\kappa)_{\rm sing} & = &
{\Bigg \{}
\frac{d_{2,1}^{\rm IR}(\tkap)}{(2 - u)^{\tg_2}} \left[ 1 + \frac{(b_1^{(4)}+{\cal C}_{1,1}^{(4)})}{\beta_0 (\tg_2-1)} (2-u) + \frac{(b_2^{(4)}+{\cal C}_{1,1}^{(4)} b_1^{(4)} +{\cal C}_{2,1}^{(4)})}{ \beta_0^2 (\tg_2-1)(\tg_2-2)} (2-u)^2 \right]
\nonumber\\ &&
+ \frac{d_{2,1}^{\rm IR}(\tkap) \alpha}{(2 - u)^{\tg_2-1}}  \left[ 1 + \frac{(b_1^{(4)}+{\cal C}_{1,0}^{(4)})}{\beta_0 (\tg_2-2)} (2-u) \right]
+ \frac{d_{2,-1}^{\rm IR}(\tkap)}{(2 - u)^{\tg_2-2}}
{\Bigg \}}
\nonumber\\ &&
{\Bigg \{}
+ \frac{d_{3,2}^{\rm IR}(\tkap)}{(3 - u)^{\tg_3+1}} \left[ 1 + \frac{(b_1^{(6)}+{\cal C}_{1,2}^{(6)})}{\beta_0 \tg_3} (3-u) + \frac{(b_2^{(6)}+{\cal C}_{1,2}^{(6)} b_1^{(6)} +{\cal C}_{2,2}^{(6)})}{ \beta_0^2 \tg_3 (\tg_3-1)} (3 - u)^2 \right]
\nonumber\\ &&
+ \frac{d_{3,1}^{\rm IR}(\tkap)}{(3 - u)^{\tg_3}} \left[ 1 + \frac{(b_1^{(6)}+{\cal C}_{1,1}^{(6)})}{\beta_0 (\tg_3-1)} (3-u) \right]
{\Bigg \}}
\nonumber\\ &&
{\Bigg \{}
+ \frac{d_{1,2}^{\rm UV}(\tkap)}{(1+ u)^{\bg_1+1}} \left[ 1 + \frac{(b_1^{(-2)}+{\cal C}_{1,2}^{(-2)})}{(-\beta_0) \bg_1} (1+u) + \frac{(b_2^{(-2)}+{\cal C}_{1,2}^{(-2)} b_1^{(-2)} +{\cal C}_{2,2}^{(-2)})}{ (-\beta_0)^2 \bg_1 (\bg_1-1)} (1+ u)^2 \right]
\nonumber\\ &&
+ \frac{d_{1,1}^{\rm UV}(\tkap)}{(1+ u)^{\bg_1}} \left[ 1 + \frac{(b_1^{(-2)}+{\cal C}_{1,1}^{(-2)})}{(-\beta_0) (\bg_1-1)} (1+u) \right]
{\Bigg \}}.
\label{Bd5Pkap}
\eea
The truncation here consists of not including the terms of higher powers of $(p-u)$ [$(2-u)^{-\tg_2+3}$, $(3-u)^{-\tg_3+2}$ and $(1+u)^{-\bg_1+2}$ and higher].
It is this (singular) Borel transform contribution that we use in the evaluation of the PV of the inverse Borel integration, Eq.~(\ref{PV1}). We recall that the power indices $\tg_p$ and $\bg_p$ ($p=1,2,\ldots$) are given in Eqs.~(\ref{tgbg}).  The values of the parameters ${\cal C}_{j,k}^{(D)}$ and the ratios $d_{p,k}^X/{\td}_{p,k}^X$ are given in Table \ref{tabCd} (cf.~also Table II in Ref.~\cite{renmod}); the latter ratios are independent of the renormalisation scale parameter $\kappa$ (thus independent of $\tkap$). We will use the central values given in Table \ref{tabCd}. For example, $d_{2,1}^{\rm IR}(\tkap)=1.7995 \; \td_{2,1}^{\rm IR}(\tkap)$, etc.

The expression for the evaluation of the $D=0$ part of the Adler function in the described PV-approach thus acquires the form
\be
{\bigg (} d(Q^2)_{(D=0)} {\bigg )}^{({\rm PV}, [N])} = \frac{1}{\beta_0} \frac{1}{2} \left( \int_{{\cal C}_{+}} + \int_{{\cal C}_{-}} \right) d u \exp \left[ - \frac{u}{\beta_0 a(\kappa Q^2)} \right] {\cal B}[d](u; \kappa)_{\rm sing} + \delta d(Q^2; \kappa)_{(D=0)}^{[N]},
\label{PV2}
\ee
where the ${\cal B}[d](u; \kappa)_{\rm sing}$ is given in Eq.~(\ref{Bd5Pkap}) and the correction polynomial $\delta d(Q^2; \kappa)_{(D=0)}^{[N]}$ is given in Eq.~(\ref{deld}). In practice, it turns out that the correction polynomial values are large as are also the PV integral values in the expression (\ref{PV2}). However, the two terms for the $(D=0)$ Adler function give to the sum rules in general contributions with opposite sign, and the sum of the two terms there is smaller (by about two orders of magnitude) then each term. We point out that the correction polynomial expression (\ref{deld}) appears in the expression (\ref{PV2}) because in the sums in the singular part of the Borel transform, Eq.~(\ref{Bd5Pkap}), truncations were made.
To understand this more clearly, the Borel transform ${\cal B}[d](u;\kappa)$ Eq.~(\ref{Bd5Pkap}), but in its nontruncated form, implies that the coefficients $d_n(\kappa)$ have the form
\bea
\frac{d_n(\kappa)}{\pi} & = &
{\Bigg \{}
\frac{d_{2,1}^{\rm IR}(\tkap)}{2^{\tg_2} \Gamma(\tg_2)} \Gamma(\tg_2+n) \left( \frac{\beta_0}{2} \right)^n
\nonumber\\ && \times
\left[ 1 + (b_1^{(4)}+{\cal C}_{1,1}^{(4)}) \left( \frac{2}{\beta_0} \right) \frac{1}{(\tg_2-1+n)} + (b_2^{(4)}+{\cal C}_{1,1}^{(4)} b_1^{(4)} +{\cal C}_{2,1}^{(4)}) \left( \frac{2}{\beta_0} \right)^2 \frac{1}{(\tg_2-1+n)(\tg_2-2+n)} + {\cal O}\left( \frac{1}{n^3} \right) \right]
\nonumber \\ &&
+ \frac{d_{2,1}^{\rm IR}(\tkap) \alpha}{2^{\tg_2-1} \Gamma(\tg_2-1)} \Gamma(\tg_2-1+n)  \left( \frac{\beta_0}{2} \right)^n \left[ 1 + (b_1^{(4)}+{\cal C}_{1,0}^{(4)}) \left( \frac{2}{\beta_0} \right) \frac{1}{(\tg_2-2+n)} + {\cal O}\left( \frac{1}{n^2} \right) \right]
\nonumber \\ &&
+ \frac{d_{2,-1}^{\rm IR}(\tkap)}{2^{\tg_2-2} \Gamma(\tg_2-2)} \Gamma(\tg_2-2+n)  \left( \frac{\beta_0}{2} \right)^n \left[ 1 + {\cal O}\left( \frac{1}{n} \right) \right]
{\Bigg \}}
\nonumber \\ &&
{\Bigg \{}
+ \frac{d_{3,2}^{\rm IR}(\tkap)}{3^{\tg_3+1} \Gamma(\tg_3+1)} \Gamma(\tg_3+1+n) \left( \frac{\beta_0}{3} \right)^n
\nonumber\\ && \times
\left[ 1 + (b_1^{(6)}+{\cal C}_{1,2}^{(6)}) \left( \frac{3}{\beta_0} \right) \frac{1}{(\tg_3+n)} + (b_2^{(6)}+{\cal C}_{1,2}^{(6)} b_1^{(6)} +{\cal C}_{2,2}^{(6)}) \left( \frac{3}{\beta_0} \right)^2 \frac{1}{(\tg_3+n)(\tg_3-1+n)} + {\cal O}\left( \frac{1}{n^3} \right) \right]
\nonumber \\&&
+ \frac{d_{3,1}^{\rm IR}(\tkap)}{3^{\tg_3} \Gamma(\tg_3)} \Gamma(\tg_3+n) \left( \frac{\beta_0}{3} \right)^n \left[ 1 +  (b_1^{(6)}+{\cal C}_{1,1}^{(6)}) \left( \frac{3}{\beta_0} \right) \frac{1}{(\tg_3-1+n)} + {\cal O}\left( \frac{1}{n^2} \right) \right] 
{\Bigg \}}
\nonumber\\ &&
{\Bigg \{}
+ \frac{d_{1,2}^{\rm UV}(\tkap)}{\Gamma(\bg_1+1)} \Gamma(\bg_1+1+n) ( -\beta_0 )^n
\nonumber\\ && \times
\left[ 1 +  (b_1^{(-2)}+{\cal C}_{1,2}^{(-2)})  \frac{1}{(-\beta_0)} \frac{1}{(\bg_1+n)} +  (b_2^{(-2)}+{\cal C}_{1,2}^{(-2)} b_1^{(-2)} +{\cal C}_{2,2}^{(-2)}) \frac{1}{(-\beta_0)^2}  \frac{1}{(\bg_1+n)(\bg_1-1+n)} + {\cal O}\left( \frac{1}{n^3} \right) \right]
\nonumber\\ &&
+ \frac{d_{1,1}^{\rm UV}(\tkap)}{\Gamma(\bg_1)} \Gamma(\bg_1+n) ( -\beta_0 )^n \left[ 1 + (b_1^{(-2)}+{\cal C}_{1,1}^{(-2)}) \frac{1}{(-\beta_0)}  \frac{1}{(\bg_1-1+n)} + {\cal O}\left( \frac{1}{n^2} \right) \right] 
{\Bigg \}}.
\label{dnexp}
\eea 
The truncation consists of neglecting the indicated relative corrections ${\cal O}(1/n^k)$ at the end of the brackets in Eq.~(\ref{dnexp}), which then gives us the ``singular'' parts $d_n(\kappa) \mapsto (d_n(\kappa))_{\rm sing}$, and the correction coefficients $\delta d(\kappa)_n = d_n(\kappa)- (d_n(\kappa))_{\rm sing}$ appearing in the correction polynomial Eq.~(\ref{deld}). It turns out that the series in the brackets of Eq.~(\ref{dnexp}), in inverse powers of $n$, are relatively slowly converging for $n < 10$-$20$, because of the relativey large values of the numerators there. Therefore, the truncation effect and the correction coefficients $\delta d(\kappa)_n$ are relatively large for such $n$. Specifically, when $\kappa=1$, we have $|{\delta d}_n(\kappa)/d_n(\kappa)| \sim 10^1$ for $0 \leq n \leq 3$, and  $|{\delta d}_n(\kappa)/d_n(\kappa)| \lesssim 1$ when $4 \leq n \leq 10$.
However, for very large values of $n$ ($n > 15$), the coefficients ${\delta d}_n(\kappa)$ in the correction polynomial  (\ref{deld}) become relatively negligible: ${\delta d}_n(\kappa)/d_n(\kappa) \to 0$ when $n \to \infty$.\footnote{In fact, this is valid for each renormalon part $(d_n)^{\rm X}_{p}$ of the contributions to $d_n$, such as $(d_n)^{\rm UV}_1$, etc.: $({\delta d}_n(\kappa))^{\rm X}_{p}/(d_n(\kappa))^{\rm X}_p \to 0$ when $n \to \infty$, for $X =$ UV1, IR2, IR3.}
Specifically, when $\kappa=1$, we have  $|{\delta d}_n(\kappa)/d_n(\kappa)| < 0.05$ when $n>15$, and $|{\delta d}_n(\kappa)/d_n(\kappa)| < 0.025$ for $n>25$.
Despite the fact that the first few terms in the polynomial (\ref{deld}) give large contributions, the sum (\ref{deld}) [and thus the sum Eq.~(\ref{PV2})] is better behaved when the truncation index $N$ there is $N \geq 5$, as it does not have the leading parts of the renormalon contributions in its coefficients ${\delta d}_n$.

\section{Results}
\label{sec:res}

\subsection{Double-pinched Finite Energy Sum Rules with high index}
\label{subs:a25}

\begin{figure}[htb] 
\begin{minipage}[b]{.49\linewidth}
\includegraphics[width=80mm,height=50mm]{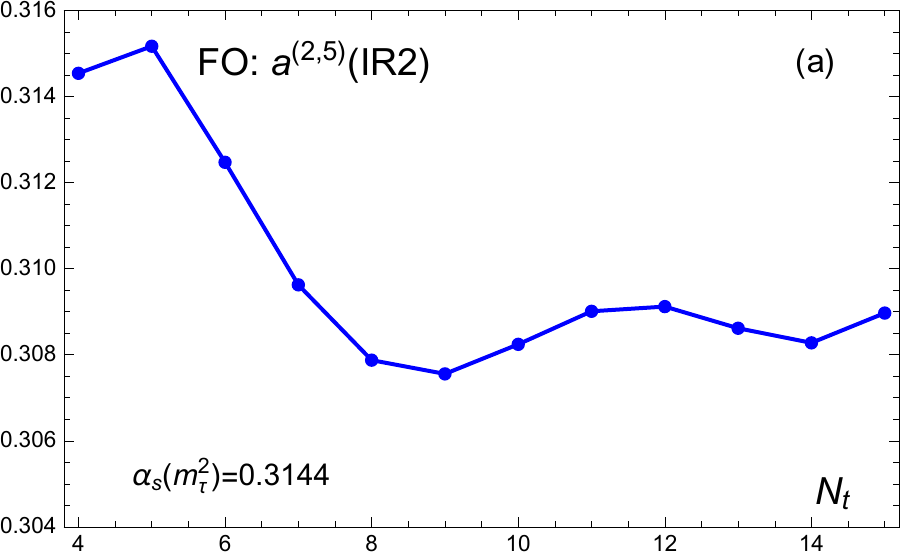}
\end{minipage}
\begin{minipage}[b]{.49\linewidth}
\includegraphics[width=80mm,height=50mm]{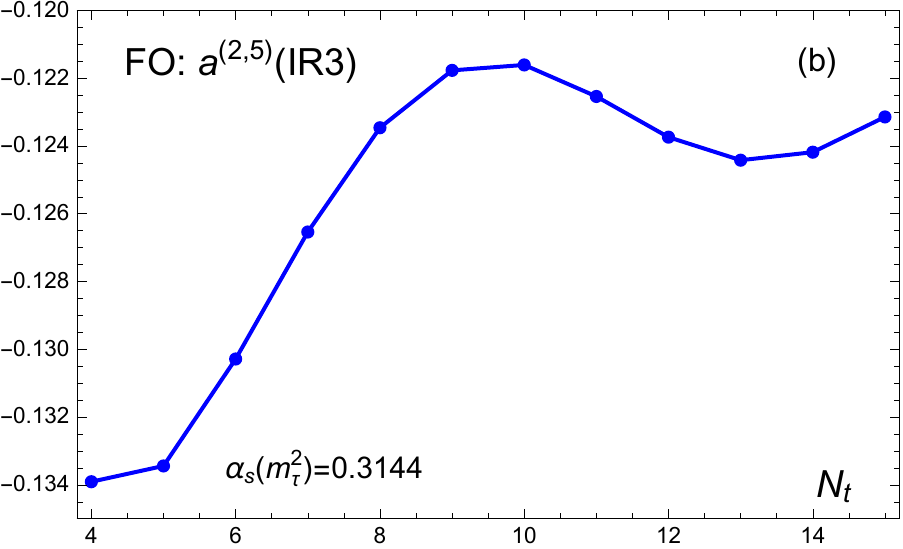}
\end{minipage}
\begin{minipage}[b]{.49\linewidth}
\includegraphics[width=80mm,height=50mm]{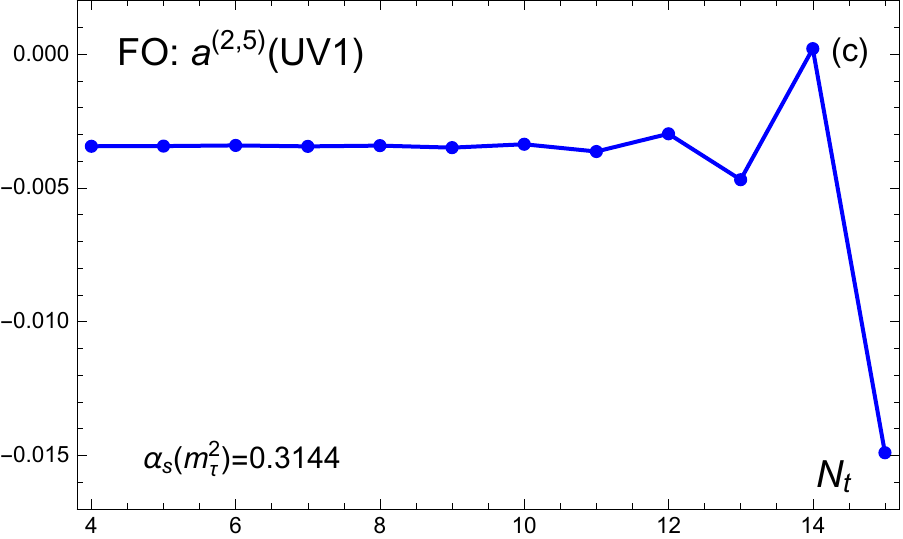}
\end{minipage}
\begin{minipage}[b]{.49\linewidth}
\includegraphics[width=80mm,height=50mm]{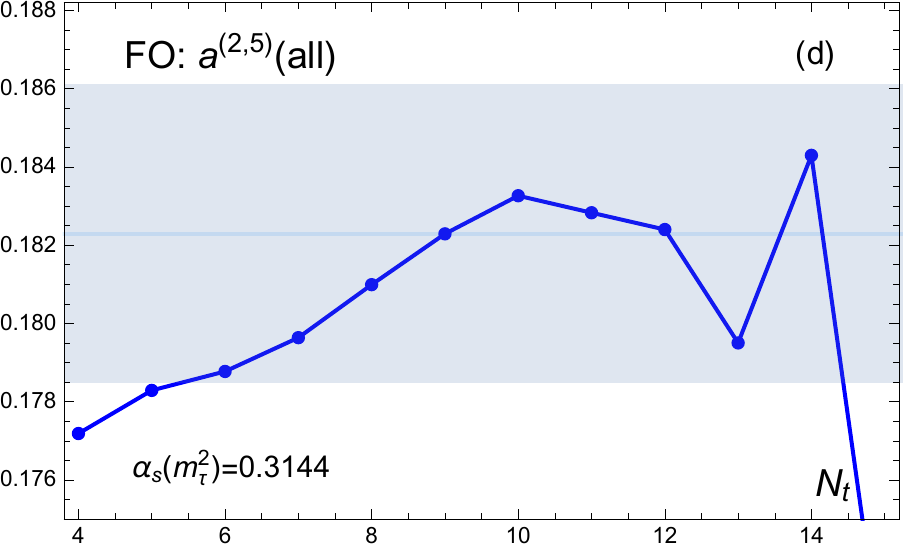}
\end{minipage}
\caption{The various contributions to the FESR moment $a^{(2,5)}(\sm)$ in the FOPT ('FO') approach, as a function of the truncation index $N_t$, for $\alpha_s(m_{\tau}^2)=0.3144$ (and $\kappa=1$): (a) the contribution from the $u=2$ IR renormalon part of the Adler function; (b) from the $u=3$ IR renormalon part; (c) from the $u=-1$ UV renormalon part; (d) the sum of all three contributions, i.e., the full theoretical FOPT value of  $a^{(2,5)}(\sm)$; the blue vertical lines in (d) represent the band of the experimental values.}
\label{a25FOavar}
\end{figure}

When we apply FESR with the moments $a^{(2,n)}(\sm)$, cf.~Eqs.~(\ref{a2n}), we can see from Eq.~(\ref{a2nth})  that the $D>0$ part of the moment $a^{(2,n)}(\sm)$ depends only on two condensate values, $\langle O_{2 n+4} \rangle$ and $\langle O_{2 n+6} \rangle$, which are of high dimension when $n$ increases. We will assume that these high-dimension condensates give a negligible contribution (cf.~also Ref.~\cite{Pich}) when $n$ is large. When equating the theoretical and the ALEPH experimental values of these moments, we can extract the QCD coupling value. It turns out that, when $n$ increases, this value appears to stabilise reasonably at $n \approx 5$ (however, see the discussion at the end of this Subsection). This is true for each of the previously described evaluation methods (FOPT, CIPT, PV). As a result, we extract the following values:
\bes
\label{a25res}
\bea
\alpha_s(m_{\tau}^2)^{\rm (FO)} & = & 0.3144 \pm 0.0036 ({\rm exp})^{+0.0024}_{-0.0031}(\kappa)^{-0.0042}_{+0.0048}(d_4)^{-0.0009}_{+0.0027}(N_t)^{+0.0027}_{-0.0069}(n)
\label{a25FOa}
\\
& = &  0.3144^{+0.0075}_{-0.0094} \approx 0.314^{+0.008}_{-0.009},
\label{a25FOb}
\\
\alpha_s(m_{\tau}^2)^{\rm (CI)} & = & 0.3282 \pm 0.0049({\rm exp})^{+0.0006}_{-0.0113}(\kappa)^{-0.0054}_{+0.0060}(d_4)^{+0.0035}_{+0.0008}(N_t)^{+0.0015}_{-0.0130}(n)
\label{a25CIa}
\\
& = &  0.3282^{+0.0087}_{-0.0187} \approx 0.328^{+0.009}_{-0.018},
\label{a25CIb}
\\
\alpha_s(m_{\tau}^2)^{\rm (PV)} & = & 0.3189 \pm 0.0041({\rm exp})^{-0.0000}_{+0.0047}(\kappa)^{-0.0040}_{+0.0043}(d_4)^{+0.0002}_{+0.0007}(N_t)^{+0.0026}_{-0.0093}(n)^{-0.0009}_{+0.0010}({\rm amb})
\label{a25PVa}
\\
& = &  0.3189^{+0.0081}_{-0.0110} \approx 0.319^{+0.008}_{-0.011}.
\label{a25PVb}
\eea
\ees 
These values were obtained with the truncation orders $N_t=9, 10, 10$ for FOPT, CIPT, PV. The truncation order in the PV approach refers to the maximum power $a^{N_t}$ in the correction polynomial (\ref{deld}). The truncation order in CIPT and PV approaches was chosen in the following way: it is such order that, when we increase the truncation index from $N_t-1$ to $N_t$, the variation of the theoretical moment $a^{(2,5)}_{\rm th}$ is minimal.\footnote{
We evaluate the differences $|a^{(2,5)}_{\rm th}(N_t) - a^{(2,5)}_{\rm th}(N_t-1)|$ with increasing $N_t=4,5,\ldots$ and look for such $N_t$ where this difference is minimal.}  On the other hand, the choice of the truncation order $N_t=9$ in the FOPT approach was chosen by looking at the stability of separate renormalon contributions (see below and Figs.~\ref{a25FOavar}). 
In the results (\ref{a25res}) we separated the various uncertainties according to their sources. The symbol ($\kappa$) indicates the variation when the renormalisation scale parameter $\kappa$ ($=\mu^2/Q^2$) is varied around $\kappa=1$, up to $\kappa_{\rm max}=2$ and down to $\kappa_{\rm min}=0.5$. The symbol ($N_t$) indicates the variation when the truncation number is varied around its central value $N_t$ to $N_t \pm 2$; ($n$) indicates the variation when we extract $\alpha_s$ from $a^{(2,n)}$ with $n=5 \pm 3$ (variation of $\alpha_s$ when $n$ varies is not weak, cf.~Table \ref{taba2n}, therefore we take $\delta n = \pm 3$). On the other hand, the symbol ($d_4$) in the results (\ref{a25res}) indicates the uncertainty due to the $d_4$ coefficient, where we vary the coefficient $d_4$ around its central value (as predicted by the considered renormalon-motivated extension) $d_4=338.19$, where we chose this variation to be $d_4=338.19 \pm 338.19$; the resulting variation of this type is obtained by using $N_t=5$, i.e., when the last term in the truncation includes $d_4$ and keeping all the other parameters of the model unchanged.\footnote{In the PV approach, we varied ${\delta d}_4$ of the correction polynomial (\ref{deld})  (with $N_t=5$) around its central value by $\pm 338.19$.}
Finally, the symbol (amb) in Eq.~(\ref{a25PVa}) represents an estimate of uncertainty due to the Borel integration ambiguity for the Adler function.\footnote{
This variation was obtained from the variation of the Adler function (\ref{PV2}) by  $\pm \delta d(Q^2) = \pm (1/(2 \pi \beta_0 i)) \left( \int_{{\cal C}_{+}} - \int_{{\cal C}_{-}} \right) \ldots$, where the integrand is the same as in Eq.~(\ref{PV2}) [cf.~also Eq.~(\ref{Bd5Pkap})].}

We mentioned earlier, in the text after Eq.~(\ref{paramrenmod}), how the value $d_4=338.19$ was obtained. We recall that the six parameters (\ref{paramrenmod}) were then determined by requiring that the expression ${\cal B}[\td](u)$ Eq.~(\ref{Btd5P}) leads to the known values of the five $d_j$ coefficients ($j=0,\ldots,4$) and to the correct value of the subleading Wilson coefficient ${\hat c}_1^{D=4}$. When the coefficient $d_4$ is varied as $d_4 = 338.19 \pm 338.19$, the uncertainties of the extracted values of $\alpha_s(m^2_{\tau})$ in Eqs.~(\ref{a25res}) [at the symbol '($d_4$)' there] would then be obtained by using the values $d_4=0$ and $d_4= 2 \times 338.19$ as input values for the determination the six parameters (\ref{paramrenmod}) of the expression ${\cal B}[\td](u)$, and then repeating the entire analysis of extraction of the values of $\alpha_s(m^2_{\tau})$ in these two new cases of the renormalon-motivated model, again with $N_t=9, 10, 10$ for the FOPT, CIPT and PV approaches. However, this requires a lot of work, for each case of $d_4$ new values of parameters (\ref{paramrenmod}) and possibly of those of Table \ref{tabCd} would be extracted. Therefore, we decided to estimate such uncertainties of $\alpha_s(m^2_{\tau})$ from the $d_4$-variation in the simpler way as described in the previous paragraph. This means that we kept the (exactly) known coefficients $d_j$ ($j=0,\ldots,3$) unchanged and took (artificially) $N_t=5$; in PV approach the values of the parameters of the renormalon model, Eqs.~(\ref{paramrenmod}) and Table \ref{tabCd}, were kept (artificially) unchanged, and in the correction polynomial (\ref{deld}) we took (artificially) $N_t=5$ and the coefficient $\delta d_4 \equiv d_4 - (d_4)_{\rm sing}$ was varied via the variation of $d_4$  (i.e., by $\pm 338.19$).

In Figs.~\ref{a25FOavar}(a)-(d) we present the various contributions to the moment $a^{(2,5)}(\sm)$ in the FOPT approach, for a fixed value of $\alpha_s(m_{\tau}^2)$, as a function of the truncation index $N_t$: in Figs.~(a), (b) and (c) the contributions of the $u=2$ IR (IR2) renormalon part, $u=3$ IR (IR3) renormalon part, and $u=-1$ UV (UV1) renormalon part of the Adler function, respectively.\footnote{We separate these parts in the coefficients $\td_n = \td_n^{\rm IR2} + \td_n^{\rm IR3} +\td_n^{\rm UV1}$, cf. Eqs.~(\ref{Btdexp}) and (\ref{Btd5P}).} In Fig.~(d)  we finally represent the entire  $a^{(2,5)}(\sm)$ which is the sum of these three contributions. While in the latter Figure one might consider $N_t=6$ as the first possible case of relative stability, Figs.~(a) and (b) show that around this index ($N_t=6 \pm 1$) we have partial cancellation of the strong instabilities from the $u=2$ and $u=3$ IR renormalons. On the other hand, $N_t= 9$ is approximately the index where both of these contributions give stationary points (minimum and maximum, respectively). This is the reason why in FOPT approach we chose $N_t=9$. The $u=-1$ UV contribution is suppressed in $a^{(2,5)}(\sm)$, as seen in Fig.~\ref{a25FOavar}(c), and this behaviour is expected on theoretical grounds as explained in the Appendix \ref{app:ta2n}.  

\begin{figure}[htb] 
\begin{minipage}[b]{.49\linewidth}
\includegraphics[width=80mm,height=50mm]{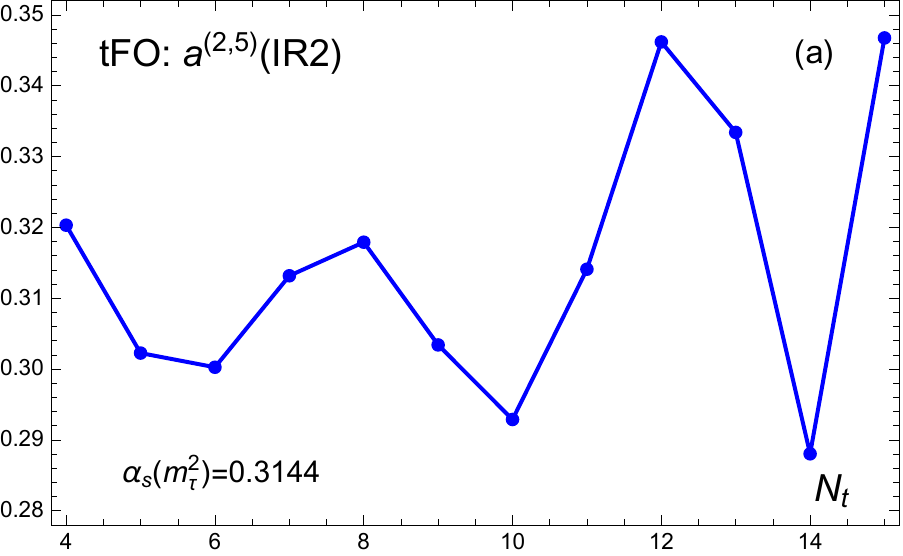}
\end{minipage}
\begin{minipage}[b]{.49\linewidth}
\includegraphics[width=80mm,height=50mm]{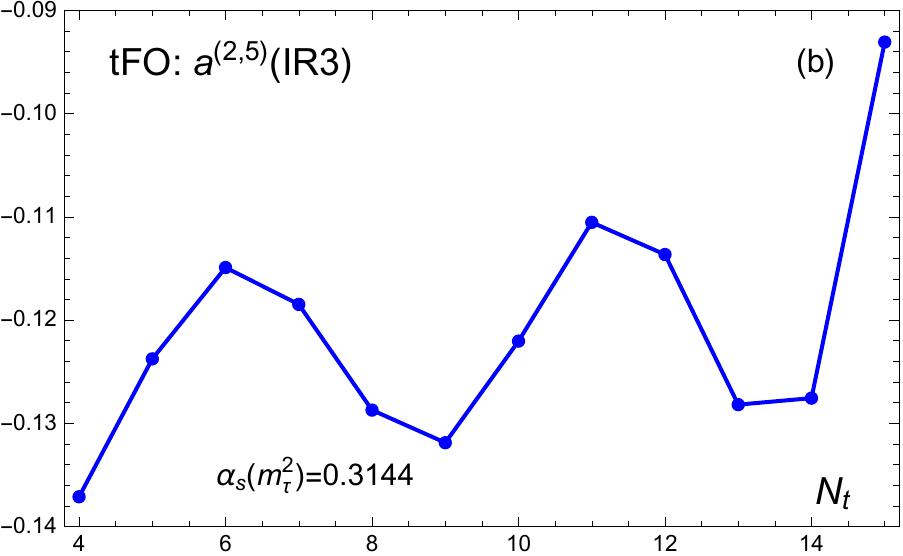}
\end{minipage}
\begin{minipage}[b]{.49\linewidth}
\includegraphics[width=80mm,height=50mm]{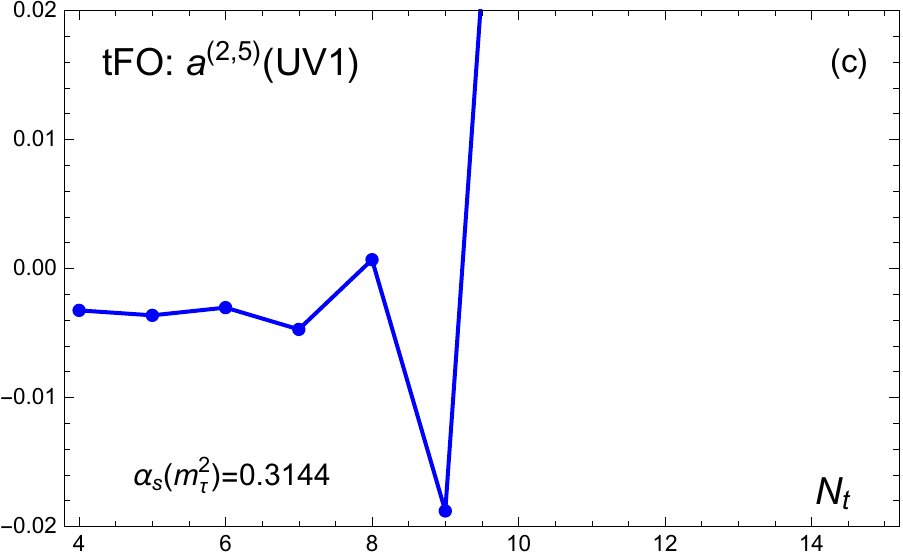}
\end{minipage}
\begin{minipage}[b]{.49\linewidth}
\includegraphics[width=80mm,height=50mm]{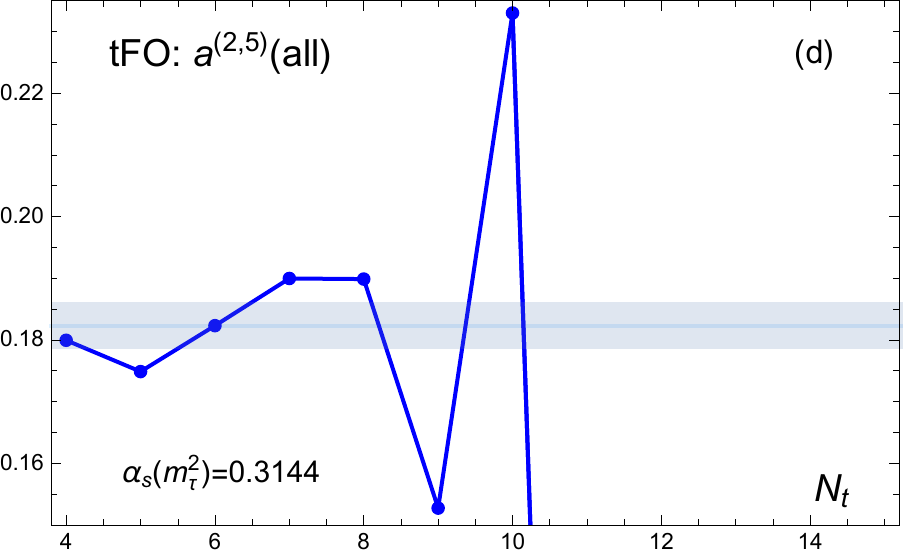}
\end{minipage}
\caption{As in Figs.~\ref{a25FOavar}, but now for the ${\widetilde{\rm FOPT}}$ ('tFO') approach.}
\label{a25tFOavar}
\end{figure}

Instead of the FOPT method Eq.~(\ref{sr2FO}), we could apply the tilde-variant $({\widetilde{\rm FOPT}})$ method Eq.~(\ref{sr2tFO}). The results are similar to those of the usual FOPT method, Eqs.~(\ref{a25FOa})-(\ref{a25FOb}): $\alpha_s(m_{\tau}^2)^{({\widetilde{\rm FO}})} = 0.314^{+0.011}_{-0.031}$, now with $N_t=6$ the optimal truncation index (the $u=2$ and $u=3$ IR contributions have local extremes at $N_t=6$, minimum and maximum, respectively; the $u=-1$ UV renormalon contribution is negligible). The uncertainties are significantly larger, though, due to the larger uncertainties of the type $(\kappa)$ and $(d_4)$.\footnote{\label{ft:tFO} This has to do primarily with the fact that $d_4 = 338.19 \pm 338.19$ implies ${\td}_4 = 37.77 \pm 338.19$, but the corresponding term $a^5 \sim \ta_5$ in the Adler function $d(\sm)_{D=0}$ (with $\kappa=1$) in the ${\widetilde{\rm FOPT}}$ approach is $\td_4 {\ta}_5(\sm) \approx 4.9 \times 10^{-5} \td_4$ while in the FOPT approach it is $d_4 a(\sm)^5 \approx 10^{-5} d_4$ [we took here $a(\sm) = 0.1$]. The effect of the variation of $d_4$ is then in the ${\widetilde{\rm FOPT}}$ of the Adler function by about a factor of $5$ stronger. Further, the fact that the central value of $\td_4$ ($37.77$) is  much smaller than the corresponding value of $d_4$ ($338.19$) may indicate that the taken uncertainty $\delta {\td}_4 (\equiv \delta d_4) = \pm 338.19$ of the coefficient $d_4$ is possibly too large.} This has to do with the fact that for $Q^2 \sim \sm$ ($\sim 1 \ {\rm GeV}^2$) the ratio $\ta_n(Q^2)/a(Q^2)^n$ [$=1 + {\cal O}(a)$] becomes large\footnote{We believe that this is related with the vicinity of the Landau singularities of the pQCD coupling $a(Q^2)$ at such low $|Q^2| \sim 1 \ {\rm GeV}^2$; such numerical problems do not occur in holomorphic variants of QCD in which the coupling $a(Q^2)$ has no Landau singularities, cf.~Ref.~\cite{renmod}.}  for $n \geq 4$, and thus the series of terms $\td_{n} \ta_{n+1}$ behaves in pQCD at such energies considerably worse than the power series of terms $d_n a^{n+1}$. For this reason, we will use the ${\widetilde{\rm FOPT}}$ method in this work only for illustrative and comparative purposes. In Figs.~\ref{a25tFOavar}(a)-(d) we present the result of the ${\widetilde{\rm FOPT}}$ approach, in analogy with the previous Figs.~\ref{a25FOavar}(a)-(d) which were for the FOPT approach. We can see that the  ${\widetilde{\rm FOPT}}$ approach gives the results which vary as a function of the truncation index considerably more strongly than the FOPT approach, although the extracted central value of $\alpha_s$ is almost the same.

An important question that arises in this part of the work is whether the described extraction of the numerical values of $\alpha_s$ from specific moments $a^{(2,n)}(\sm)$ with high $n$ (such as $n=5 \pm 1$), under the assumption that the corresponding high-dimension condensate contributions are negligible [cf.~Eq.~(\ref{a2nth})], is a realistic approach. For this reason, in Table \ref{taba2n} we present the values of $\alpha_s(m_{\tau}^2)$ extracted from moments $a^{(2,n)}(\sm)$ (with the condensate contributions neglected) for a wide range of $n$, in the FOPT, CIPT and PV approaches, with the corresponding truncation indices $N_t=9, 10, 10$.
\begin{table}
  \caption{The extracted values of $\alpha_s(m_{\tau}^2)$ from FESR moments $a^{(m,n)}(\sm)$ under the assumption of zero condensate contributions, in the FOPT (FO), CIPT (CI) and PV approach, with the corresponding truncation indices $N_t=9, 10, 10$. Only the experimental uncertainties (which are not dominant) were included. In the last column we present the experimental values of these moments as extracted from ALEPH data. For $(m,n)=(2,1)$, the CIPT approach cannot achieve the experimental values of $a^{(2,1)}(\sm)$ for any $\alpha_s$.}
\label{taba2n}
\begin{ruledtabular}
\begin{tabular}{r|ccc|c}
  $(m,n)$ & $\alpha_s(m_{\tau}^2)$ $(FO, N_t=9)$ & $\alpha_s(m_{\tau}^2)$ $(CI, N_t=10)$  & $\alpha_s(m_{\tau}^2)$ $(PV, N_t=10)$ & $a^{(m,n)}_{\rm exp}(\sm)$
  \\
\hline
(2,0)       &  $0.289 \pm 0.001$ & $0.290 \pm 0.001$ &  $0.300 \pm 0.002$ & $0.240 \pm 0.003$ \\
(2,1)       &  $0.300 \pm 0.002$ & --  &  $0.310 \pm 0.002$ & $0.206 \pm 0.003$ \\
(2,2)       &  $0.308 \pm 0.003$ & $0.315 \pm 0.002$ &  $0.310 \pm 0.002$ & $0.196 \pm 0.003$ \\
(2,3)       &  $0.311 \pm 0.003$ & $0.332 \pm 0.005$ &  $0.317 \pm 0.003$ & $0.190 \pm 0.003$ \\
(2,4)       &  $0.313 \pm 0.003$ & $0.326 \pm 0.004$ &  $0.317 \pm 0.004$ & $0.185 \pm 0.003$ \\
(2,5)       &  $0.314 \pm 0.004$ & $0.328 \pm 0.005$ &  $0.319 \pm 0.004$ & $0.182 \pm 0.004$ \\
(2,6)       &  $0.315 \pm 0.004$ & $0.328 \pm 0.006$ &  $0.320 \pm 0.005$ & $0.180 \pm 0.004$ \\
(2,7)       &  $0.316 \pm 0.005$ & $0.329 \pm 0.006$ &  $0.321 \pm 0.005$ & $0.179 \pm 0.005$ \\
(2,8)       &  $0.317 \pm 0.005$ & $0.330 \pm 0.007$ &  $0.321 \pm 0.006$ & $0.177 \pm 0.005$ \\
(2,9)       &  $0.318 \pm 0.006$ & $0.330 \pm 0.007$ &  $0.322 \pm 0.006$ & $0.176 \pm 0.005$ \\
(2,10)     &  $0.318 \pm 0.006$ & $0.331 \pm 0.007$ &  $0.323 \pm 0.006$ & $0.176 \pm 0.005$ \\
(2,20)      &  $0.322 \pm 0.008$ & $0.335 \pm 0.010$ &  $0.327 \pm 0.009$ & $0.173 \pm 0.007$ \\
\hline
(0,0)      &  $0.324 \pm 0.014$ & $0.336 \pm 0.019$  &  $0.329 \pm 0.016$ & $0.168 \pm 0.012$ \\
\end{tabular}
\end{ruledtabular}
\end{table} 
Only the experimental uncertainties are included in the Table. We can see in the Table that the extracted values of $\alpha_s$ continue to grow when $n$ increases beyond $n=5$. In the last line of the Table we present the values of $\alpha_s(m_{\tau}^2)$ extracted in these approaches from the moment $a^{(0,0)}(\sm)$, i.e., the moment which has the weight function a simple constant $g^{(0,0)}(Q^2)=1/\sm$
\bes
\label{a00}
\bea
a^{(0,0)}_{\rm exp}(\sm) & = & \frac{1}{\sm} \int_0^{\sm} d \sg \; \omega_{\rm (exp)}(\sg) - 1,
\label{a00exp} \\
a^{(0,0)}_{\rm th}(\sm) & = &  \frac{1}{2 \pi} \int_{-\pi}^{+\pi} d \phi \;
(1 + e^{i \phi}) d \left( \sm e^{i \phi} \right)_{(D=0)}.
\label{a00th}
\eea \ees
This moment has no condensate contributions. In the Table we can see that the values of  $\alpha_s(m_{\tau}^2)$ extracted from $a^{(2,n)}(\sm)$ at very high $n$ ($> 20$) appear to increase toward the values extracted from the moment $a^{(0,0)}(\sm)$. This can be understood in the following way. If we denote $x \equiv q^2/\sm = - Q^2/\sm$, the weight function $g^{(2,n)}$ of the moment $a^{(2,n)}$, Eq.~(\ref{g2n}), in the limit $n \to \infty$ and for $|x| <1$ converges to the constant $1/\sm = g^{(0,0)}$
\be
\lim_{n \to \infty} \frac{1}{\sm} (1 - x)^2 \sum_{k=0}^{n} (k+1) x^k =  \frac{1}{\sm}.
\label{g2nlim} \ee
As we see, the double-pinch factor $(1-x)^2$ in the weight functions $g^{(2,n)}$, which was there in order to suppress the duality violation effects, tends to disappear in the large $n$ limit. This, together with the results in Table \ref{taba2n}, indicates that the extracted values Eqs.~(\ref{a25res}) at $n=5$ should not be regarded as reliable, and that the results at even larger values of $n$ are not reliable because we lose the suppression of the (unaccounted for) quark-hadron duality violation effects there. Nonetheless, the results (\ref{a25res}) are illustrative and useful for comparison of the three applied methods (FOPT, CIPT, PV), and will represent an important element for our conclusions about the reliability of these three methods.

\subsection{Double-pinched Borel-Laplace sum rules}
\label{subs:BL}

In this Section we fit of the values of $\alpha_s$ and of the first few condensates to the double-pinched Borel-Laplace sum rules (for various truncation indices $N_t$), using again the ALEPH data. The truncation index $N_t$ of the applied evaluations is then fixed by considering the relative stability of the resulting first two double-pinched FESRs $a^{(2,0)}(\sm)$ and $a^{(2,1)}(\sm)$ under the variation of $N_t$.
  
The (double-pinched) Borel-Laplace sum rules to the ALEPH data, as described in Sec.~\ref{subs:weight}, cf.~Eqs.~(\ref{BL})-(\ref{BthD468}), take the form
\be
{\rm Re} B_{\rm exp}(M^2;\sm) = {\rm Re} B_{\rm th}(M^2;\sm),
\label{BLsr} \ee
where in the theoretical part we included, in addition to the $D=0$ contribution, also the $D=4,6,8$ contributions, cf.~Eq.~(\ref{BthD468}). One of the advantages of using these sum rules, in comparison to the (high-index) FESRs of the previous Section \ref{subs:a25}, is that we have now an additional continuous complex parameter $M^2$. As argued in Sec.~\ref{subs:weight}, the argument ${\rm Arg}(M^2)=\Psi$ of these parameters [$M^2=|M^2| \exp(i \Psi)$] is preferrably in the range $0 \leq \Psi < \pi/2$, and we use specifically $\Psi = 0, \pi/6, \pi/4$, and $|M^2|$ in the range $[0.9, 1.5] \ {\rm GeV}^2$ [cf.~Eq.~(\ref{rangeM2})]. In practice, we cannot ensure the equality (\ref{BLsr}) for a continuous set of values of $M^2$. Therefore, we decided to minimise the following sum of the squares of deviations between the theoretical and experimental values:
\be
\chi^2 = \sum_{\alpha=0}^n \left( \frac{ {\rm Re} B_{\rm th}(M^2_{\alpha};\sm) - {\rm Re} B_{\rm exp}(M^2_{\alpha};\sm) }{\delta_B(M^2_{\alpha})} \right)^2 ,
\label{xi2} \ee
where $M_{\alpha}^2$ is a specific sufficiently dense set of points along the rays with $\Psi=0, \pi/6, \pi/4$  and $0.9 \ {\rm GeV}^2 \leq |M|^2 \leq 1.5 \ {\rm GeV}^2$. In practice, we chose 11 equidistant points along each of the three rays, i.e., the sum (\ref{xi2}) contains 33 terms.\footnote{The fit results are practically unchanged if the number of points $M^2_{\alpha}$ is increased beyond 33.} Further, $\delta_B(M^2_{\alpha})$ is the experimental standard deviation of $B_{\rm exp}(M^2_{\alpha};\sm)$.\footnote{The construction of $\delta_B(M^2_{\alpha})$ involves the covariance matrix of the ALEPH data and the weight function $f(\sigma_j; M^2)$ (where $\sigma=\sigma_j$ is in the $j$'th bin of ALEPH data) corresponding to the Borel-Laplace sum rule, as explained in App.~C of Ref.~\cite{3dAQCD}. In the present case of the real part of the double-pinched Borel transform, we have  $f(\sigma_j; M^2)={\rm Re} \; g_{M^2}(Q^2=-\sigma_j)$ $= {\rm Re} [ (1/M^2) (1 - \sigma_j/M^2) \exp(-\sigma_j/M^2) ]$.}

\begin{figure}[htb] 
\begin{minipage}[b]{.49\linewidth}
  \centering\includegraphics[width=85mm]{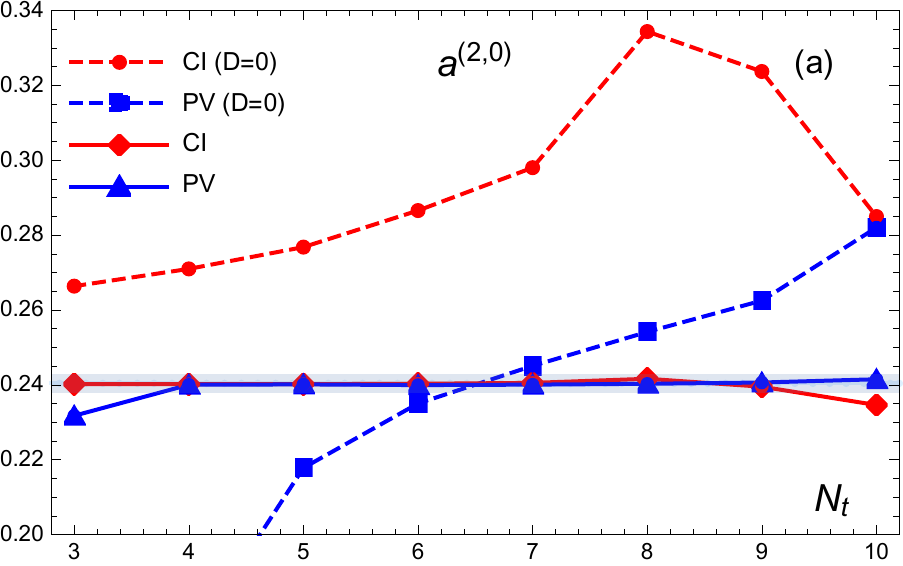}
  \end{minipage}
\begin{minipage}[b]{.49\linewidth}
  \centering\includegraphics[width=85mm]{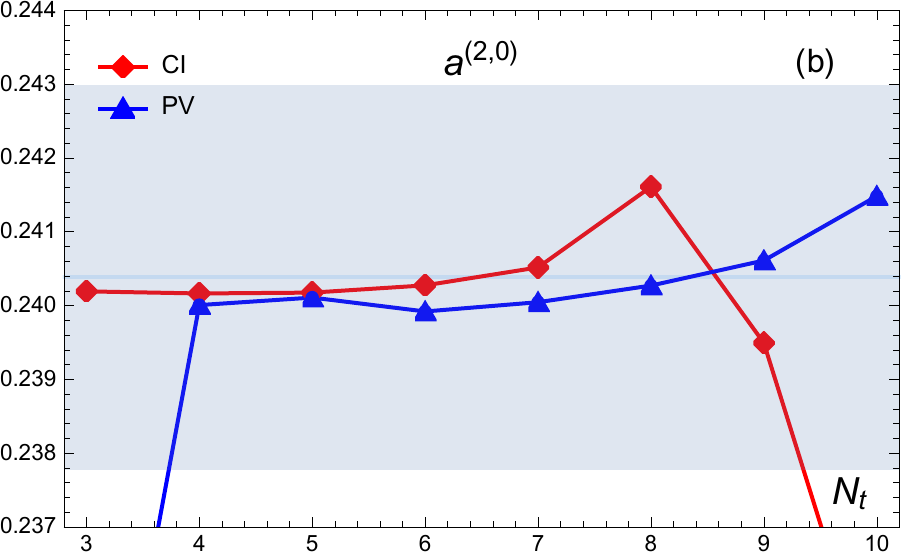}
\end{minipage}
\vspace{-0.2cm}
\caption{\footnotesize  (Coloured online) (a) The moment $a^{(2,0)}(\sm)$ as a function of the truncation index $N_t$, in the CIPT ('CI') and PV approaches. The red diamonds and the black triangles are the CIPT and PV full results, i.e., with the corresponding condensate values; the red circles and the black squares are the results where the condensate values are set equal to zero (but $\alpha_s$ values are those used for the full results).  (b) The zoomed version for better visibility. The blue band is the experimental band.}
\label{a20CIPVavar}
\end{figure}

\begin{figure}[htb] 
\begin{minipage}[b]{.49\linewidth}
  \centering\includegraphics[width=85mm]{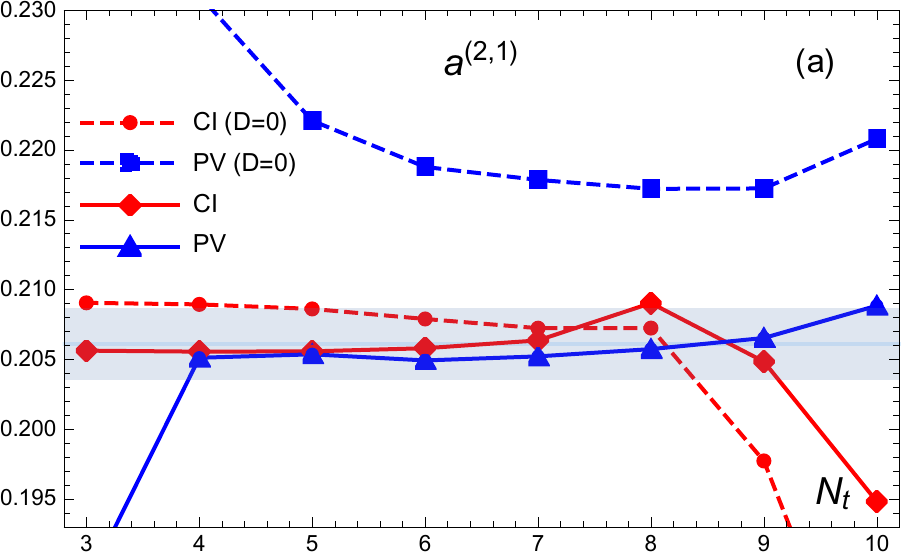}
  \end{minipage}
\begin{minipage}[b]{.49\linewidth}
  \centering\includegraphics[width=85mm]{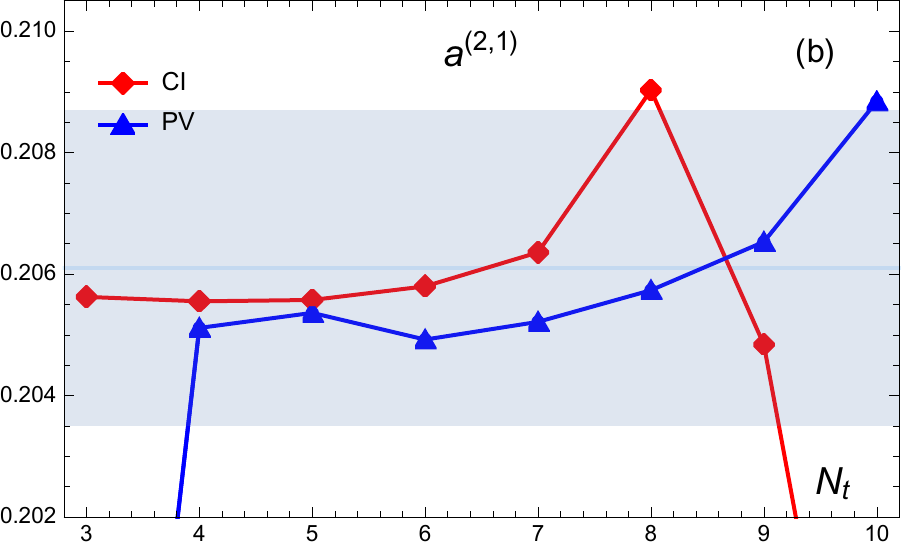}
\end{minipage}
\vspace{-0.2cm}
\caption{\footnotesize  The same as in Figs.~\ref{a20CIPVavar}, but for the moment $a^{(2,1)}(\sm)$ ($= r_{\tau}(\sm)^{(D=0)}$).}
\label{a21CIPVavar}
\end{figure}

\begin{figure}[htb] 
\begin{minipage}[b]{.49\linewidth}
  \centering\includegraphics[width=85mm]{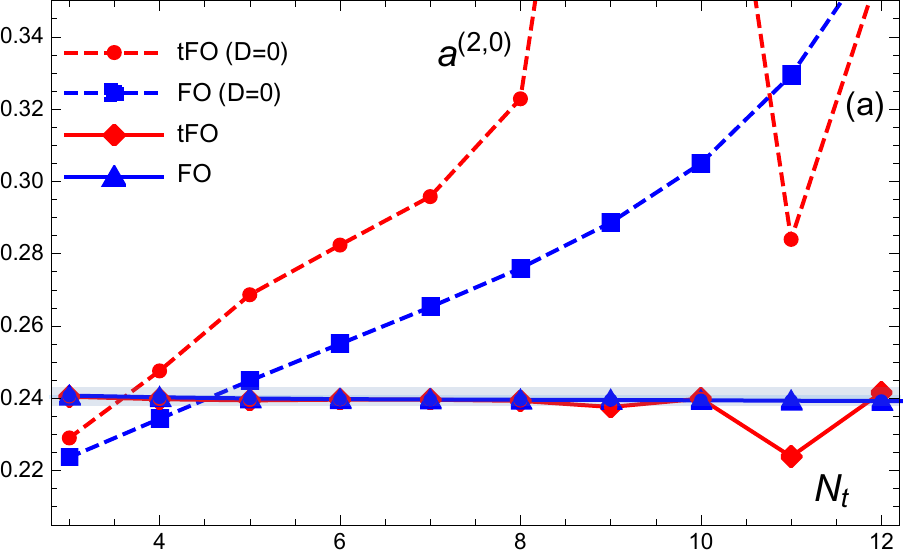}
  \end{minipage}
\begin{minipage}[b]{.49\linewidth}
  \centering\includegraphics[width=85mm]{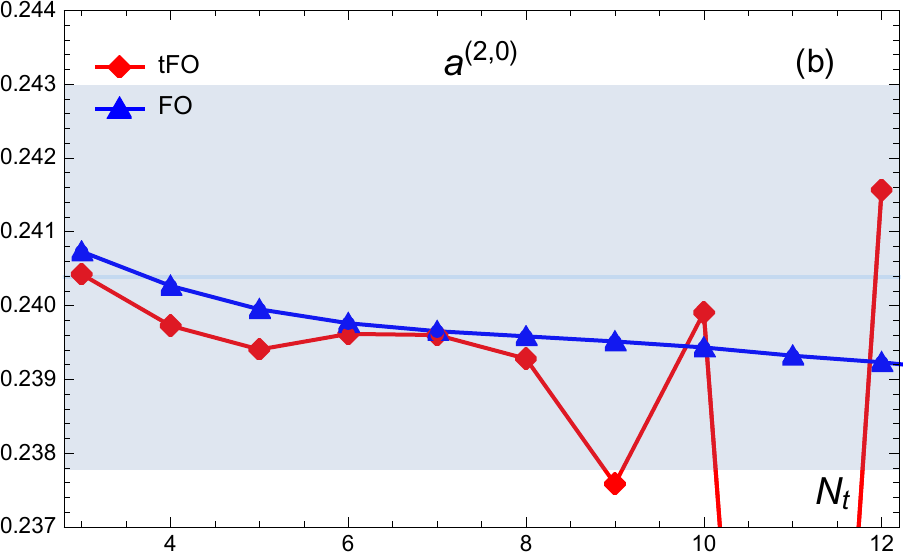}
\end{minipage}
\vspace{-0.2cm}
\caption{\footnotesize  The results for the moments $a^{(2,0)}(\sm)$ analogous to those of Figs.~\ref{a20CIPVavar}, but using the results of the global fits of the FOPT ('FO') and  ${\widetilde{\rm FOPT}}$ ('tFO') approaches instead of the CIPT and PV approaches.}
\label{a20FOtFOavar}
\end{figure}

\begin{figure}[htb] 
\begin{minipage}[b]{.49\linewidth}
  \centering\includegraphics[width=85mm]{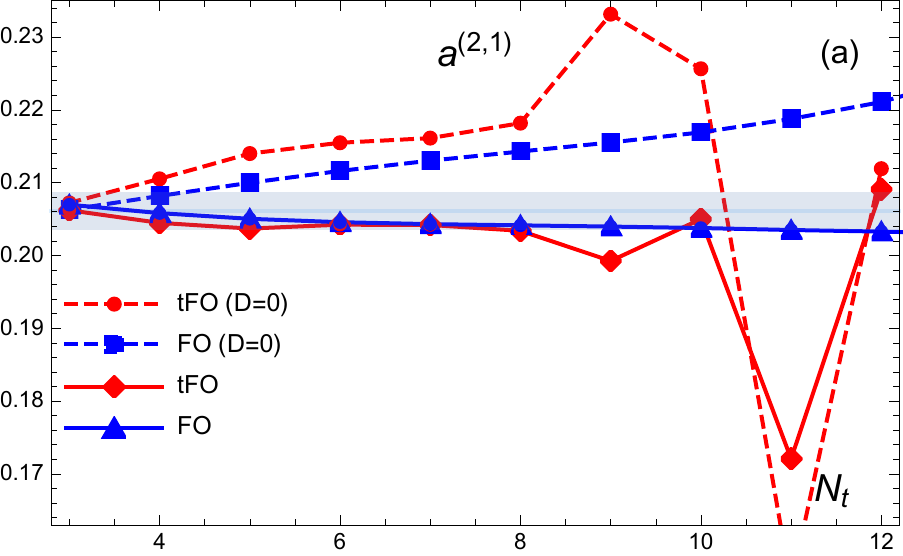}
  \end{minipage}
\begin{minipage}[b]{.49\linewidth}
  \centering\includegraphics[width=85mm]{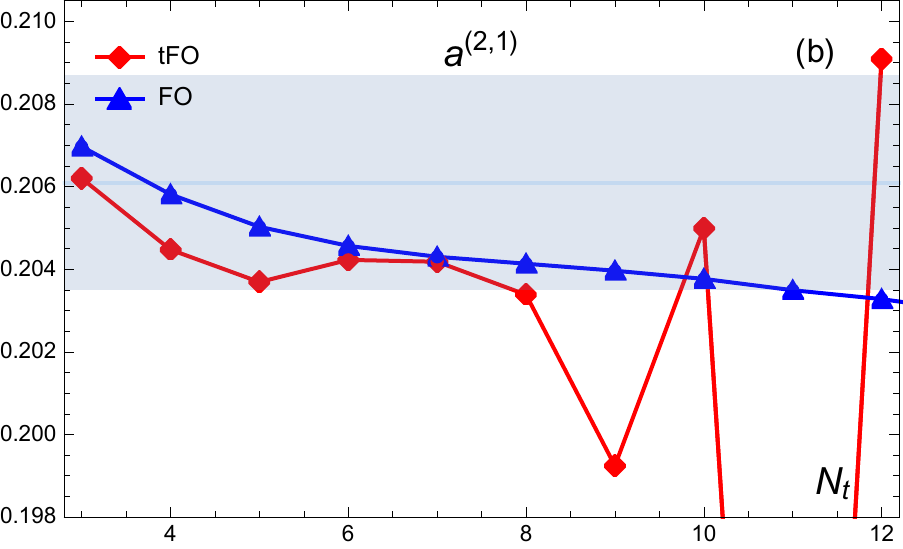}
\end{minipage}
\vspace{-0.2cm}
\caption{\footnotesize  The same as in Figs.~\ref{a20FOtFOavar}, but for the moment $a^{(2,1)}(\sm)$ ($= r_{\tau}(\sm)^{(D=0)}$).}
\label{a21FOtFOavar}
\end{figure}

The theoretical expression ${\rm Re} B_{\rm th}(M^2_{\alpha};\sm)$ now depends on four different parameters: $\alpha_s$ and $\langle O_D \rangle$ ($D=4,6,8$). The minimisation is performed in the global sense, i.e., simultaneously with respect to all these four parameters. It turns out that in many evaluation cases, the achieved minimum is very small, $\chi^2 \lesssim 10^{-3}$, i.e., the fits are good. The extracted values for $\alpha_s$ are
\bes
\label{BLresal}
\bea
\alpha_s(m_{\tau}^2)^{\rm (FO)} & = & 0.3075 \pm 0.0003({\rm exp})^{+0.0036}_{-0.0044}(\kappa)^{-0.0031}_{+0.0034}(d_4)^{-0.0031}_{+0.0036}(N_t)
\label{BLalFOa}
\\
& = &  0.3075^{+0.0061}_{-0.0062} \approx 0.308 \pm 0.006,
\label{BLalFOb}
\\
\alpha_s(m_{\tau}^2)^{\rm (CI)} & = & 0.3349 \pm 0.0004({\rm exp})^{+0.0060}_{+0.0029}(\kappa)^{-0.0055}_{+0.0059}(d_4)^{-0.0045}_{+0.0059}(N_t)
\label{BLalCIa}
\\
& = &  0.3349^{+0.0103}_{-0.0071} \approx 0.335^{+0.010}_{-0.007},
\label{BLalCIb}
\\
\alpha_s(m_{\tau}^2)^{\rm (PV)} & = & 0.3157^{+0.0012}_{-0.0014}({\rm exp})^{+0.0039}_{-0.0019}(\kappa)^{-0.0045}_{+0.0039}(d_4)^{-0.0010}_{+0.0051}(N_t)^{-0.0021}_{+0.0032}({\rm amb})
\label{BLalPVa}
\\
& = &  0.3157^{+0.0083}_{-0.0056} \approx 0.316^{+0.008}_{-0.006}.
\label{BLalPVb}
\eea
\ees
The various uncertainties are of the same type as those explained in the previous Sec.~\ref{subs:a25}, Eqs.~(\ref{a25res}). The truncation numbers $N_t$ were chosen in a somewhat similar way as in Sec.~\ref{subs:a25}. Namely, we consider the moments $a^{(2,0)}(\sm)$ and $a^{(2,1)}(\sm)$ as functions of $N_t$, using in these moments, at each $N_t$, the corresponding $\alpha_s$ and $\langle O_D \rangle$ values obtained from the mentioned global fits of Borel-Laplace sum rules with the same $N_t$. The optimal truncation index $N_t$ is then determined to be such at which the best stability of these moments is achieved (cf.~also the discussion of Figs.~\ref{a20CIPVavar}-\ref{a21FOtFOavar} later on). This time the truncation numbers turn out to be $N_t=8, 5, 6$ for FOPT, CIPT, PV, respectively; and the variation of $N_t$ around these values we take in general as $N_t \to N_t \pm 2$.\footnote{For the FO method, we take $N_t=8 \pm 2$. For the CI method, we take $N_t=5^{+2}_{-1}$; the case $N_t=3$ is not included as it does not use all the exactly known coefficients $d_j$, and the corresponding extracted value of $\alpha_s$ is significantly higher than in the cases of higher $N_t > 3$. For the PV method, we take $N_t=6^{+2}_{-1}$; the case $N_t=4$ is not included because the fit quality is much worse there ($\chi^2 = 2.2 \sim 10^0$).}

The various uncertainties are obtained in the same way as in the previous Section, with the exception of the experimental uncertainty which can be regarded here only as an estimate. Namely, for various values of $M_{\alpha}^2$, the quantities ${\rm Re} B_{\rm exp}(M_{\alpha}^2;\sm)$ are correlated with each other in a complicated manner, i.e., their covariance matrix is complicated and its inversion becomes numerically unstable when the set of the $M^2_{\alpha}$ values increases.\footnote{Cf.~the discussion in App.~C of \cite{3dAQCD}, where unpinched Borel-Laplace was used, in the context of a QCD with holomorphic coupling.} This is also the reason why for the minimisation we used the simple sum of squares Eq.~(\ref{xi2}) and not a sum involving the inverse of the covariance matrix of the Borel-Laplace sum rules. Therefore, for the estimate of the experimental uncertainties, we proceeded in the following way. We evaluated the sum of squares of the type of Eq.~(\ref{xi2}), for a small number of points $M^2_{\alpha}$  (two points along each ray, i.e., the initial and final; the sum has thus six terms), and varied each of the quantities $\alpha_s$ and $\langle O_D \rangle$ separately around the value of the minimum of $\chi^2$, until this value of $\chi^2$ increased by unity. For example, $\chi^2(\langle O_4 \rangle \pm \delta \langle O_4 \rangle_{\rm exp}) = \chi^2_{\rm min} +1$ ($\approx 1$). The number of terms (six) in these sums of squares was taken so low in order to not subestimate the experimental uncertainties. Nonetheless, as we can see, the estimates of the experimental uncertainties obtained in this way are still significantly lower than the various theoretical uncertainties.

\begin{table}
  \caption{The results for $\alpha_s(m_{\tau}^2)$ and the three condensates $\langle O_D \rangle_{V+A}$ ($D=4, 6,8$) as obtained by the Borel-Laplace sum rule. Included are the optimal truncation numbers ($N_t$) and the values of the fit quality $\chi^2$ [cf.~the text and Eq.~(\ref{xi2})].}
 \label{tabBL}
\begin{ruledtabular}
\begin{tabular}{r|rrrr|r|r}
  method & $\alpha_s(m_{\tau}^2)$ &  $\langle O_4 \rangle_{V+A}$ ($10^3 \ {\rm GeV}^4$) & $\langle O_6 \rangle_{V+A}$ ($10^3 \ {\rm GeV}^6$)  & $\langle O_8 \rangle_{V+A}$ ($10^3 \ {\rm GeV}^8$) & $N_t$ & $\chi^2$ \\
\hline
FOPT       &  $0.3075^{+0.0052}_{-0.0056}$      & $-2.8^{+1.4}_{-2.1}$   &  $+2.1 \pm 0.6$  &  $-0.8^{+0.2}_{-0.1}$ & 8 & $4. \times 10^{-3}$ \\
CIPT      &   $0.3349^{+0.0103}_{-0.0067}$ &  $-2.6^{+1.0}_{-1.9}$ & $+0.8^{+0.5}_{-0.2}$ & $-0.8^{+0.4}_{-2.8}$ & 5 &  $2. \times 10^{-4}$ \\
PV      &  $0.3157^{+0.0083}_{-0.0056}$ & $-0.2^{+1.8}_{-1.3}$ & $+2.9 \pm 0.7$ & $ -1.2^{+0.7}_{-2.4}$ & 6 &  $1. \times 10^{-3}$
\end{tabular}
\end{ruledtabular}
\end{table}
In Table \ref{tabBL} we present the results for $\alpha_s$ and the condensates.\footnote{
Instead of $\langle O_4 \rangle_{V+A}$ we present the corresponding values for the gluon condensate, $\langle a GG \rangle = 6  \langle O_4 \rangle_{V+A} + 6 f^2_{\pi} m^2_{\pi}$, where $6 f^2_{\pi} m^2_{\pi} \approx 0.00199 \ {\rm GeV}^4$.} 
The final uncertainties in the condensate values are obtained in the same way as for $\alpha_s(m_{\tau}^2)$, i.e., by combining various theoretical uncertainties and the experimental uncertainty.

\begin{figure}[htb] 
\begin{minipage}[b]{.49\linewidth}
  \centering\includegraphics[width=85mm]{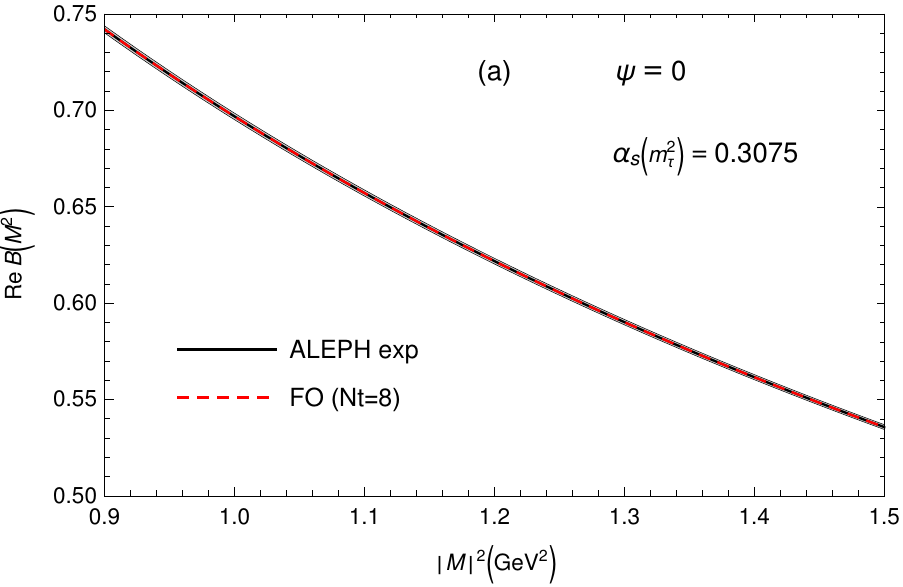}
  \end{minipage} 
\begin{minipage}[b]{.49\linewidth}
  \centering\includegraphics[width=85mm]{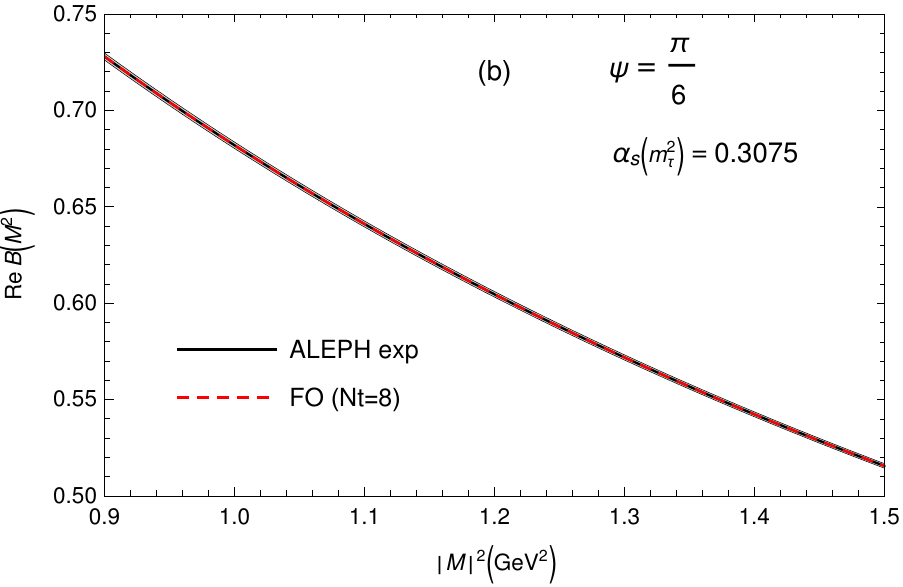}
\end{minipage}
\vspace{-0.2cm}
\caption{\footnotesize  (Coloured online) (a) and (b) The values of ${\rm Re} B(M^2;\sm)$ along the rays $M^2=|M^2| \exp( i \Psi)$ with $\Psi=0$, $\pi/6$, respectively. The narrow grey band are the experimental predictions. The red dashed line is the result of the FOPT global fit with truncation index $N_t=8$; this line is virtually indistinguishable from the central experimental line.}
\label{FigPsiPFO}
\end{figure}

In Figs.~\ref{FigPsiPFO}(a), (b), we present the quantities ${\rm Re} B(M^2;\sm)$ along the rays $M^2 = |M^2|$ and $M^2= |M^2| \exp(i \pi/6)$. The grey experimental band represents the values ${\rm Re} B_{\rm exp}(M^2;\sm) \pm  \delta_B(M^2)$ and is rather narrow.
The FOPT ($N_t=8$) theoretical prediction (global fit) is the red dashed line, which is virtually indistinguishable from the central experimental line. The results for the ray  $M^2= |M^2| \exp(i \pi/4)$ are similar.

In Figs.~\ref{a20CIPVavar} and \ref{a21CIPVavar} we present the moments $a^{(2,0)}$ and $a^{(2,1)}$ [$=r_{\tau}(\sm)^{(D=0)}$] as a function of the truncation index $N_t$, in the CIPT and PV approaches. At each order $N_t$ we employed the corresponding central values of the parameters $\alpha_s$ and $\langle O_D \rangle_{V+A}$ ($D=4,6,8$) obtained by the global approach (fit by the Borel-Laplace at $N_t$). E.g., for $N_t=5$ the corresponding CIPT central values are those in Table \ref{tabBL}.\footnote{We recall that  $a^{(2,0)}$ depends on the condensates $D=4,6$ and $a^{(2,1)}$ on the condensates $D=6,8$, cf.~Eq.~(\ref{a2nth}).} 
In the Figures we included, for comparison, the values of these moments when the contributions of the condensates are set equal to zero (but $\alpha_s$ values are those used in the full moments). Further, the experimental band (based on ALEPH data) is included. We can see that the full moments (i.e., those with the condensates included) are rather stable under the variation of $N_t$ (especially at $N_t=4$-$7$) and are consistent with the experimental values. In Figs.~\ref{a20CIPVavar}(b) and \ref{a21CIPVavar}(b) we can see that the relatively best stability of these results under the variation of $N_t$ is at $N_t \approx 5$ for CIPT and $N_t \approx 6$ for PV. On the other hand, the results without the condensate contributions are unstable under the variation of $N_t$, and in general deviate significantly from the experimental band. We point out that the values of $\alpha_s$ and of the condensate values $\langle O_D \rangle_{V+A}$ ($D=4,6,8$) were obtained from a global anaylsis involving fits of the theoretical Borel-Laplace quantities ${\rm Re} B(M^2; \sm)$ to the corresponding experimental bands, i.e., quantities with a significantly different structure than those of the FESR moments $a^{(2,n)}(\sm)$.

The behaviour of the moments $a^{(2,0)}(\sm)$ and $a^{(2,1)}(\sm)$ in the case of the FOPT methods shows qualitatively similar behaviour as in the case of the CIPT and PV methods presented in Figs.~\ref{a20CIPVavar} and \ref{a21CIPVavar}. Again, as in the previous Sec.~\ref{subs:a25}, we can apply here, instead of the FOPT approach Eq.~(\ref{sr2FO}), the tilde-variant $({\widetilde{\rm FOPT}})$ Eq.~(\ref{sr2tFO}). The results turn out to be very similar to those of the FOPT approach Eqs.~(\ref{BLalFOa})-(\ref{BLalFOb}):  $\alpha_s(m_{\tau}^2)^{({\widetilde{\rm FO}})} = 0.307^{+0.024}_{-0.021}$, and now $N_t=6$ is the optimal truncation index (also $N_t=5,7$ appear to be acceptable). Again, as in Sec.~\ref{subs:a25}, the uncertainties are significantly larger than in the FOPT method, principally because  of the larger uncertainty of the type $(d_4)$, cf.~footnote \ref{ft:tFO}. There, we will use the $({\widetilde{\rm FOPT}})$ results only for illustrative and comparative purposes.

The results for the moments $a^{(2,0)}$ and $a^{(2,1)}$ as a function of the truncation index $N_t$, in the  FOPT and ${\widetilde{\rm FOPT}}$ approaches, are presented in Figs.~\ref{a20FOtFOavar} and \ref{a21FOtFOavar}, i.e., the results analogous to those of Figs.~\ref{a20CIPVavar} and \ref{a21CIPVavar}.
Figures \ref{a20FOtFOavar}(b) and \ref{a21FOtFOavar}(b) indicate that the stability is achieved at $N_t \approx 8$ for FOPT and $N_t \approx 6$ for ${\widetilde{\rm FOPT}}$.

\begin{figure}[htb] 
\begin{minipage}[b]{.49\linewidth}
  \centering\includegraphics[width=85mm]{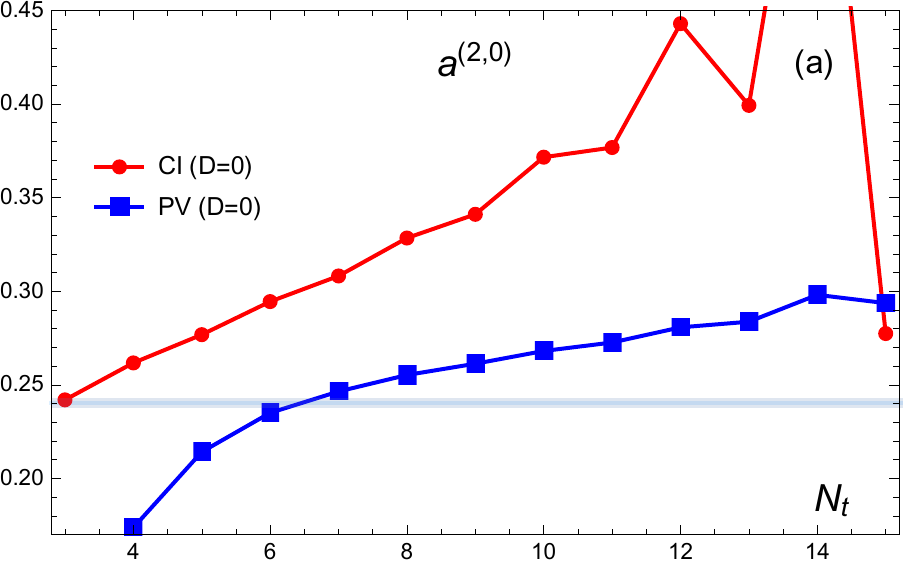}
  \end{minipage}
\begin{minipage}[b]{.49\linewidth}
  \centering\includegraphics[width=85mm]{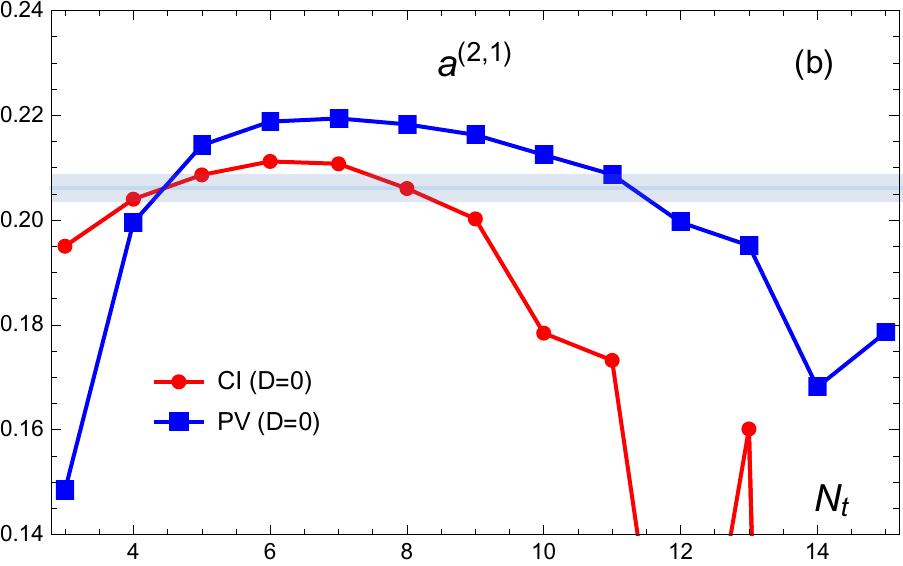}
\end{minipage}
\vspace{-0.2cm}
\caption{\footnotesize  (Coloured online) (a) The moment $a^{(2,0)}(\sm)_{(D=0)}$ as a function of the truncation index $N_t$, in the CIPT ('CI') and PV approaches.   (b) The same, but for the moment $a^{(2,1)}(\sm)_{(D=0)}$. The QCD coupling is kept fixed this time, $\alpha_s(m^2_{\tau})=0.3349$ and $0.3157$ for CIPT and PV, respectively. The experimental (ALEPH) values of $a^{(2,n)}(\sm)$ ($n=0,1$) are denoted as blue bands.}
\label{aCIPVaf}
\end{figure}

\begin{figure}[htb] 
\begin{minipage}[b]{.49\linewidth}
  \centering\includegraphics[width=85mm]{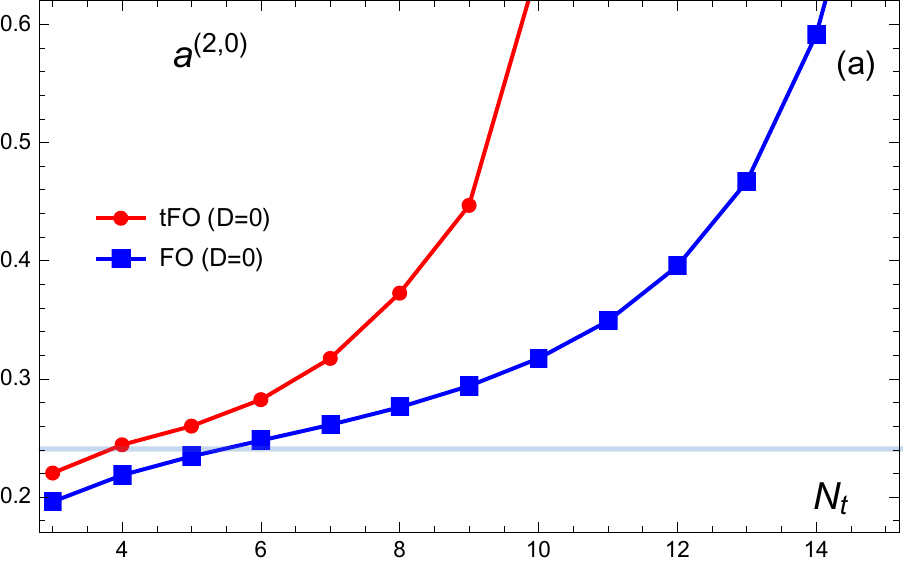}
  \end{minipage}
\begin{minipage}[b]{.49\linewidth}
  \centering\includegraphics[width=85mm]{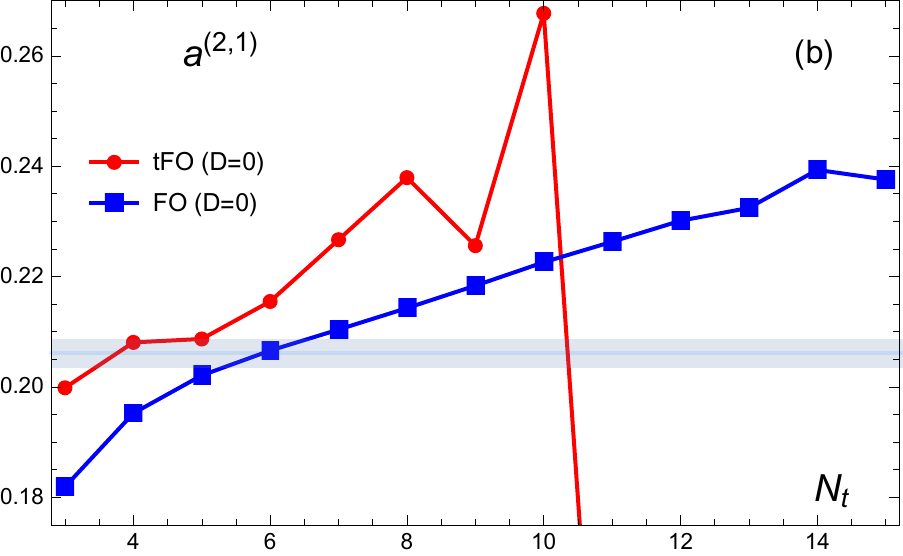}
\end{minipage}
\vspace{-0.2cm}
\caption{\footnotesize  (Coloured online) The same as in Figs.~\ref{aCIPVaf}, but this time for the methods FOPT ('FO') and  ${\widetilde{\rm FOPT}}$ ('tFO').
  The fixed QCD coupling values are $\alpha_s(m^2_{\tau})=0.3075$ and $0.3074$, respectively.}
\label{aFOtFOaf}
\end{figure}

Concerning the (local) stability of the results for the momenta $a^{(2,0)}(\sm)$ and $a^{(2,1)}(\sm)$ under the variation of the truncation index $N_t$, one question that appears is whether we get such a stability also when the values of the fit parameters ($\alpha_s$ and $\langle O_D \rangle_{V+A}$) are not fitted at each $N_t$ but are kept fixed. For this, it is sufficient to consider the $D=0$ contributions $a^{(2,0)}(\sm)_{(D=0)}$ and $a^{(2,1)}(\sm)_{(D=0)}$ as a function of $N_t$ at a fixed value of $\alpha_s$. In Figs.~\ref{aCIPVaf}-\ref{aFOtFOaf} we present these results, for the corresponding fixed central value of $\alpha_s(m^2_{\tau})$ which is chosen as the central value of each corresponding method - cf.~Table \ref{tabBL}; and $\alpha_s(m^2_{\tau})=0.3074$ for ${\widetilde{\rm FOPT}}$. We can see that these contributions in general show no local stability under the variation of $N_t$, although Fig.~\ref{aCIPVaf}(b) indicates that $N_t \approx 6, 7$ might be a reasonable value for the CI and PV method, respectively.

We also notice that Figs.~\ref{a20CIPVavar}(a)-\ref{a21FOtFOavar}(a) show that the condensate contributions (corrections) are numerically significant in $a^{(2,0)}(\sm)$ and $a^{(2,1)}(\sm)$, which may cast doubt on the (illustrative) analysis in Sec.~\ref{subs:a25} where the central values of $\alpha_s(m^2_{\tau})$ were extracted from $a^{(2,5)}(\sm)$ in Eqs.~(\ref{a25res}) when neglecting the condensate corrections. Nonetheless, closer inspection of Figs.~\ref{a20CIPVavar}(a)-\ref{a21FOtFOavar}(a) reveals that the consensate corrections are in general significantly smaller for $a^{(2,1)}(\sm)$ than for $a^{(2,0)}(\sm)$ (the only partial exception being the PV approach). This indicates that high dimension condensates probably do not contribute significantly [cf.~Eq.~(\ref{a2nth})], and that consequently $a^{(2,5)}(\sm)$ momentum sum rule has only small corrections from condensates (of dimension $D=14, 16$).
  
We point out that the Borel-Laplace QCD sum rules were first introduced in \cite{SVZ}, and later applied in the literature, e.g.~in Refs.~\cite{IoffeBL,3dAQCD}; these Borel-Laplace sum rules had no pinch factor $(1+Q^2/\sm)^n$. Part of the analysis in the work of Ref.~\cite{Pich} uses single-pinched Borel-Laplace sum rules, for $M^2 >0$; the condensate contributions are not included (but they are included in FESRs), and the extraction of $\alpha_s$ with the Borel-Laplace there is always for a specific chosen value of $M^2>0$ at a time.

Concerning the described global fit with Borel-Laplace sum rules, the following question may be raised. The IR renormalon structure of the used extended Adler function includes only the IR renormalons at $u=2$ and $u=3$, but not $u=4$, cf.~Eqs.~(\ref{Btd5P}) and (\ref{Bd5P}). This would at first suggest that only the first two condensates, $\langle O_D \rangle_{\rm V+A}$ with $D=4$ and $D=6$, should be used in the Adler function to counter the corresponding renormalon ambiguities. But we used the first three condensates ($D=4,6,8$) in the Adler function for our global fit, i.e., one more. The main reason is that the use of only $D=4$ and $6$ condensates is not enough because of the simplifying assumptions that we made for the OPE structure Eq.~(\ref{DOPE}). Namely, we assumed in Eq.~(\ref{DOPE}) that the condensates are $Q^2$-independent.\footnote{As far as we are aware, all numerical analyses of the semihadronic $\tau$-decays in the literature use these assumptions. Further, the ${\cal O}(a)$ terms in the OPE (\ref{PiOPE}) are considered negligible.} However, for the condensate term with $D=6$ (and those with $D \geq 8$) this is not correct, as indicated by the structure of the Borel transform of the Adler function in the LB approximation \cite{LB1,LB2,ren}, ${\cal B}[d](u)^{\rm (LB)}$, where the IR poles at $u \geq 3$ are not single, but double poles. This is reflected also by the double pole at $u=3$ in the Borel ${\cal B}[\td](u)$ in Eq.~(\ref{Btd5P}), and by the corresponding fact that in ${\cal B}[d](u)$ in Eq.~(\ref{Bd5P}) the most singular pole structure at $u=3$ is $\sim 1/(3 - u)^{\tg_3+1}$ [and not $\sim 1/(3-u)^{\tg_3}$]. In order to counter the renormalon ambiguity originating from such a singularity, the corresponding $D=6$ operator should have nonzero one-loop anomalous dimension coefficient, $-\gamma^{(1)}_{O_6}/\beta_0=-1$ (cf.~\cite{Btes,ren,renmod}), i.e., $\langle O_6(Q^2) \rangle = \langle {\bar O}^{(2)}_6 \rangle/a(Q^2)$ where $\langle {\bar O}^{(2)}_6 \rangle$ is $Q^2$-independent. The $D=6$ condensate contributions to the Adler function then have the form
\be
d(Q^2)_{(D=6)} = 6 \pi^2 \left[ \frac{\langle {\bar O}^{(2)}_6 \rangle}{a(Q^2) (Q^2)^3} + \frac{\langle {\bar O}^{(1)}_6 \rangle}{(Q^2)^3} \right],
\label{dD6} \ee
which is different from the $D=6$ ($k=3$) form in Eq.~(\ref{DOPE}). Here, $\langle {\bar O}^{(2)}_6 \rangle$ and $\langle {\bar O}^{(1)}_6 \rangle$ are $Q^2$-independent. The first term and the second term then counter the renormalon ambiguity originating from the terms $\td^{\rm IR}_{3,2}/(3 - u)^2$ and $\td^{\rm IR}_{3,1}/(3-u)$ in ${\cal B}[\td](u)$ of Eq.~(\ref{Btd5P}), respectively [i.e., from the terms $\sim d^{\rm IR}_{3,2}/(3 - u)^{\tg_3+1}$ and $\sim d^{\rm IR}_{3,1}/(3 - u)^{\tg_3}$ in the Borel transform of the Adler function, Eq.~(\ref{Bd5P})]. There are indications that the effects of the first term on the right-hand side of Eq.~(\ref{dD6}) are reasonably well approximated in the sum rules by two condensate terms ($k=3$ and $k=4$) of the simple OPE type Eq.~(\ref{DOPE}), i.e., by the $Q^2$-independent condensate contributions with $D=6$ and $D=8$. This is what we used in our global fit.

On the other hand, if we use in the global fits with Borel-Laplace sum rules only two condensate terms, $D=4$ and $D=6$ (without $D=8$) with $Q^2$-independent condensates, then it turns out that the (two) condensates do not stabilise the resulting moment $a^{(2,1)}(\sm)$ as a function of the truncation index $N_t$. In fact, they make the variation of $a^{(2,1)}(\sm)$ with $N_t$ even worse than for the pure $D=0$ parts $a^{(2,1)}(\sm)_{(D=0)}$, in stark contrast with the results in Figs.~(\ref{a21CIPVavar}) and (\ref{a21FOtFOavar}).\footnote{On the other hand, $a^{(2,0)}(\sm)$ does get reasonably stabilised as a function of $N_t$ when only the $D=4$ and $D=6$ $Q^2$-independent condensates of the Adler function are used in the global fit with Borel-Laplace sum rules. We recall that, in the approximation of $Q^2$-independent condensates, $a^{(2,0)}(\sm)$ depends on the condensates $D=4$ and $D=6$ (but not $D=8$), and the $D=4$ condensate contribution in $a^{(2,0)}(\sm)$ is numerically probably more important than the $D=6$ contribution.}  We recall that $a^{(2,1)}(\sm)$ depends on $D=6$ and $D=8$ condensates (when these condensates are considered $Q^2$-independent), and our Adler extension formally does not require $D=8$ condensate (which would counter the $u=4$ IR renormalon pole ambiguity effects). Therefore, the numerical results of our global fits suggest that the $D=8$ condensate ($Q^2$-independent) in our analysis simulates the role of the effects of the running of the $1/a(Q^2)$ factor in the first term on the right-hand side of Eq.~(\ref{dD6}), in the Borel-Laplace sum rules and in $a^{(2,1)}(\sm)$. A global fit analysis using the more explicit form (\ref{dD6}) remains outstanding, but we expect it to give results similar to those presented here.

These questions notwithstanding, our global fit analysis with OPE with condensates assumed to be $Q^2$-independent, Eqs.~(\ref{PiOPE})-(\ref{DOPE}), can be repeated by including one more condensate term, of dimension $D (\equiv 2 k)=10$. This then gives us the results presented in Table \ref{tabBLO10}.
\begin{table}
  \caption{As Table \ref{tabBL}, but now $\langle O_{10} \rangle_{V+A}$ is included in the global fit. The condensates $\langle O_D \rangle_{V+A}$ are presented in units of $10^3 \ {\rm GeV}^D$.}
 \label{tabBLO10}
\begin{ruledtabular}
\begin{tabular}{r|rrrrr|r|r}
  method & $\alpha_s(m_{\tau}^2)$ &  $\langle O_4 \rangle_{V+A}$  & $\langle O_6 \rangle_{V+A}$  & $\langle O_8 \rangle_{V+A}$  & $\langle O_{10} \rangle_{V+A}$ & $N_t$ & $\chi^2$ \\
\hline
FOPT       &  $0.3115^{+0.0051}_{-0.0057}$      & $-3.7^{+1.7}_{-2.6}$   &  $+3.3^{+1.0}_{-0.9}$  &  $-2.1^{+0.6}_{-0.5}$ &  $+0.8 \pm 0.3$ & 8 & $6. \times 10^{-5}$ \\
CIPT      &   $0.3362^{+0.0112}_{-0.0099}$ &  $-2.7^{+0.8}_{-0.6}$ & $+1.0^{+0.2}_{-1.3}$ & $-1.1^{+0.4}_{-0.2}$  & $+0.2^{+0.2}_{-1.7}$& 5 &  $1. \times 10^{-5}$ \\
PV      &  $0.3180^{+0.0089}_{-0.0069}$ & $-0.4^{+1.8}_{-1.2}$ & $+3.4^{+1.3}_{-2.1}$ & $-1.8^{+1.0}_{-0.2}$ & $+0.4^{+0.4}_{-2.8}$ & 6 &  $7. \times 10^{-5}$
\end{tabular}
\end{ruledtabular}
\end{table}
The values of index $N_t$ were kept unchanged in comparison to Table \ref{tabBL}. If we determined $N_t$, as in the previous case of $\langle O_{10} \rangle_{V+A}=0$, as the value at which the resulting momenta $a^{(2,0})(\sm)$ and $a^{(2,1)}(\sm)$ are least $N_t$-sensitive, then we would obtain in the present case (when $\langle O_{10} \rangle_{V+A}$ is varied): for FOPT $N_t=8$ (unchanged); for CIPT $N_t=4$-$5$, and for PV $N_t=5$. Nonetheless, here we kept $N_t$ unchanged ($N_t=8, 5, 6$ for FOPT, CIPT, PV, respectively), so that the comparison of the results of Tables \ref{tabBL} and \ref{tabBLO10} gives us the effects of the OPE-truncation change only ($D_{\rm max}=8 \mapsto 10$), without interference of the effects of the $N_t$-change. The resulting uncertainties of the extracted values of $\alpha_s(m_{\tau}^2)$ under the variation of $N_t$ are anyway similar in the two cases $D_{\rm max}=8$ and $D_{\rm max}=10$.\footnote{In the case of $D_{\rm max}=8$ these variations of extracted values of $\alpha_s(m_{\tau}^2)$ are given in Eqs.~(\ref{BLalFOa}),(\ref{BLalCIa}), (\ref{BLalPVa}), at the symbol '($N_t$)'. In the case of $D_{\rm max}=10$, these variations are: $\mp 0.0015$ (FOPT, $N_t=8 \pm 1$); $^{-0.0049}_{+0.0062}$ (CIPT, $N_t=5 \pm 1$); $^{-0.0015}_{+0.0050}$ (PV, $N_t=6 \pm 1$).}

Comparison of Tables \ref{tabBLO10} and \ref{tabBL} shows that the OPE-truncation effects are moderate: the values of $\alpha_s(m_{\tau}^2)$ change by less than the uncertainties given in Table \ref{tabBL}, and even the values of the condensates in most cases change by less than 50 percent. The values of condensates of dimension $D=8$ and $D=10$ in Table \ref{tabBLO10} indicate that their contributions to sum rules are small and tend to cancel each other.\footnote{The large variation (uncertainty) of $\langle O_{10} \rangle_{V+A}$ into negative values in Table \ref{tabBLO10} (and of  $\langle O_{8} \rangle_{V+A}$ in Table \ref{tabBL}), in the PV and CIPT approaches, exists due to the instability of these extracted condensates when the renormalization scale parameter is varied from $\kappa=1$ to $\kappa=0.5$. This indicates that the RGE-running of the pQCD running coupling $a(\mu^2)$ along the contour $\mu^2= 0.5 \; \sm \exp(i \phi)$ (where $0.5 \; \sm \approx 1.4 \ {\rm GeV}^2$ is low) is unreliable, due to the vicinity of the unphysical Landau singularities of the pQCD coupling at such low values of $|\mu^2|$. We note that in the PV approach, the polynomial correction part Eq.~(\ref{deld}) is treated as in the CIPT approach when integrated along the contour.} We can estimate the OPE-truncation uncertainty in the extracted values of $\alpha_s(m_{\tau}^2)$ as the difference between the corresponding central values in Tables \ref{tabBLO10} and \ref{tabBL}, and add this uncertainty in quadrature to the results of Eqs.~(\ref{BLresal}). This then gives us
\bes
\label{BLresalF}
\bea
\alpha_s(m_{\tau}^2)^{\rm (FO)} & = &  0.3075^{+0.0061}_{-0.0062} \pm 0.0040 =
 0.3075^{+0.0073}_{-0.0074}  \; \left( \approx 0.308 \pm 0.007 \right),
\label{BLalFOF}
\\
\alpha_s(m_{\tau}^2)^{\rm (CI)} & = &  0.3349^{+0.0103}_{-0.0071} \pm 0.0013 =
 0.3349^{+0.0104}_{-0.0071} \; \left( \approx 0.335^{+0.010}_{-0.007} \right),
\label{BLalCIF}
\\
\alpha_s(m_{\tau}^2)^{\rm (PV)} & = & 0.3157^{+0.0083}_{-0.0056} \pm 0.0023 =
0.3157^{+0.0086}_{-0.0061} \; \left( \approx 0.316^{+0.008}_{-0.006} \right).
\label{BLalPVF}
\eea
\ees
We can see that, due to the OPE truncation effects, the uncertainty of $\alpha_s(m_{\tau}^2)$ increases moderately in the FOPT case, and remains practically unchanged in the CIPT and PV cases.

As in the case of $D_{\rm max}=8$, we can evaluate in the case $D_{\rm max}=10$ the FESR momenta $a^{(2,0)}(\sm)$ and $a^{(2,1)}(\sm)$ as a function of $N_t$, and obtain results analogous to those in Figs.~\ref{a20CIPVavar}-\ref{a21FOtFOavar}. We will not present such Figures, but it turns out that now the stability of these momenta under the variation of $N_t$ is even stronger, and the agreement with the experimental values is even better. 

The behaviour of the extracted values of condensates with increasing dimension $D$ is qualitatively similar to that in the work of \cite{Pich} where various pinched FESRs were applied to the ALEPH $\tau$-decay spectral functions. The dependence on the OPE-truncation variation (variation of $D_{\rm max}$) is somewhat milder in our analysis, though. The fact that the extracted values of condensates with $D \geq 8$  in our analysis are small, or tend to cancel each other, is possibly related with the fact that our $D=0$ Adler function extension contains the first two IR renormalons ($u=2$ and $u=3$), cf.~Eq.~(\ref{Btd5P}) [cf.~also Eqs.~(\ref{Bd5P}) and (\ref{Bd5Pkap})]. However, also the FESR fit results of \cite{Pich}, in the $V+A$ channel, show a similar trend when $D_{\rm max}=8$ or 10. 

The works of \cite{Bo2015,Bo2017} (cf.~also \cite{BoRee2019} where $R_{ee}(s)$ is used), on the other hand, give for (sufficiently pinched) FESRs the solutions of OPE with considerably larger absolute values of the condensates $\langle O_D \rangle_{V+A}$, for many terms (up to $D=16$).\footnote{
The works \cite{Bo2015,Bo2017} and \cite{Pich} use for the $D=0$ Adler function the series truncated at $\sim a^5$ (i.e., $N_t=5$), with $d_4 = 283 \pm 283$ \cite{Bo2015,Bo2017} and $d_4=275 \pm 400$ \cite{Pich}.}
This shows that there are at least two very different sets of OPE solutions to the $\tau$-decay data, which correspond to the same or approximately same spectral function $\omega(\sigma)_{V+A}$ (ALEPH) for (sufficiently pinched) FESRs. The results of the works \cite{Pich} and \cite{Bo2015,Bo2017} suggest that the duality violations are well suppressed in the $V+A$ channel for FESRs which are at least doubly-pinched.\footnote{The $n$-pinched FESR weight functions $g(Q^2)$ behave in the timelike limit, i.e., when $Q^2 \to -\sm$, as: $g(Q^2) \sim (1 + Q^2/\sm)^n$.} In our global fit analysis, we used doubly-pinched Borel-Laplace sum rules, whose weight functions $g_{M^2}(Q^2)$, Eq.~(\ref{gM2}), are additionally suppressed in the timelike limit by the exponential factor $\exp(\cos(\Psi) Q^2/|M^2|) \to \exp(- \cos(\Psi) \sm/|M^2|)$ (where we took: $\Psi \equiv {\rm arg}(M^2) = 0, \pi/6, \pi/4$).

\subsection{FESRs and Borel-Laplace sum rules with resummation based on inverse Mellin transform}
\label{subs:resiM}

In Ref.~\cite{renmod} a resummation of the Adler function was performed using an approach of characteristic functions (related with the approach of \cite{Neubert}).  It has the form
\be
d(Q^2)_{(D=0); \rm res} = \int_0^1 \frac{dt}{t} G^{(-)}_D(t) a(t e^{-\tK} Q^2) +
\int_1^{\infty} \frac{dt}{t} G^{(+)}_D(t) a(t e^{-\tK} Q^2) + \int_0^1 \frac{dt}{t} G^{\rm (SL)}_D(t) \left[ a(t e^{-\tK} Q^2) - a(e^{-\tK} Q^2) \right],
\label{Dres2b}
\ee
where the characteristic functions $ G^{(\pm)}_D(t)$ and $G^{\rm (SL)}_D(t)$ are inverse Mellin transforms of different parts of the Borel transform ${\cal B}[\td](u)$ Eq.~(\ref{Btd5P}); they involve simple positive or negative powers of $t$ and $\ln t$ (cf.~\cite{renmod} for details). The main difference between the FOPT, CIPT and PV methods, on one hand, and this evaluation method, on the other hand, is that this method does not involve truncation. However, when this resummed version, $d(\sm e^{i \phi})_{(D=0); \rm res}$, is used in the sum rules, e.g. in the contour integrals (\ref{a2nth}) and (\ref{Bth}), the integrations over $t$ at $0<t<1$ involve the (pQCD) coupling $a(t e^{-\tK} \sm e^{i \phi})$ at low momenta. For small $\phi \approx 0$ and small $t$ this means that the integrations are performed close to the Landau cuts of $a(Q^2)$ in the complex $Q^2$-plane, i.e., at $0 < Q^2 < {\Lambda}^2_{\rm Lan.}$ ($\sim 0.1 \ {\rm GeV}^2$), and this makes the evaluation numerically unreliable. The extracted values of the parameters also indicate this problem. Namely, the central values of $\alpha_s$ in the $a^{(2,5)}$-approach and in the global fit approach with this resummation method are disparate, $\alpha_s(m_{\tau}^2) = 0.377$ and $0.246$, respectively. We will not use these results, as they are significantly affected by the mentioned problem of Landau singularities.\footnote{A general discussion of the Landau singularity problems in pQCD couplings is given, e.g., in \cite{LandauC}.} In this context, we mention that this resummation approach works well when the QCD coupling has no Landau singularities \cite{renmod}.

\section{Summary of the results and comparison with literature}
\label{sec:summ}

The main results of the paper are in Eqs.~(\ref{BLresalF}). For the purpose of additional comparison of different methods (FOPT, PV, CIPT), the results in Eqs.~(\ref{a25res}) are also important.

We can argue that the FOPT and PV methods have the following feature in common: (a) the FOPT [or $({\widetilde{\rm FOPT}})$] perturbation series for the sum rules, as argued in the Appendix, explicitly have the leading renormalon contribution  of the Adler function $d(Q^2)_{(D=0)}$ suppressed in them;\footnote{
  We recall that the leading renormalon contribution is the double-pole $u=1$ UV renormalon in the perturbation series of the auxiliary quantity ${\td}(Q^2;\kappa)_{(D=0)}$, and its analog in the perturbation series of $d(Q^2)_{(D=0)}$. We point out that this suppression of the leading renormalon contribution in the sum rules is true not just in the large-$\beta_0$ approximation, but in the exact approach, as shown in the Appendix.}; (b) the PV approach in the sum rules isolates the dominant parts of the contributions from the renormalon singularities (UV and IR) of the Adler function, and resums them with the PV convention, while the perturbation series of the correction part in this approach is largely free of the renormalon contributions.

On the other hand, the CIPT approach to the sum rules keeps unchanged the entire coefficients $d_n$ of the perturbation series of the Adler function in the resummation of the sum rules, thus importing the strong renormalon-dominated divergence of $d_n$'s (when $n$ increases) in the sum rule evaluation. It is true that the CIPT approach also transforms the powers $a(Q^2)^n$ [or the log derivatives $\ta_n(Q^2)$] of the Adler function into different functions via contour integration with specific weight functions, but this change in general does not account for the renormalon cancellations which are neither reflected in the (unchanged) expansion coefficients $d_n$ of the CIPT series. We believe that these aspects are the main reason why the extracted values of $\alpha_s$ from  the (truncated) CIPT approach  differ significantly from the (truncated) FOPT and PV methods (while the latter two methods give mutually similar results). These conclusions are valid not just in the analysis of Sec.~\ref{subs:a25} of the moments $a^{(2,5)}(\sm)_{(D=0)}$ where the condensate contributions were neglected, but also in the analysis of Sec.~\ref{subs:BL} where the first three condensates were included.

In the Appendix we argued that the FOPT (and ${\widetilde {\rm FOPT}}$) expansion of the moments $a^{(2,n)}_{(D=0)}$ (with $n$ large, such as $n=5$) has the UV renormalons (at $u=-1, -2, \ldots$) as well as some of the IR renormalons (at $u \geq 2$) suppressed by one power, in comparison to the Adler function $d(Q^2)_{(D=0)}$. Specifically for the first IR renormalon (at $u=2$) of the Adler function this implies that it is almost entirely supressed in the FOPT evaluation of the moments $a^{(2,n)}_{(D=0)}$ for $n \geq 1$.\footnote{This means, the renormalon contribution $\sim 1/(2 -u)^{\tg_2}$ in the Borel transform of the Adler function $d(Q^2)_{(D=0)}$ is suppressed to $~\sim 1/(2-u)^{\tg_2-1}$ in the Borel transform of $a^{(2,n)}_{(D=0)}$ for $n \geq 1$, cf.~also Eq.~(\ref{Bd5P}) and (\ref{Bd5Pkap}). We recall that in our considered renormalon-motivated Adler function extension there are only renormalons UV1 (at $u=-1$), IR2 (at $u=2$) and IR3 (at $u=3$).}

On the other hand, the FOPT (and the ${\widetilde {\rm FOPT}}$) expansion of the Borel-Laplace sum rules ${\rm Re} B(M^2)$ has only the UV renormalons suppressed by one power, in comparison to the Adler function $d(Q^2)_{(D=0)}$; but the IR renormalons are not suppressed. Therefore, one might expect that the FOPT (and the ${\widetilde {\rm FOPT}}$) global fit analysis with Borel-Laplace sum rules would give us more unstable results and less reliable value of the extracted $\alpha_s$ than such an analysis with the moments $a^{(2,n)}_{(D=0)}$ ($n=5$). However, the inclusion of the condensate contributions (with $D=4, 6, 8$) in such an analysis takes care of the fact that the IR-renormalon contributions of the Adler function are not suppressed in the FOPT  (and the ${\widetilde {\rm FOPT}}$) Borel-Laplace sum rules [cf.~also the more detailed discussion around Eq.~(\ref{dD6})]. Our analysis with Borel-Laplace was indeed performed with the (low-dimension) condensate contributions included. Therefore, we can argue that the FOPT (and the ${\widetilde {\rm FOPT}}$) global fit with Borel-Laplace sum rules gives us reliable extraction of the values of $\alpha_s$ (and of the condensates).

The PV global fit with Borel-Laplace sum rules and condensates is also expected to give reliable results, because the renormalon structure of the Adler function is taken into account correctly (in an isolated, resummed form) in such sum rules.

However, the CIPT global fit with Borel-Laplace sum rules and condensates is again expected to present problems, because the truncated CIPT approach does not suppress the leading UV renormalon contributions and deteriorates in a significant way the interplay between the IR renormalon effects (in the $D=0$ Borel-Laplace part) and the condensate contributions. In this context, we recall that the truncated CIPT is neither a perturbation power series nor does it represent a resummation of the evaluated quantity (because it is truncated).\footnote{Somewhat related arguments for the preference of the FOPT methods over the CIPT methods, in FESRs of semihadronic $\tau$-decays, were presented in Refs.~\cite{BJ,BJ2,BoiOl,HoangR}.} 

The numerical results presented in this work [principally Eqs.~(\ref{a25res}), (\ref{BLresal}), (\ref{BLresalF}), Table \ref{tabBL}] appear to confirm the arguments given above. Namely, the extracted values of $\alpha_s$ are closer to each other when the FOPT (and ${\widetilde {\rm FOPT}}$) and PV evaluation methods are used, while the extracted value of $\alpha_s$ becomes significantly larger when the (truncated) CIPT evaluation method is used. This is true in the analysis of Sec.~\ref{subs:a25} where the high order moments $a^{(2,n)}$ were considered, and in the analysis of Sec.~\ref{subs:BL} where Borel-Laplace sum rules with condensates were considered. Therefore, in our main predictions for $\alpha_s$ we will include the (truncated) FOPT and PV evaluation methods, but not the (truncated) CIPT method. Further, as argued at the end of Sec.~\ref{subs:a25}, the results (\ref{a25res}) are not reliable because of the unaccounted nonperturbative effects (from condensates and the quark-hadron duality violation effects). However, the results (\ref{a25res}) serve principally as an additional comparison of the three methods (FOPT, PV and CIPT) as mentioned above.  

Furthermore, we believe that the fact that we used a renormalon-motivated extension of the coefficients $d_n$ ($n \geq 4$) of the Adler function does not introduce large model ambiguities. One reason is that the extension is motivated on the known renormalon structure of the Adler function, and simultaneously reproduces correctly the first four coefficients $d_n$ ($n=0,1,2,3$). The other reason is that our methods used truncation indices $N_t$ (i.e., truncation at $a^{N_t}$) which were often low ($N_t=5, 6$).\footnote{We recall that the truncation index $N_t$ was determined in each method in such a way that a relative stability of the full moments $a^{(2,0)}(\sm)$ and $a^{(2,1)}(\sm)$ is achieved under the variation of $N_t$.}

On the gounds mentioned above, our main results are represented by the global fit results (with the double-pinched Borel Laplace sum rules) of the truncated FOPT and PV approaches, Eqs.~(\ref{BLalFOF}), (\ref{BLalPVF}). We obtain our central result by averaging between these two results. This then gives the following averaged results of the global fits:
\bes
\label{2av}
\bea
\alpha_s(m_{\tau}^2) &=& 0.3116 \pm 0.0073 \qquad ({\rm FOPT+PV, \; global }) 
\label{2ava} \\
\Rightarrow \;
\alpha_s(M_Z^2) &=& 0.1176 \pm 0.0010.  
\label{2avb} \eea \ees
It turned out that the central result of the ${\widetilde {\rm FOPT}}$ approach is practically equal to that of the FOPT approach, but the uncertainties are higher. We did not include the  ${\widetilde {\rm FOPT}}$ approach result in the average (\ref{2av}).

The uncertainty $\pm 0.0073$ in Eq.~(\ref{2ava}) was obtained by adding in quadrature the deviation between the average value $0.3116$ and the the central value $0.3075$ of Eq.~(\ref{BLalFOF}), and the minimal uncertainty $\pm 0.0061$ of the two results Eqs.~(\ref{BLalFOF}) and  (\ref{BLalPVF}) (cf.~a similar reasoning in Ref.~\cite{Pich}). The result Eq.~(\ref{2avb}), at the canonical scale $Q^2=M_Z^2$ (and $N_f=5$) was then obtained by RGE-evolution using the five-loop $\MSbar$ $\beta$-function \cite{5lMSbarbeta} and the corresponding four-loop quark threshold matching \cite{4lquarkthresh}.\footnote{The threshold matching was performed at the scales $Q^2_{\rm thr} = \kappa {\bar m}_q^2$ with $\kappa=2$, and ${\bar m}_q \equiv {\bar m}_q( {\bar m}_q^2)$ equal to $4.2$ GeV ($q=b$) and $1.27$ GeV ($q=c$).}

If, however, we included in the average also the CIPT result (\ref{BLalCIF}), the average central value and the uncertainties would significantly increase
\bes
\label{3av}
\bea
\alpha_s(m_{\tau}^2) &=& 0.3194 \pm 0.0167 \qquad ({\rm FOPT+PV+CIPT, \; global }) 
\label{3ava} \\
\Rightarrow \;
\alpha_s(M_Z^2) &=& 0.1186 \pm 0.0021.  
\label{3avb} \eea \ees
In this context, we note that the relations and differences between the FOPT and CIPT approach in FESRs of the semihadronic $\tau$ decays were investigated in the works \cite{HoangR} from the point of view of Borel transforms and Borel sums. A Borel transform was constructed for the CIPT of FESRs, by first rewriting the CIPT series of such FESRs formally as a (FOPT-type) series in powers of $a(\sm)$, $\sum r_n^{\rm (CI)} a(\sm)^{n+1}$, using for the Adler function either the large-$\beta_0$ approximation \cite{LB1,LB2} or the renormalon-motivated model of Ref.~\cite{BJ}.\footnote{We note that this sum $\sum r_n^{\rm (CI)} a(\sm)^{n+1}$, strictly speaking, is not a (FOPT-type) perturbation series, because each coefficient $r_n^{\rm (CI)}$ in this sum is itself a series in powers of $a(\sm)$.}  It was shown that the resulting Borel transform has a significantly different structure of nonanalyticity than the Borel transform of the FOPT FESRs (for the latter, cf.~Appendix \ref{app:renSR}). For example, in the large-$\beta_0$ approximation the $u=2$ IR renormalon of the Adler function is completely suppressed in the Borel transform of the FOPT FESRs momenta $a^{(2,n)}(\sm)$ (with $n \geq 1$),\footnote{In the large-$\beta_0$ approximation, the same type of relations are valid for the Borel transform of $a^{(2,n)}(\sm)$ as in Eq.~(\ref{Bta2n}) for the Borel transform of $\ta^{(2,n)}(\sm)$. Further, the $u=2$ IR renormalon residue is in our considered model numerically significant in ${\cal B}[\td](u)$, Eqs.~(\ref{Btd5P})-(\ref{paramrenmod}), as well as in ${\cal B}[d](u)$, cf.~Eqs.~(\ref{Bd5P}), (\ref{Bd5Pkap}) and Table \ref{tabCd} (first line, last column).} while this is not the case for the $u=2$ IR renormalon in the Borel transform of the CIPT FESRs momenta $a^{(2,n)}(\sm)$. The Borel transforms of the CIPT FESRs do not reflect the $D > 0$ structure of the OPE of the Adler function Eq.~(\ref{DOPE}), or equivalently, the corresponding OPE of the FESRs Eq.~(\ref{a2nth}). This is in contrast with the Borel transform of the FOPT FESRs which do respect this $D>0$ OPE structure as explained in Appendix \ref{app:renSR}. The authors of \cite{HoangR} suggest that the CIPT FESRs would require different, nonstandard OPE corrections, i.e., corrections which would not correspond to the ($D > 0$) OPE corrections (\ref{DOPE}) and (\ref{dD6}) of the Adler function. In the present work, we did not try to implement such nonstandard OPE in the CIPT evaluations. For these reasons, and for the reasons explained earlier in this Section, we consider it correct to include only the FOPT and PV results, leading to Eqs.~(\ref{2av}), and not to include the values of $\alpha_s$ extracted from the CIPT evaluations Eqs.~(\ref{3av}). The question of how to treat correctly the CIPT evaluations of the sum rules, in particular the related nonstandard OPE corrections, is left open in this work.

If we perform the truncation in all methods at the index $N_t=5$ ($\sim a^5$),\footnote{This truncation was used in \cite{Pich} (cf.~also \cite{Pich3,Pich4}) where $N_t=5$ FOPT and CIPT methods were used and $d_4 = 275 \pm 400$ (at $\kappa=1$). We use $d_4 \approx 338 \pm 338$, i.e., the central value $d_4 \approx 338$ as suggested by the described renormalon-motivated extension of the $D=0$ Adler function.} the central results do not change very much; e.g., the average of the central values of the FOPT+PV methods (and of FOPT+CIPT+PV methods) of the global fits, when always $N_t=5$ is taken, is $\alpha_s(m_{\tau}^2)=0.3172$ (and $0.3231$), respectively, not very far away from the respective central value $0.3116$ Eq.~(\ref{2ava}) (and $0.3194$), respectively.

For comparison, we present in Table \ref{tabreslit} the values of $\alpha_s(m_{\tau}^2)$ extracted from ALEPH $\tau$-decay data by various groups, using various sum rules and various methods of evaluation.
\begin{table}
  \caption{The values of $\alpha_s(m_{\tau}^2)$, extracted from ALEPH $\tau$-decay data, as obtained by various groups applying  sum rules and various methods. 'BL' stands for (double-pinched) Borel Laplace, and `DV' stands for a duality violation model.}
 \label{tabreslit}
\begin{ruledtabular}
\begin{tabular}{l|l|lll|l}
group &  sum rule & FOPT & CIPT & PV & average \\
\hline
Baikov et al.~\cite{BCK} & $a^{(2,1)}=r_{\tau}$ & $0.322 \pm 0.020$ & $0.342 \pm 0.011$ & --- & $0.332 \pm 0.016$ \\
Beneke \& Jamin \cite{BJ} & $a^{(2,1)}=r_{\tau}$ &  --- & --- & $0.316 \pm 0.016$ & $0.316 \pm 0.016$ \\
Caprini \cite{Caprini2020} & $a^{(2,1)}=r_{\tau}$ &  --- & --- & $0.314 \pm 0.006$ &  $0.314 \pm 0.006$ \\
Davier et al.~\cite{Davetal} & $a^{(i,j)}$ & $0.324$ & $0.341 \pm 0.008$ & --- & $0.332 \pm 0.012$ \\
Pich \& R.-S.~\cite{Pich}   &  $a^{(i,j)}$     & $0.320 \pm 0.012$ &  $0.335 \pm 0.013$ & --- & $0.328 \pm 0.013$  \\
this work  & BL & $0.308 \pm 0.007$ & $0.335^{+0.010}_{-0.007}$ & $0.316^{+0.008}_{-0.006}$ & $0.312 \pm 0.007$ (FOPT+PV) \\
Boito et al.~\cite{Bo2015} & DV in $a^{(i,j)}$ & $0.296 \pm 0.010$ & $0.310 \pm 0.014$ & --- & $0.303 \pm 0.012$ \\
Pich \& R.-S.~\cite{Pich}   &   DV in $a^{(i,j)}$ & $0.298 \pm 0.031$ & $0.312 \pm 0.047$ & --- & $0.302 \pm 0.032$ 
\end{tabular}
\end{ruledtabular}
\end{table}
In the Table the results are presented to three digits.
The result of Ref.~\cite{Caprini2020} is an update of the result of Ref.~\cite{CF2}, and uses a (PV) summation of a renormalon-motivated Borel transform with a conformal mapping. The results from Ref.~\cite{Davetal} in the Table are given for their V+A channel analysis. We can see in the Table that the results of the exhaustive analysis of Ref.~\cite{Pich} gave a result for CIPT approach very similar to ours, while their FOPT analysis gave a result significantly higher than ours. The latter occurred principally because in our case of FOPT evaluation the optimal truncation index turned out to be relatively high ($N_t=8$) which indicates that the (renormalon-motivated) extension of the Adler function beyond the order $a^5$ ($N_t=5$) plays a role in our case of FOPT evaluations. Nonetheless, we recall that the ${\widetilde {\rm FOPT}}$ method in our global fits gave a lower index value $N_t=6$ and a similarly low central value $\alpha_s(m_{\tau}^2)=0.307$ (but higher uncertainties). We can see in Table \ref{tabreslit} that the duality violation analysis of Ref.~\cite{Bo2015} (cf.~also \cite{Bo2011,PerisPC1,Bo2017}) gives even significantly lower values of $\alpha_s$. On the other hand, it was argued in Ref.~\cite{Pich} that the uncertainties in this DV-model should be larger.

In a recent work \cite{Bo2021}, the mentioned DV-model strategy was used in FOPT-evaluated FESRs of semihadronic $\tau$-decays, using an experimental spectral V-channel function based on data from various experiments (ALEPH, OPAL, BABAR, and supplemented by $e^+ e^- \to$ hadrons data), and they obtained the result $\alpha_s(m_{\tau}^2) = 0.3077 \pm 0.0075$, which is very close to our result $\alpha_s(m_{\tau}^2) = 0.3075^{+0.0066}_{-0.0069}$ obtained from Borel-Laplace sum rule global fit with FOPT method, cf.~Eq.~(\ref{BLalFOF}) and Tables \ref{tabBL} and \ref{tabreslit}.

In this context, we point out that our analysis used the combined V+A channel of ALEPH data, and involved double-pinched sum rules ($a^{(2,n)}$ and double-pinched Borel-Laplace). We believe that both of these aspects suppress significantly the possible duality violation effects in our analysis. 

The Mathematica programs on which our calculations were based are available from the www page Ref.~\cite{prgs}.

\begin{acknowledgments}
This work was supported in part by FONDECYT (Chile) Grants No.~1200189 and No.~1180344, and Project PIIC (UTFSM). 
\end{acknowledgments}

\appendix

\section{Renormalon structure of Adler function-related sum rules}
\label{app:renSR}

In this Appendix we will present relations between the renormalon structure of the Adler function $d(Q^2)_{(D=0)}$ and the considered FESRs $a^{(2,n)}(\sm)$ and double-pinched Borel-Laplace sum rules.

The $(D=0)$ parts of the theoretical side of the FESRs $a^{(2,n)}(\sm)$, cf.~Eqs.~(\ref{G2n}) and (\ref{a2nth}), consist of the following elements:
\be
\delta_{x^n}^{(d)} \equiv - \frac{i}{2 \pi} \oint_{|x|=1} \frac{d x}{x} x^n d(Q^2=-\sm x)_{(D=0)}
= (-1)^n \frac{1}{2 \pi} \int_{-\pi}^{+\pi} d \phi \; e^{i n \phi} d(\sm e^{i \phi})_{(D=0)} \qquad (n=0,1,\ldots),
  \label{dxnd} \ee
where $x \equiv q^2/\sm = - Q^2/\sm = - e^{i \phi}$, and the perturbation expansion of $d(Q^2)_{(D=0)}$ in powers $a^{n+1}$ is given in Eq.~(\ref{dpt}), and in logarithmic derivatives $\ta_{n+1}$ in Eq.~(\ref{dlpt}) [cf.~Eq.~(\ref{tan})]. The auxiliary quantity $\td(Q^2;\kappa)_{(D=0)}$ is defined then via the expansion Eq.~(\ref{tdpt}) as expansion in powers $a^{n+1}$. This auxiliary quantity is independent of the renormalisation scale parameter $\kappa$ only when $a(\kappa Q^2)$ runs according to the one-loop RGE, due to the (exact) identities (\ref{tdnkap}).

On the basis of the sum rule quantity $\delta_{x^n}^{(d)}$ of Eq.~(\ref{dxnd}) we define the corresponding sum rule quantity with $d \mapsto \td$
\be
\delta_{x^n}^{(\td)}(\kappa) \equiv - \frac{i}{2 \pi} \oint_{|x|=1} \frac{d x}{x} x^n \td(Q^2=-\sm x; \kappa)_{(D=0)}
= (-1)^n  \frac{1}{2 \pi} \int_{-\pi}^{+\pi} d \phi \; e^{i n \phi} \td(\sm e^{i \phi};\kappa)_{(D=0)}.
\label{dxntd} \ee
It is $\kappa$-independent only in the case of one-loop running of $a(\kappa Q^2)$.

The Borel transform of $\td_{(D=0)}$, ${\cal B}[\td](u;\kappa)$ of Eq.~(\ref{Btdexp}), has the exact $\kappa$-dependence as given in Eq.~(\ref{Btdkappa}). The auxiliary quantity $\td_{(D=0)}$ is obtained from the Borel transform ${\cal B}[\td]$ by the inverse Borel transformation
\be
\td(Q^2;\kappa)_{(D=0)}=\frac{1}{\beta_0} \int_{0}^{+\infty} du \exp \left[- \frac{u}{\beta_0 a(\kappa Q^2)} \right] {\cal B}[\td](u;\kappa) = {\td}_0 a(\kappa Q^2)  + {\td}_1(\kappa) a(\kappa Q^2)^2 + \ldots + {\td}_n(\kappa)  a(\kappa Q^2)^{n+1} + \ldots
\label{invB} \ee
Further, if we apply in $\td(\sm e^{i \phi};\kappa)_{(D=0)}$ in the sum rule (\ref{dxntd}) the one-loop $(1 \ell)$ RGE running of $a(\kappa \sm e^{i \phi})$ around $\phi=0$
\bes
\label{a1l}
\bea
\frac{1}{a^{(1 \ell)}(\kappa \sm e^{i \phi})} &=&    \frac{1}{a^{(1 \ell)}(\kappa \sm)} + i \beta_0 \phi \; \Rightarrow
\label{a1la} \\
\exp \left[ - \frac{u}{ \beta_0 a^{(1 \ell)}(\kappa \sm e^{i \phi}) } \right]  &=&
\exp \left[ - \frac{u}{ \beta_0 a^{(1 \ell)}(\kappa \sm) } \right] \exp( - i u \phi),
\label{a1lb} \eea \ees
then the quantity $\delta_{x^n}^{(\td)}$ Eq.~(\ref{dxntd}), in this $(1 \ell)$-approximation, turns out to be
\be
\delta_{x^n}^{(\td; 1 \ell)} =  
\frac{1}{\beta_0} \int_{0}^{+\infty} du \exp \left[- \frac{u}{\beta_0 a^{(1 \ell)}(\kappa \sm)} \right] \frac{\sin (\pi u)}{\pi} \frac{1}{(u-n)} {\cal B}[\td](u;\kappa),
\label{dxntd1l} \ee
which means that the Borel transform of $\delta_{x^n}^{(\td; 1 \ell)}$ is\footnote{Cf.~\cite{Boitoetal,BoiOl,Caprini2019} where the notation is slightly different; cf.~also \cite{BJ,Btes}.}
\be
{\cal B}[ \delta_{x^n}^{(\td; 1 \ell)}](u; \kappa) = \frac{\sin (\pi u)}{\pi} \frac{1}{(u-n)} {\cal B}[\td](u;\kappa).
\label{Btd1l} \ee   
The expression (\ref{dxntd1l}) was obtained by using in the definition (\ref{dxntd}) the $(1 \ell)$ version of the identity (\ref{invB})\footnote{i.e., in Eq.~(\ref{invB}) we replace everywhere $a(\kappa Q^2) \mapsto a^{(1 \ell)}(\kappa Q^2)$.}, exchange the order of integration over $du$ and $d \phi$, and use the $(1 \ell)$-identity (\ref{a1lb}); the integration over $d \phi$ is then trivial
\be
\int_{-\pi}^{+\pi} d \phi \; e^{i (n-u) \phi} = (-1)^{n+1} 2 \; \frac{\sin (\pi u)}{(n-u)},
\label{dphi} \ee
leading to the identity (\ref{dxntd1l}). The quantity $\delta_{x^n}^{(\td; 1 \ell)}$ Eq.~(\ref{dxntd1l}) is $\kappa$-independent because the Adler function auxiliary quantity $\td(\sm e^{i \phi}; \kappa)_{(D=0)}$ in the integrand in Eq.~(\ref{dxntd}) is $\kappa$-independent when $a(\kappa Q^2)$ runs according to the one-loop RGE
\be
\td(Q^2)^{(1 \ell)}_{(D=0)}=\frac{1}{\beta_0} \int_{0}^{+\infty} du \exp \left[- \frac{u}{\beta_0 a^{(1 \ell)}(\kappa Q^2)} \right] {\cal B}[\td](u;\kappa) = {\td}_0 a^{(1 \ell)}(\kappa Q^2)  + \ldots + {\td}_n(\kappa)  a^{(1 \ell)}(\kappa Q^2)^{n+1} + \ldots
\label{invB1l} \ee
We point out that the coefficients ${\td}_n(\kappa)$ remain unaffected by this replacement $a(\kappa Q^2) \mapsto a^{(1 \ell)}(\kappa Q^2)$, leading from Eq.~(\ref{invB}) to (\ref{invB1l}).
We can see also explicitly that the expression Eq.~(\ref{dxntd1l}) is $\kappa$-independent, because ${\cal B}[{\td}](u; \kappa) = \kappa^u {\cal B}[{\td}](u)$ [Eq.~(\ref{Btdkappa})] and therefore by Eq.~(\ref{Btd1l})
\be
{\cal B}[ \delta_{x^n}^{(\td; 1 \ell)}](u; \kappa) = \exp( u \ln \kappa) {\cal B}[ \delta_{x^n}^{(\td; 1 \ell)}](u)
\label{Btd1lkappa} \ee
and
\be
a^{(1 \ell)}(\kappa \sm) = \frac{ a(\sm) }{(1 + a(\sm) \beta_0 \ln \kappa)}.
\label{a1lsm} \ee
When combining Eqs.~(\ref{Btd1lkappa})-(\ref{a1lsm}), we see that the integrand on the right-hand side of Eq.~(\ref{dxntd1l}) [cf.~also Eq.~(\ref{Btd1l})] is $\kappa$-independent and thus $\delta_{x^n}^{(\td; 1 \ell)}$ is $\kappa$-independent.

\subsection{One-loop Borel transform of FESR auxiliary moments $\ta^{(2,n)}$}
\label{app:ta2n}

Using the identity (\ref{dxntd1l}) we can now obtain directly the one-loop Borel transform of the $(D=0)$-part of the FESR moments $a^{(2,n)}$. The $(D=0)$ part  of $a^{(2,n)}$, at any loop order, is [cf.~Eq.~(\ref{a2nth})]
\be
a^{(2,n)}(\sm)_{(D=0)} =   \frac{1}{2 \pi} \int_{-\pi}^{+\pi} d \phi \;
G^{(2,n)} \left(\sm e^{i \phi} \right) d \left( \sm e^{i \phi} \right)_{(D=0)},
\label{a2nD0} \ee
and the auxiliary (tilde) version is defined to be
\be
\ta^{(2,n)}(\sm; \kappa)_{(D=0)} =   \frac{1}{2 \pi} \int_{-\pi}^{+\pi} d \phi \;
G^{(2,n)} \left(\sm e^{i \phi} \right) \td \left( \sm e^{i \phi}; \kappa \right)_{(D=0)},
\label{ta2nD0} \ee
in complete analogy with the definitions (\ref{dxnd}) and (\ref{dxntd}).
When taking into account the explicit expression (\ref{G2n}) for the integrated weight function $G^{(2,n)}(Q^2)$, we obtain
\bes
\label{ata2ndel}
\bea
a^{(2,n)}(\sm)_{(D=0)} &=& \delta_{x^0}^{(d)} - \left( \frac{n+3}{n+1} \right) \delta_{x^1}^{(d)} + \left( \frac{n+3}{n+1} \right) \delta_{x^{n+2}}^{(d)} - \delta_{x^{n+3}}^{(d)},
\label{a2ndel} \\
\ta^{(2,n)}(\sm; \kappa)_{(D=0)} &=& \delta_{x^0}^{(\td)}(\kappa) - \left( \frac{n+3}{n+1} \right) \delta_{x^1}^{(\td)}(\kappa) + \left( \frac{n+3}{n+1} \right) \delta_{x^{n+2}}^{(\td)}(\kappa) - \delta_{x^{n+3}}^{(\td)}(\kappa).
\label{ta2ndel} \eea \ees
When $\ta_{n+1}(\kappa Q^2)$ terms in  $\td(Q^2, \kappa)$ in Eq.~(\ref{ta2nD0}) evolve according to the one-loop RGE from $Q^2=\sm$ to $Q^2=\sm e^{i \phi}$ [cf.~also Eq.~(\ref{invB1l})], we obtain
\be
\ta^{(2,n);(1 \ell)}(\sm)_{(D=0)} = \delta_{x^0}^{(\td; 1 \ell)} - \left( \frac{n+3}{n+1} \right) \delta_{x^1}^{(\td; 1 \ell)} + \left( \frac{n+3}{n+1} \right) \delta_{x^{n+2}}^{(\td; 1 \ell)} - \delta_{x^{n+3}}^{(\td; 1 \ell)}.
\label{ta2ndel1l}
\ee
Using here the corresponding identities (\ref{dxntd1l})-(\ref{Btd1l}), the following one-loop Borel transform of the auxiliary (tilde) FESR momentum $\ta^{(2,n)}(\sm)_{(D=0)}$ is obtained:
\be
{\cal B}[\ta^{(2,n);(1 \ell)}(\sm)](u; \kappa) =
{\cal B}[\td](u; \kappa)  \frac{\sin (\pi u)}{\pi}  \left\{ \frac{1}{u} -
\left( \frac{n+3}{n+1} \right) \frac{1}{(u-1)} +  \left( \frac{n+3}{n+1} \right) \frac{1}{(u-n-2)}-  \frac{1}{(u-n-3)} \right\},
\label{Bta2n} \ee
where we recall that the $(D=0)$ auxiliary Adler function $\td(Q^2; \kappa) \equiv \td(Q^2; \kappa)_{(D=0)}$ was defined in Eq.~(\ref{tdpt}) as a power series, and ${\cal B}[\td](u; \kappa)$ is given as expansion in Eq.~(\ref{Btdexp}) and as the renormalon-motivated ansatz in Eq.~(\ref{Btd5P}).  We point out that the expansion of the expression (\ref{Bta2n}) in powers of $u$ generates the coefficients of the (one-loop) FOPT expansion of the sum rule $\ta^{(2,n);(1 \ell)}(\sm)$, i.e., in powers of $a^{(1 \ell)}(\kappa \sm)$, as implied by the relation (\ref{dxntd1l}). We argue here (and later on) that the superscript $(1 \ell)$ on the left-hand side of the relation (\ref{Bta2n}) can be omitted, because the coefficients of the (FOPT) expansion in powers of $a(\kappa \sm)$ of the quantity $\ta^{(2,n)}(\sm;\kappa)$ are unchanged when in the expansion we replace $a(\kappa \sm) \mapsto a^{(1 \ell)}(\kappa \sm)$ [and thus obtain $\ta^{(2,n);(1 \ell)}(\sm)$], in complete analogy with Eqs.~(\ref{invB}) and (\ref{invB1l}) for $\td$ and $\td^{(1 \ell)}$. Thus we can write ${\cal B}[\ta^{(2,n);(1 \ell)}(\sm)](u; \kappa) = {\cal B}[\ta^{(2,n)}(\sm)](u; \kappa)$.

We note that the identity (\ref{Bta2n}) implies, due to the factor $\sin (\pi u)$, that for all $n \geq 0$ we have suppression of the leading UV $u=-1$ double-pole renormalon of ${\cal B}[\td](u)$ [cf.~Eq.~(\ref{Btd5P})] into single-pole renormalon in the Borel transform ${\cal B}[\ta^{(2,n)}](u)$.  Further, for $n \geq 1$ the leading IR $u=2$ (single-pole) of ${\cal B}[\td](u)$ [$\sim 1/(2-u)$] disappears in ${\cal B}[\ta^{(2,n)}]$.

The IR renormalons which do not get suppressed in ${\cal B}[\ta^{(2,n)}]$ are $u=(n+2)$ and $u=(n+3)$. This is reflected in the survival of the condensates $\langle O_D \rangle$ with $D=2(n+2)$ and $D=2(n+3)$ in $a^{(2,n)}$, cf.~Eq.~(\ref{a2nth}) where we assumed that the condensates in the OPE (\ref{DOPE}) are $Q^2$-independent [however, see the related discussion around Eq.~(\ref{dD6})]. 

\subsection{One-loop Borel transform of auxiliary Borel Laplace $\tb$}
\label{app:tb}

We will denote the $(D=0)$ part of the Borel-Laplace sum rule $B_{\rm th}(M^2;\sm)$ as $B_{\rm th}(M^2;\sm)_{(D=0)} \equiv b(M^2;\sm)$, i.e., the contour integral in Eq.~(\ref{Bth}) involving $d(\sm e^{i \phi})_{(D=0)}$. The corresponding auxiliary (tilde) quantity is then obtained by replacement $d_{(D=0)} \mapsto \td_{(D=0)}$
\bea
\lefteqn{
  \tb(M^2;\kappa) = \frac{1}{2 \pi}  \int_{-\pi}^{+\pi} d \phi
}
\nonumber\\ && \times
 \left\{ \left[ \left(1 + e^{i \phi} \right)^2 - 2 \frac{M^2}{\sm}  \left(1 + e^{i \phi} \right) +  2 \left( \frac{M^2}{\sm} \right)^2 \right] \exp \left( \frac{\sm}{M^2} e^{i \phi} \right) -  2 \left( \frac{M^2}{\sm} \right)^2 \exp \left( - \frac{\sm}{M^2} \right) \right\} \td \left( \sm e^{i \phi}; \kappa \right)_{(D=0)}.
\label{tbdef} \eea  
In an analogous way as in the previous Sec.~\ref{app:ta2n}, we then obtain the relation between the Borel transform of this quantity in the one-loop case and the Borel transform of $\td_{(D=0)}$
\bea
\lefteqn{
{\cal B}[\tb^{(1 \ell)}(M^2;\sm)](u; \kappa) =
{\cal B}[\td](u; \kappa)  \frac{\sin (\pi u)}{\pi}  \times \sum_{n=0}^{\infty} \frac{(-1)^n}{n!} \left(\frac{\sm}{M^2} \right)^n 
}
\nonumber\\ && \times
 \left\{ \frac{1}{(u-2 -n)} - 2 \left( 1 - \frac{M^2}{\sm} \right) \frac{1}{(u - 1 -n)} + \left( 1 - 2 \frac{M^2}{\sm} \right) \frac{1}{(u-n)} + 2 \left( \frac{M^2}{\sm} \right)^2 \frac{n}{u (u -n)} \right\} .
\label{Btb} \eea
Again, as in Eq.~(\ref{Bta2n}), we see that, due to the factor $\sin (\pi u)$, we have suppression of the leading UV $u=-1$ double-pole renormalon of ${\cal B}[\td](u)$ [cf.~Eq.~(\ref{Btd5P})] into single-pole renormalon in the Borel ${\cal B}[\tb^{(1 \ell)}(M^2)](u)$.  We point out that the superscript $(1 \ell)$ on the left-hand side of the relation (\ref{Btb}) can be omitted, in analogy with the argument about Eq.~(\ref{Bta2n}) (in the paragraph after that equation), i.e., we have ${\cal B}[\tb^{(1 \ell)}(M^2;\sm)](u; \kappa) = {\cal B}[\tb(M^2;\sm)](u; \kappa)$.

In order to obtain the identity (\ref{Btb}), the following integrations over angle $\phi$ were performed [in addition to the integration (\ref{dphi}) with $n=0$]:
\bea
{\cal J}_m(u; {\cal A}) & \equiv & \frac{1}{2 \pi} \int_{-\pi}^{+\pi} d \phi \; e^{i (m -u) \phi} \exp \left( {\cal A} e^{i \phi} \right) = \frac{1}{2 \pi} \sum_{n=0}^{\infty} \frac{ {\cal A}^n}{n!} \int_{-\pi}^{+\pi} d \phi \; e^{i (n+m-u) \phi},
\label{Jmdef} \eea 
where $m=0,1,2$, and ${\cal A}=\sm/M^2$ is a complex constant. The resulting expressions for these integrals are
\bes
\label{Jmres}
\bea
{\cal J}_m(u; {\cal A}) & = &
(-1)^m  \frac{\sin (\pi u)}{\pi} \sum_{n=0}^{\infty} \frac{(-1)^n}{n!} {\cal A}^n \frac{1}{(u-m-n)} \quad (u>0 \; \& \; u \not= m, m+1, \ldots),
\label{Jm1} \\
& = & \frac{1}{(N-m)!} {\cal A}^{N-m} \qquad (u = N = m, m+1, \ldots)
\label{Jm2}
\eea \ees
The last identity (\ref{Jm2}) is obtained as the limit of the expression (\ref{Jm1}) when $u=N+\epsilon$ and $\epsilon \to 0$ (with $N \geq m$ integer). 
    
\subsection{Relation with the full expansion coefficients}
\label{app:coeffs}

We will argue here that the results (\ref{Bta2n}) and (\ref{Btb}) give us information about the full (i.e., beyond one-loop approximation) FOPT expansion coefficients of the $(D=0)$ sum rule quantities $a^{(2,n)}(\sm)_{(D=0)}$ and $B_{\rm th}(M^2; \sm)_{(D=0)} \equiv  b(M^2;\sm)$.\footnote{
This is in contrast to the usual arguments in the literature which refer to the large-$\beta_0$ (LB) approximations of physical quantities. The latter approximations give us information only on the LB-parts $d_N^{\rm (LB)}$ of the full expansion coefficients $d_n$ ($\not= d_n^{\rm (LB)}$) of the Adler function $d(Q^2)_{(D=0)}$, and on the LB-parts $r_n^{\rm (LB)}$ of the full FOPT expansion coefficients $r_n$  ($\not= r_n^{\rm (LB)}$) of the sum rule quantities ${\cal R}(\sm)$.}

Here we will denote, for simplicity and generality, the sum rule quantity as ${\cal R}(\sm)_{(D=0)} \equiv r(\sm)$, i.e., this can be $a^{(2,n)}(\sm)_{(D=0)}$ or any FESR moment $a^{(j,n)}(\sm)_{(D=0)}$, or Borel-Laplace sum rule $b(M^2;\sm)$. Further, we will denote the corresponding auxiliary quantity as $\tr(\sm; \kappa)$, in analogy with the function $\td(Q^2;\kappa)_{(D=0)}$ Eq.~(\ref{tdpt}) which is auxiliary to the Adler function $d(Q^2)_{(D=0)}$ Eqs.~(\ref{dpt})-(\ref{dlpt}).

Starting with the renormalon-motivated expression for the Borel transform ${\cal B}[\td](u)$ of Eq.~(\ref{Btd5P}) [cf.~also Eqs.~(\ref{Btdkappa}) and (\ref{Btdexp})], the relations  (\ref{Bta2n}) and (\ref{Btb}) generate the coefficients of the auxiliary quantity $\tr(\sm; \kappa)$ which is related with the original sum rule $r(\sm)$ in the same way as the auxiliary Adler function $\td(Q^2; \kappa)_{(D=0)}$ is related with the Adler function $d(Q^2)_{(D=0)}$ [cf.~Eqs.~(\ref{dpt})-(\ref{tdpt})]
\be
   {\cal B}[\tr^{(1 \ell)}(\sm)](u; \kappa) = \tr_0 + \frac{\tr_1(\kappa)}{1! \beta_0} u + \ldots + \frac{\tr_n(\kappa)}{n! \beta_0^n} u^n + \ldots
   = {\cal B}[\tr(\sm)](u; \kappa),
\label{Btr} \ee
where
\bes
\label{tr1ltr}
\bea
\tr^{(1 \ell)}(\sm) &=& \tr_0 a^{(1 \ell)}(\kappa \sm) + \tr_1(\kappa) a^{(1 \ell)}(\kappa \sm)^2 + \ldots + \tr_n(\kappa) a^{(1 \ell)}(\kappa \sm)^{n+1} + \ldots,
\label{tr1l} \\
\tr(\sm; \kappa) &=& \tr_0 a(\kappa \sm) + \tr_1(\kappa)  a(\kappa \sm)^2 + \ldots + \tr_n(\kappa) a(\kappa \sm)^{n+1} + \ldots.
\label{tr} \eea \ees
In Eq.~(\ref{tr}) we explicitly wrote down the auxiliary quantity $\tr(\sm; \kappa)$ at any loop level, in order to point out that it has identical coefficients as the one-loop version $\tr^{(1 \ell)}(\sm)$ in Eq.~(\ref{tr1l}). These generated coefficients  $\tr_n(\kappa)$ contain the full (i.e., beyond one-loop) information about the original sum rule quantity
$r(\sm)$ whose two variants of the perturbation expansion ['lpt' and 'pt', in analogy with Eqs.~(\ref{dlpt}) and (\ref{dpt}) for the Adler function] are
\bes
\label{rlptrpt}
\bea
r(\sm)_{\rm lpt} &=& \tr_0 a(\kappa \sm) + \tr_1(\kappa) \ta_2(\kappa \sm) + \ldots + \tr_n(\kappa) \ta_{n+1}(\kappa \sm) + \ldots,
\label{rlpt} \\
r(\sm)_{\rm pt} &=& r_0 a(\kappa \sm) + r_1(\kappa) a(\kappa \sm)^2 + \ldots + r_n(\kappa) a(\kappa \sm)^{n+1} + \ldots,
\label{rpt} \eea \ees
where we recall the definition (\ref{tan}) of the logarithmic derivatives $\ta_{n+1}$. Here, $\tr_n(\kappa)$ and $r_n(\kappa)$ are interpreted as coefficients of the ${\widetilde {\rm FOPT}}$ expansion and of the FOPT expansion of the sum rule quantity $r(\sm)$, respectively, with the renormalisation scale parameter $\kappa$. The two sets are related in the same way Eq.~(\ref{rntrk}) as the corresponding coefficients of the Adler function expansion coefficients Eq.~(\ref{dntdk}).

It can be checked that the described construction of expansions (\ref{rlptrpt}) represents the  ${\widetilde {\rm FOPT}}$ expansion and of the FOPT expansion of the sum rules, by comparing the obtained coefficients with those obtained in the direct application (via Taylor expansions) of the  ${\widetilde {\rm FOPT}}$ and the FOPT expansion as described in Sec.~\ref{ssubs:FO}, Eqs.~(\ref{sr2tFO}) and (\ref{sr2FO}).

Furthermore, in order to understand better why the construction, leading via Eq.~(\ref{Btr}) to expansions Eqs.~(\ref{rlptrpt}), gives the usual ${\widetilde {\rm FOPT}}$ and FOPT expansions, we note that the logarithmic derivatives $\ta_{n+1}$ as defined in Eq.~(\ref{tan}), although being quantities which contain the information on the RGE running to any chosen loop level, simulate in all Taylor expansions the powers $(a^{(1 \ell)})^{n+1}$ of the one-loop coupling because of the relations [cf.~Eq.~(\ref{dkta})]
\bes
\label{dtada1l}
\bea
\left(  \frac{d}{d \ln Q^2} \right)^k  \ta_{n+1}(\kappa Q^2) &=& (-\beta_0)^k \frac{(n+k)!}{n!} \; {\ta}_{n+k+1}(\kappa Q^2),
\label{dta} \\
\left(  \frac{d}{d \ln Q^2} \right)^k  \left( a^{(1 \ell)}(\kappa Q^2) \right)^{n+1}  &=& (-\beta_0)^k \frac{(n+k)!}{n!} \left( a^{(1 \ell)}(\kappa Q^2) \right)^{n+k+1}
\quad (n,k=0,1, \ldots).
\label{da1l} \eea \ees
We can regard the powers $(a^{(1 \ell)})^{n+1}$ in the construction described in this Appendix as an instrument of provisional replacement: (a) in the full physical quantities we replace the (full) couplings $\ta_{n+1}$: $\ta_{n+1} \mapsto (a^{(1 \ell)})^{n+1}$; (b) thus the  Borel transforms of the auxiliary power series can be constructed and the simple one-loop RGE running can be used in the relations involving such Borel transforms; (c) at the end, the inverse replacements $(a^{(1 \ell)})^{n+1} \mapsto \ta_{n+1}$ are made [cf.~Eqs.~(\ref{tr1l}) and (\ref{rlpt})].

The above arguments also show that the sum rules (which are timelike quantities) have FOPT perturbation expansions (\ref{rpt}) for which the same renormalon-related arguments \cite{renmod} can be applied as for the perturbation expansions of spacelike quantities such as the Adler function Eq.~(\ref{dpt}), except that now, instead of in general complex (and nonnegative) $Q^2$, we have $Q^2 = \sm >0$.\footnote{In Ref.~\cite{renmod} these arguments were presented for spacelike quantities, but are valid also for (FOPT expansion) of timelike quantities, as shown here.} Specifically, if $\tr(\sm)$ has a double-pole (DP) or a single-pole (SP) $u=-1$ UV renormalon, the 'lpt'-expansion coefficients for large $n$ behave as: $\tr_n \sim (n+1)! (- \beta_0)^n$ (DP) and   $\tr_n \sim n! (- \beta_0)^n$ (SP). And the usual perturbation expansion ('pt') cefficients $r_n$ behave as $r_n \sim \Gamma(\bg_1 + 1 +n) (-\beta_0)^n [1 + {\cal O}(1/n)]$ (DP) and  $r_n \sim \Gamma(\bg_1 +n) (-\beta_0)^n  [1 + {\cal O}(1/n)]$ (SP) where $\bg_1=1 - c_1/\beta_0$ [cf.~Eqs.~(\ref{tgbg}) and (\ref{Jn})]. To illustrate this, we present in Table \ref{tabtrnrn}, which is analogous to Table \ref{tabtdndn} made for the coefficients of the Adler function, the expansion coefficients $r_n$ for the FESR moment $r(\sm) = a^{(2,1)}(\sm)$ at increasing $n$, and we can see that at large $n$ they are dominated by single-pole (SP) $u=-1$ UV renormalon contribution (columns 4 and 5), not double-pole (DP, column 6) where no convergence of the corresponding ratio is seen when $n$ increases.
\begin{table}
  \caption{The ($\MSbar$) coefficients $\tr_n$ and $r_n$ (with $\kappa=1$) of the FOPT expansion in powers of $a(\sm)$ for the FESR $r(\sm) = a^{(2,1)}(\sm)_{(D=0)}$ [$=r_{\tau}(\sm)^{(D=0)}$], where the considered renormalon-motivated Adler function extension was used. The values of the first four coefficients ($n=0,1,2,3$) coincide with the exactly known values. See the text for details.}
\label{tabtrnrn}
\begin{ruledtabular}
\begin{tabular}{r|rr|rrr}
 $n$ & $\tr_n$ & $r_n$ & $\tr_n/(n! (-\beta_0)^n)$  & $r_n/(\Gamma(\bg_1+n)(-\beta_0)^n)$ & $r_n/(\Gamma(\bg_1+1+n)(-\beta_0)^n)$ 
\\
\hline
0       &  1       & 1         &  1  & 0.229221  & 1.09217 \\
1       &  5.20232 & 5.20232   &  -2.31214  & -2.52525   & -2.0872 \\
2       &  17.1174 & 26.3659 &  1.6906  &  4.7014   & 2.12745 \\
3       &  27.7416 & 127.079 &  -0.405912  & -4.55729  & -1.41977 \\
4       &  12.3144 & 645.972 &  0.0200203  & 3.20758  & 0.761917 \\
5       &  753.119 & 4177.38 & -0.108835 & -2.18985  & -0.420328 \\
6       &  6687.87 & 34981.1 & 0.0715913 & 1.56435  & 0.251914 \\
7       & 35360.5 & 353440. & -0.0240331 & -1.13123 &  -0.1569 \\
8       & -199635. & $3.63992 \times 10^{6}$ & -0.00753801  & 0.718154 & 0.0874744 \\
9       & $3.45803 \times 10^{6}$ & $3.76036 \times 10^{7}$ & -0.00644797 & -0.401639  & -0.0436096 \\
10      & $-7.23142 \times 10^{7}$ & $3.25358 \times 10^{8}$ & -0.00599288 & 0.167699  & 0.0164252 \\
11      & $2.05973 \times 10^{9}$ & $3.28947 \times 10^{9}$ & -0.00689679 & -0.0738062  & -0.00658403 \\
12      & $-5.25891 \times 10^{10}$ & $1.98763 \times 10^{10}$ & -0.00652182 & 0.0176815  & 0.00144813 \\
13      & $1.56285 \times 10^{10}$ & $5.80646 \times 10^{11}$ & -0.00662621 & -0.0188019  & -0.00142332 \\
14      & $-4.89715 \times 10^{10}$ & $-1.45412 \times 10^{12}$ & -0.00659144 & -0.0015842  & -0.000111486 \\
15      & $1.65649 \times 10^{15}$ & $3.06976 \times 10^{14}$ & -0.00660619 & -0.0104602 & -0.000687724 \\
\hline
20      & $-1.77596 \times 10^{23}$ & $-1.86613 \times 10^{22}$ & -0.00660174  & -0.00742934 & -0.000367609 \\
\hline
25      & $6.52936 \times 10^{31}$ & $5.81784 \times 10^{30}$ & -0.0066018  & -0.00750835 & -0.000297834 \\
\hline
30      & $-6.43869 \times 10^{40}$ & $-4.97698 \times 10^{39}$ & -0.00660179 & -0.0075187 & -0.000248882 
\end{tabular}
\end{ruledtabular}
\end{table}
For better visualisation of the behaviour of the various contributions (X=UV1, IR2, IR3) to the expansion coefficients, we present in Table \ref{tabctdndn}, for $\kappa=1$ the ratios of the lpt-coefficients $\td_n^{\rm X}/\td_n$ and the pt-coefficients $d_n^{\rm X}/d_n$ of the renormalon-motivated Adler function extension $d(Q^2)_{(D=0)}$, and in Table \ref{tabctrnrn} the corresponding ratios of the ($\widetilde{\rm FOPT}$) lpt-coefficients $\tr_n^{\rm X}/\tr_n$ and the (FOPT) pt-coefficients $r_n^{\rm X}/r_n$ of the moment $a^{(2,1)}(\sm)_{(D=0)}$.
\begin{table}
  \caption{The ratios of $\td_n^{\rm X}/\td_n$ and $d_n^{\rm X}/d_n$ for the separate renormalon contributions X= UV1, IR2, IR3 ($\kappa=1$). See the text for details.}
\label{tabctdndn}
\begin{ruledtabular}
\begin{tabular}{r|rrr|rrr}
  $n$ & $\td_n^{\rm UV1}/\td_n$ &  $\td_n^{\rm IR2}/\td_n$ &  $\td_n^{\rm IR3}/\td_n$ &
 $d_n^{\rm UV1}/d_n$ &  $d_n^{\rm IR2}/d_n$ &  $d_n^{\rm IR3}/d_n$   
\\
\hline
0       &  -0.0369777  & 1.74086     &   -0.703878   & -0.0369777 & 1.74086 & -0.703878 \\
1       &  0.0751416   & 1.82491     &  -0.900055  & 0.0751416 & 1.82491 & -0.900055 \\
2       &  -0.227154 & 2.28546 &  -1.0583  &  -0.0888303 & 2.07472 & -0.985892 \\
3       &  0.257563 & 1.15757 &  -0.415136  & 0.0786098 & 1.61026 & -0.688871 \\
4       &  -1.97777 & 4.03101 &  -1.05324  & -0.100583 & 1.75565 & -0.65507 \\
5       &  0.573134 & 0.526712 & -0.0998467 & 0.0988345 & 1.31603 & -0.414865 \\
6       &  1.55361 & -0.642222 & 0.0886165 & -0.170178 & 1.57339 & -0.403212 \\
7       & 0.856263 & 0.159776 & -0.0160393 & 0.170568 & 1.04224 & -0.212812 \\
8       & 1.08529 & -0.0919885 &  0.0066947  & -0.445023 & 1.72205 & -0.277025 \\
9       & 0.964573 & 0.0373903 & -0.00196364 & 0.325567 & 0.771334 & -0.096902 \\
10      & 1.01747 & -0.0181542 & 0.000684715 & -3.6304 & 5.12945 & -0.499046 \\
11      & 0.992026 & 0.00819475 & -0.000220967 & 0.56694 & 0.468045 & -0.0349855 \\
12      & 1.00378 & -0.00385884 & 0.0000740791 & 1.81027 & -0.859241 & 0.0489724 \\
13      & 0.99823 & 0.00179409 & -0.0000244279 & 0.795075 & 0.214165 & -0.00924029 \\
14      & 1.00084 & -0.000844392 & $8.12647 \times 10^{-6}$  & 1.17141 & -0.177161 & 0.00574714 \\
15      & 0.999605 & 0.000397336 & $-2.69455 \times 10^{-6}$ & 0.924117 & 0.0777677 & -0.00188487 \\
\hline
20      & 1.00001 & $-9.63217 \times 10^{-6}$ & $1.09306 \times 10^{-8}$  & 1.00407 & -0.00409122 & 0.0000213327 \\
\hline
25      & 1. & $2.4567 \times 10^{-7}$ & $-4.45651 \times 10^{-11}$  & 0.999823 & 0.000177578 & $-1.80582 \times 10^{-7}$ \\ 
\hline
30      & 1. & $-6.48368 \times 10^{-9}$ & $1.82225 \times 10^{-13}$ & 1.00001 & $-7.25984 \times 10^{-6}$ & $1.34644 \times 10^{-9}$
\end{tabular}
\end{ruledtabular}
\end{table}
\begin{table}
  \caption{The ratios of $\tr_n^{\rm X}/\tr_n$ and $r_n^{\rm X}/r_n$ for the separate renormalon contributions X= UV1, IR2, IR3 ($\kappa=1$) to the expansion coefficients of the moment $a^{(2,1)}(\sm)_{(D=0)}$. See the text for details.}
\label{tabctrnrn}
\begin{ruledtabular}
\begin{tabular}{r|rrr|rrr}
  $n$ & $\tr_n^{\rm UV1}/\tr_n$ &  $\tr_n^{\rm IR2}/\tr_n$ &  $\tr_n^{\rm IR3}/\tr_n$ &
 $r_n^{\rm UV1}/r_n$ &  $r_n^{\rm IR2}/r_n$ &  $r_n^{\rm IR3}/r_n$   
\\
\hline
0      &  -0.0369777  & 1.74086  & -0.703878  & -0.0369777 & 1.74086 & -0.703878 \\
1       &  -0.0016366 & 1.76735   &  -0.765715  & -0.0016366 & 1.76735 & -0.765715 \\
2       &  0.00115689 & 1.90818 & -0.909336 & 0.000176997 & 1.85878 & -0.858957 \\
3       &  0.0292924 & 1.94063 & -0.969924 & 0.00678763 & 1.88949 & -0.896275 \\
4       & -0.336112 & -9.25836 & 10.5945 & 0.00356353 & 1.68224 & -0.685802 \\
5       &  0.0544734 & -0.674834 & 1.62035 & 0.00768777 & 1.06683 & -0.0745162 \\
6       &  -0.0935855 & 0.492648 & 0.600937 & -0.00304746 & 0.433505 & 0.569543 \\
7       & 0.276281 & 0.688188 & 0.0355313 & 0.00586736 & 0.132549 & 0.861583 \\
8       & 0.874355 & 0.0718354 & 0.0538095 & -0.0117557 & 0.134466 & 0.87729 \\
9       & 1.02372 & -0.149521 & 0.125801 & 0.0172694 & 0.245631 & 0.7371 \\
10      & 1.1017 & -0.0243494 & -0.0773549 & -0.044734 & 0.375478 & 0.669256 \\
11      & 0.957225 & 0.0227348 & 0.0200405 & 0.100847 & 0.35506 & 0.544093 \\
12      & 1.01226 & -0.00625002 & -0.00601012 & -0.419918 & 0.484553 & 0.935364 \\
13      & 0.996316 & 0.00154863 & 0.00213505 & 0.396335 & 0.177134 & 0.426531 \\
14      & 1.00157 & -0.00075326 & -0.000818179 & 4.7062 & -1.19453 & -2.51167 \\
15      & 0.999335 & 0.000372428 & 0.000292517 & 0.713473 & 0.108541 & 0.177986 \\
\hline
20      & 1.00001 & $-6.22701 \times 10^{-6}$ & $-1.53368 \times 10^{-6}$ & 1.00816 & -0.00481066 & -0.00334754 \\
\hline
25      & 1. & $1.21648 \times 10^{-7}$ & $7.73617 \times 10^{-9}$ & 0.99982 & 0.000145671 & 0.0000345216 \\
\hline
30      & 1. & $-2.596 \times 10^{-9}$ & $-3.75139 \times 10^{-11}$ & 1. & $-4.50729 \times 10^{-6}$ & $-3.03692 \times 10^{-7}$
\end{tabular}
\end{ruledtabular}
\end{table}
The three types of the coefficients $\td_n^{\rm X}$ are generated by the corresponding X-parts of the Borel transform ${\cal B}[\td^{\rm X}](u)$ of Eq.~(\ref{Btd5P}):
\bes
\label{BtdX}
\bea
{\cal B}[\td^{\rm IR2}](u) & = & \exp \left( \tK u \right) \pi 
\td_{2,1}^{\rm IR} \left[ \frac{1}{(2-u)} + \tal (-1) \ln \left( 1 - \frac{u}{2} \right) \right] ,
\label{BtdIR2}
\\
{\cal B}[\td^{\rm IR3}](u) & = & \exp \left( \tK u \right) \pi 
\left[ \frac{ \td_{3,2}^{\rm IR} }{(3 - u)^2} + \frac{ \td_{3,1}^{\rm IR} }{(3 - u)} \right],
\label{BtdIR3}
\\
{\cal B}[\td^{\rm UV1}](u) & = & \exp \left( \tK u \right) \pi \frac{ \td_{1,2}^{\rm UV} }{(1 + u)^2} .
\label{BtdUV1}
\eea \ees
The corresponding pt-coefficients $d_n^{\rm X}$ were obtained by applying the linear transformations (\ref{dntdk}) to $\td_k^{\rm X}$ (instead of $\td_k$).

The three types of the coefficients $\tr_n^{\rm X}$ of the moment $a^{(2,1)}(\sm)_{(D=0)}$ are generated by the Borel transform (\ref{Bta2n}) (for $n=1$ there), where in the first factor on the right-hand side of Eq.~(\ref{Bta2n}) we apply the corresponding part ${\cal B}[\td^{\rm X}](u; \kappa)$ ($\kappa=1$), and the coefficients $r_n^{\rm X}$ are obtained from $\tr_k^{\rm X}$'s by applying to them the linear tranformation (\ref{rntrk}) [cf.~Eq.~(\ref{dntdk})].

Inspection of the Tables \ref{tabctdndn} and \ref{tabctrnrn} leads to the following observations.

In Table \ref{tabctdndn} we see that the UV1 contribution becomes dominant in the (lpt-)coefficients $\td_n$ for $n \geq 7$: $0.85 < |\td_n^{\rm UV1}/\td_n| <1.09$; $|\td_n^{\rm IR2}/\td_n| < 0.16$; $|\td_n^{\rm IR3}/\td_n| <0.02$. On the other hand, the UV1 contribution becomes dominant in the (pt-) coefficients $d_n$ for $n \geq 13$: $0.79 < |d_n^{\rm UV1}/d_n| <1.17$; $|d_n^{\rm IR2}/d_n| < 0.22$; $|d_n^{\rm IR3}/d_n| <0.01$.

In Table \ref{tabctrnrn} we see that the UV1 contribution becomes dominant in the  ($\widetilde{\rm FOPT}$) lpt-coefficients $\tr_n$ for $n \geq 8$: $0.87 < |\tr_n^{\rm UV1}/\tr_n| <1.11$; $|\tr_n^{\rm IR2}/\tr_n| < 0.15$; $|\tr_n^{\rm IR3}/\tr_n| <0.13$. On the other hand, the UV1 contribution becomes dominant in the (FOPT) pt-coefficients $r_n$ for $n \geq 15$: $0.71 < |r_n^{\rm UV1}/r_n| <1.24$; $|r_n^{\rm IR2}/r_n| < 0.11$; $|r_n^{\rm IR3}/r_n| <0.18$.

We point out that the IR2 renormalon contribution is not cancelled exactly in the moment $a^{(2,1)}(\sm)_{(D=0)}$, because of the subleading IR2-term $\sim \ln(1 - u/2)$ in the Borel transform ${\cal B}[\td](u)$. Further, the aforementioned numerical behaviour of the IR2-type ratios $\tr_n^{\rm IR2}/\tr_n$ and $\td_n^{\rm IR2}/\td_n$ (as well as  $r_n^{\rm IR2}/r_n$ and $d_n^{\rm IR2}/d_n$) at high $n$ might suggest that IR2 renormalon is not suppressed in the moment $r(\sm) = a^{(2,1)}(\sm)_{(D=0)}$, in contradiction with the conclusions coming from the identity (\ref{Bta2n}). However, we should keep in mind that the entire coefficients $\tr_n$ ($r_n$) at high $n$ get significantly suppressed in comparison with $\td_n$ ($d_n$) (cf.~also Tables \ref{tabtdndn} and \ref{tabtrnrn}), due to the suppression of the dominant renormalon UV1 in the moment $a^{(2,1)}(\sm)_{(D=0)}$: i.e., as explained earlier, at large $n$ we have $\td_n \sim (n+1)! (-\beta_0)^n$ and $\tr_n \sim n! (-\beta_0)^n$; the corresponding pt-coefficients are $d_n \sim \Gamma(\bg_1 + 1 +n) (-\beta_0)^n [1 + {\cal O}(1/n)]$ and $r_n \sim \Gamma(\bg_1 +n) (-\beta_0)^n [1 + {\cal O}(1/n)]$.

In the work \cite{BKM}, similar renormalon-dominated asymptotic behaviour of perturbation coefficient was found for spacelike and timelike quantities related with the scalar current-current correlator; the analysis in \cite{BKM} was performed in the large-$\beta_0$ approximation.


\begin{thebibliography}{99}

 \bibitem{ALEPH2}
  S.~Schael {\it et al.}  [ALEPH Collaboration],
``Branching ratios and spectral functions of tau decays: final ALEPH measurements and physics implications,''
Phys.\ Rept.\  {\bf 421} (2005), 191
  [hep-ex/0506072];
  M.~Davier, A.~H\"ocker and Z.~Zhang,
  ``The Physics of hadronic tau decays,''
 Rev.\ Mod.\ Phys.\  {\bf 78} (2006), 1043
  [hep-ph/0507078].
  
\bibitem{DDHMZ}
  M.~Davier, S.~Descotes-Genon, A.~H\"ocker, B.~Malaescu and Z.~Zhang,
  ``The Determination of $\alpha_s$ from $\tau$ decays revisited,''
 Eur.\ Phys.\ J.\ C {\bf 56} (2008), 305
  [arXiv:0803.0979 [hep-ph]].
  
   \bibitem{ALEPHfin}
  M.~Davier, A.~H\"ocker, B.~Malaescu, C.~Z.~Yuan and Z.~Zhang,
  ``Update of the ALEPH non-strange spectral functions from hadronic $\tau$ decays,''
  Eur.\ Phys.\ J.\ C {\bf 74} (2014) no. 3, 2803
  [arXiv:1312.1501 [hep-ex]].

\bibitem{ALEPHwww}
  The measured data of ALEPH Collaboration, with covariance matrix corrections described in Ref.~\cite{ALEPHfin}, are available on the following web page:
http://aleph.web.lal.in2p3.fr/tau/specfun13.html

\bibitem{DBTrev}
A.~Deur, S.~J.~Brodsky and G.~F.~de Teramond,
``The QCD running coupling,''
Nucl. Phys. \textbf{90} (2016), 1
[arXiv:1604.08082 [hep-ph]].

\bibitem{alpha2019}
D.~d'Enterria, S.~Kluth, S.~Alekhin, P.~A.~Baikov, A.~Banfi, F.~Barreiro, A.~Bazavov, S.~Bethke, J.~Bl\"umlein and D.~Boito, \textit{et al.}
``$\alpha_s$(2019): Precision measurements of the QCD coupling,''
Workshop Proceedings, ECT, Trento, 11-15 Febr.~2019
[arXiv:1907.01435 [hep-ph]].

\bibitem{PDG2020}
P.A.~Zyla \textit{et al.} [Particle Data Group],
``Review of Particle Physics,''
PTEP \textbf{2020} (2020) no.8, 083C01

\bibitem{SEW}
W.~J.~Marciano and A.~Sirlin,
``Electroweak Radiative Corrections to tau Decay,''
Phys. Rev. Lett. \textbf{61} (1988), 1815-1818

\bibitem{dpEW}
E.~Braaten and C.~S.~Li,
``Electroweak radiative corrections to the semihadronic decay rate of the tau lepton,''
Phys. Rev. D \textbf{42} (1990), 3888-3891
  
\bibitem{NP88}
S.~Narison and A.~Pich,
``QCD formulation of the $\tau$ decay and determination of $\Lambda(MS)$,''
Phys. Lett. B \textbf{211} (1988), 183-188.

\bibitem{B88B89}
E.~Braaten,
``QCD predictions for the decay of the $\tau$ lepton,''
Phys. Rev. Lett. \textbf{60} (1988), 1606-1609;
E.~Braaten,
``The perturbative QCD corrections to the ratio R for $\tau$ decay,''
Phys. Rev. D \textbf{39} (1989), 1458.

\bibitem{BNP92}
E.~Braaten, S.~Narison and A.~Pich,
``QCD analysis of the $\tau$ hadronic width,''
Nucl. Phys. B \textbf{373} (1992), 581-612.

\bibitem{GCTK01}
G.~Cveti\v{c} and T.~Lee,
``Bilocal expansion of Borel amplitude and hadronic tau decay width,''
Phys. Rev. D \textbf{64} (2001), 014030
[arXiv:hep-ph/0101297 [hep-ph]].
  
\bibitem{BCK}
  P.~A.~Baikov, K.~G.~Chetyrkin and J.~H.~K\"uhn,
  ``Order $\alpha_s^4$ QCD Corrections to $Z$ and $\tau$ Decays,''
  Phys.\ Rev.\ Lett.\  {\bf 101} (2008), 012002
  [arXiv:0801.1821 [hep-ph]].

\bibitem{DP92}
F.~Le Diberder and A.~Pich,
``Testing QCD with $\tau$ decays,''
Phys. Lett. B \textbf{289} (1992), 165-175.

\bibitem{CI1}
A.~A.~Pivovarov,
``Renormalization group analysis of the $\tau$ lepton decay within QCD,''
Sov. J. Nucl. Phys. \textbf{54} (1991), 676-678
[arXiv:hep-ph/0302003 [hep-ph]].

\bibitem{CI2}
F.~Le Diberder and A.~Pich,
``The perturbative QCD prediction to $R_{\tau}$ revisited,''
Phys. Lett. B \textbf{286} (1992), 147-152.

\bibitem{CIAPT}
A.~P.~Bakulev, S.~V.~Mikhailov and N.~G.~Stefanis,
``Higher-order QCD perturbation theory in different schemes: From FOPT to CIPT to FAPT,''
JHEP \textbf{06} (2010), 085
[arXiv:1004.4125 [hep-ph]].  

\bibitem{Davetal}
M.~Davier, A.~H\"ocker, B.~Malaescu, C.~Z.~Yuan and Z.~Zhang,
``Update of the ALEPH non-strange spectral functions from hadronic $\tau$ decays,''
Eur. Phys. J. C \textbf{74} (2014) no.3, 2803
[arXiv:1312.1501 [hep-ex]].

\bibitem{Pich}
A.~Pich and A.~Rodr\'\i{}guez-S\'anchez,
``Determination of the QCD coupling from ALEPH $\tau$ decay data,''
Phys. Rev. D \textbf{94} (2016) no.3, 034027
[arXiv:1605.06830 [hep-ph]].

\bibitem{Bo2015}
D.~Boito, M.~Golterman, K.~Maltman, J.~Osborne and S.~Peris,
``Strong coupling from the revised ALEPH data for hadronic $\tau$ decays,''
Phys. Rev. D \textbf{91} (2015) no.3, 034003
[arXiv:1410.3528 [hep-ph]].

\bibitem{Cata}
O.~Cat\`a, M.~Golterman and S.~Peris,
``Unraveling duality violations in hadronic $\tau$ decays,''
Phys. Rev. D \textbf{77} (2008), 093006
[arXiv:0803.0246 [hep-ph]];
``Possible duality violations in $\tau$ decay and their impact on the determination of $\alpha_s$,''
Phys. Rev. D \textbf{79} (2009), 053002
[arXiv:0812.2285 [hep-ph]].

\bibitem{BJ}
M.~Beneke and M.~Jamin,
``$\alpha_s$ and the $\tau$ hadronic width: fixed-order, contour-improved and higher-order perturbation theory,''
JHEP \textbf{09} (2008), 044
[arXiv:0806.3156 [hep-ph]].

\bibitem{BJ2}
M.~Beneke, D.~Boito and M.~Jamin,
``Perturbative expansion of $\tau$ hadronic spectral function moments and $\alpha_s$ extractions,''
JHEP \textbf{01} (2013), 125
[arXiv:1210.8038 [hep-ph]].

\bibitem{BoiOl}
D.~Boito and F.~Oliani,
``Renormalons in integrated spectral function moments and $\alpha_s$ extractions,''
Phys. Rev. D \textbf{101} (2020) no.7, 074003
[arXiv:2002.12419 [hep-ph]].

\bibitem{renmod}
  G.~Cveti\v{c},
  ``Renormalon-motivated evaluation of QCD observables,''
  Phys.\ Rev.\ D {\bf 99} (2019) no. 1, 014028
  [arXiv:1812.01580 [hep-ph]].

\bibitem{Chib}
B.~Chibisov, R.~D.~Dikeman, M.~A.~Shifman and N.~Uraltsev,
``Operator product expansion, heavy quarks, QCD duality and its violations,''
Int. J. Mod. Phys. A \textbf{12} (1997), 2075-2133
[arXiv:hep-ph/9605465 [hep-ph]].

\bibitem{Malt}
K.~Maltman,
``Constraints on hadronic spectral functions from continuous families of finite energy sum rules,''
Phys. Lett. B \textbf{440} (1998), 367
[arXiv:hep-ph/9901239 [hep-ph]].  

\bibitem{DomSch}
C.~A.~Dominguez and K.~Schilcher,
``Chiral sum rules and duality in QCD,''
Phys. Lett. B \textbf{448} (1999), 93-98
[arXiv:hep-ph/9811261 [hep-ph]].
  
\bibitem{Cir}
V.~Cirigliano, E.~Golowich and K.~Maltman,
``QCD condensates for the light quark V-A correlator,''
Phys. Rev. D \textbf{68} (2003), 054013
[arXiv:hep-ph/0305118 [hep-ph]].


\bibitem{GonzAl}
M.~Gonz\'alez-Alonso, A.~Pich and J.~Prades,
``Pinched weights and duality violation in QCD sum rules: a critical analysis,''
Phys. Rev. D \textbf{82} (2010), 014019
[arXiv:1004.4987 [hep-ph]].


\bibitem{Dom}
C.~A.~Dominguez, L.~A.~Hernandez, K.~Schilcher and H.~Spiesberger,
``Tests of quark-hadron duality in $\tau$-decays,''
Mod. Phys. Lett. A \textbf{31} (2016) no.31, 1630036
[arXiv:1607.02048 [hep-ph]].

\bibitem{RSan}
M.~Gonz\'alez-Alonso, A.~Pich and A.~Rodr\'\i{}guez-S\'anchez,
``Updated determination of chiral couplings and vacuum condensates from hadronic $\tau$ decay data,''
Phys. Rev. D \textbf{94} (2016) no.1, 014017
[arXiv:1602.06112 [hep-ph]].

\bibitem{SVZ}
M.~A.~Shifman, A.~I.~Vainshtein and V.~I.~Zakharov,
``QCD and resonance physics. Theoretical foundations,''
Nucl. Phys. B \textbf{147} (1979), 385-447.

\bibitem{GCCV2012}
G.~Cveti\v{c} and C.~Villavicencio,
``Operator Product Expansion with analytic QCD in $\tau$ decay physics,''
Phys. Rev. D \textbf{86} (2012), 116001
[arXiv:1209.2953 [hep-ph]].


\bibitem{OPAL}
  K.~Ackerstaff {\it et al.} [OPAL Collaboration],
  ``Measurement of the strong coupling constant $\alpha_s$ and the vector and axial vector spectral functions in hadronic tau decays,''
  Eur.\ Phys.\ J.\ C {\bf 7} (1999), 571
  [hep-ex/9808019].
 
\bibitem{PerisPC1}
  D.~Boito, M.~Golterman, M.~Jamin, A.~Mahdavi, K.~Maltman, J.~Osborne and S.~Peris,
  ``An updated determination of $\alpha_s$ from $\tau$ decays,''
  Phys.\ Rev.\ D {\bf 85} (2012), 093015
  [arXiv:1203.3146 [hep-ph]].

\bibitem{Nesterenko:2016pmx} 
  A.~V.~Nesterenko,
  ``Strong interactions in spacelike and timelike domains: dispersive approach,'' Elsevier, Amsterdam, 2016, eBook ISBN: 9780128034484.

\bibitem{AKR}
S.~Eidelman, F.~Jegerlehner, A.~L.~Kataev and O.~Veretin,
``Testing nonperturbative strong interaction effects via the Adler function,''
Phys. Lett. B \textbf{454} (1999), 369-380
[arXiv:hep-ph/9812521 [hep-ph]].

\bibitem{ANR}
A.~V.~Nesterenko,
``Explicit form of the R-ratio of electron–positron annihilation into hadrons,''
J. Phys. G \textbf{46} (2019) no.11, 115006
[arXiv:1902.06504 [hep-ph]];
``Recurrent form of the renormalization group relations for the higher-order hadronic vacuum polarization function perturbative expansion coefficients,''
J. Phys. G \textbf{47} (2020) no.10, 105001
[arXiv:2004.00609 [hep-ph]].


\bibitem{amurev}
T.~Aoyama, N.~Asmussen, M.~Benayoun, J.~Bijnens, T.~Blum, M.~Bruno, I.~Caprini, C.~M.~Carloni Calame, M.~C\`e and G.~Colangelo, \textit{et al.}
``The anomalous magnetic moment of the muon in the Standard Model,''
Phys. Rept. \textbf{887} (2020), 1-166
[arXiv:2006.04822 [hep-ph]].

\bibitem{amuZoltan}
S.~Borsanyi, Z.~Fodor, J.~N.~Guenther, C.~Hoelbling, S.~D.~Katz, L.~Lellouch, T.~Lippert, K.~Miura, L.~Parato and K.~K.~Szabo, \textit{et al.}
``Leading hadronic contribution to the muon magnetic moment from lattice QCD,''
Nature 593 (2021) 51
[arXiv:2002.12347 [hep-lat]].

\bibitem{NestJPG42}
A.~V.~Nesterenko,
  ``Hadronic vacuum polarization function within dispersive approach to QCD,''
  J.\ Phys.\ G {\bf 42} (2015), 085004
  [arXiv:1411.2554 [hep-ph]].
  
\bibitem{amuO}
G.~Cveti\v{c} and R.~K\"ogerler,
``Infrared-suppressed QCD coupling and the hadronic contribution to muon g-2,''
J. Phys. G \textbf{47} (2020) no.10, 10LT01
[arXiv:2007.05584 [hep-ph]];
``Lattice-motivated QCD coupling and hadronic contribution to muon $g-2$,''
J. Phys. G \textbf{48} (2021) no.5, 055008
[arXiv:2009.13742 [hep-ph]].
  
\bibitem{5lMSbarbeta}
  P.~A.~Baikov, K.~G.~Chetyrkin and J.~H.~K\"uhn,
  ``Five-loop running of the QCD coupling constant,''
 Phys.\ Rev.\ Lett.\  {\bf 118} (2017) no. 8, 082002
  [arXiv:1606.08659 [hep-ph]].

  
\bibitem{d1}
  K.~G.~Chetyrkin, A.~L.~Kataev and F.~V.~Tkachov,
``Higher order corrections to $\sigma_T$ ($e^+ e^- \to$ Hadrons) 
in Quantum Chromodynamics,''
 Phys.\ Lett.\ B {\bf 85} (1979), 277;
  M.~Dine and J.~R.~Sapirstein,
  ``Higher order QCD corrections in $e^+e^-$ annihilation,''
  Phys.\ Rev.\ Lett.\  {\bf 43} (1979), 668;
  W.~Celmaster and R.~J.~Gonsalves,
``An analytic calculation of higher order Quantum Chromodynamic 
corrections in  $e^+e^-$ annihilation,''
  Phys.\ Rev.\ Lett.\  {\bf 44} (1980), 560.
 
\bibitem{d2}
  S.~G.~Gorishnii, A.~L.~Kataev and S.~A.~Larin,
``The ${\cal O}(\alpha_s^3)$ corrections to $\sigma_{tot} (e^+ e^- \to$ hadrons) and 
$\Gamma(\tau^- \to \nu_{\tau} + {\rm hadrons})$ in QCD,''
 Phys.\ Lett.\ B {\bf 259} (1991), 144;
  L.~R.~Surguladze and M.~A.~Samuel,
``Total hadronic cross-section in $e^+e^-$ annihilation 
at the four loop level of perturbative QCD,''
 Phys.\ Rev.\ Lett.\  {\bf 66} (1991), 560
  Erratum: [Phys.\ Rev.\ Lett.\  {\bf 66} (1991), 2416].

\bibitem{MiniMOM}
  L.~von Smekal, K.~Maltman and A.~Sternbeck,
  ``The Strong coupling and its running to four loops in a minimal MOM scheme,''
  Phys.\ Lett.\ B {\bf 681} (2009), 336
  [arXiv:0903.1696 [hep-ph]].

\bibitem{BoucaudMM}
  P.~Boucaud, F.~De Soto, J.~P.~Leroy, A.~Le Yaouanc, J.~Micheli, O.~Pene and J.~Rodr\'{\i}guez-Quintero,
  ``Ghost-gluon running coupling, power corrections and the determination of $\Lambda_{\overline {\rm MS}}$,''
  Phys.\ Rev.\ D {\bf 79} (2009), 014508
  [arXiv:0811.2059 [hep-ph]];
S.~Zafeiropoulos, P.~Boucaud, F.~De Soto, J.~Rodr\'{\i}guez-Quintero and J.~Segovia,
``Strong running coupling from the gauge sector of domain wall Lattice QCD with physical quark masses,''
Phys. Rev. Lett. \textbf{122} (2019) no.16, 162002
[arXiv:1902.08148 [hep-ph]].


\bibitem{CheRet}
  K.~G.~Chetyrkin and A.~R\'etey,
  ``Three-loop three-linear vertices and four-loop ${\widetilde {\rm MOM}}$ $\beta$ functions in massless QCD,''
 [arXiv:hep-ph/0007088 [hep-ph]].

  \bibitem{AKGCR}
A.~V.~Garkusha, A.~L.~Kataev and V.~S.~Molokoedov,
``Renormalization scheme and gauge (in)dependence of the generalized Crewther relation: what are the real grounds of the $\beta$-factorization property?,''
JHEP \textbf{02} (2018), 161
[arXiv:1801.06231 [hep-ph]].

\bibitem{ECH}
  G.~Grunberg,
  ``Renormalization group improved perturbative QCD,''
  Phys.\ Lett.\  {\bf 95B} (1980), 70
  Erratum: [Phys.\ Lett.\  {\bf 110B} (1982), 501];
  ``Renormalization scheme independent QCD and QED: the method of Effective Charges,''
  Phys.\ Rev.\ D {\bf 29} (1984), 2315.

  \bibitem{KatStar}
  A.~L.~Kataev and V.~V.~Starshenko,
  ``Estimates of the higher order QCD corrections to $R(s)$, $R_{\tau}$ and deep inelastic scattering sum rules,''
Mod. Phys. Lett. A \textbf{10} (1995), 235-250
[arXiv:hep-ph/9502348 [hep-ph]].

\bibitem{Boitoetal}
  D.~Boito, P.~Masjuan and F.~Oliani,
  ``Higher-order QCD corrections to hadronic $\tau$ decays from Pad\'e approximants,''
  JHEP {\bf 1808}, 075 (2018)
  [arXiv:1807.01567 [hep-ph]].

\bibitem{Caprini2019}
I.~Caprini,
``Higher-order perturbative coefficients in QCD from series acceleration by conformal mappings,''
Phys. Rev. D \textbf{100} (2019) no.5, 056019
[arXiv:1908.06632 [hep-ph]].


\bibitem{Pich2}
A.~Pich,
``Tau Decay Determination of the QCD Coupling,''
in {\it Proceedings of the Workshop on Precision measurements of $\alpha_s$,} pp.~18-19
[arXiv:1107.1123 [hep-ph]].

\bibitem{CF1}
I.~Caprini and J.~Fischer,
``$\alpha_s$ from $\tau$ decays: Contour-improved versus fixed-order summation in a new QCD perturbation expansion,''
Eur. Phys. J. C \textbf{64} (2009), 35-45
[arXiv:0906.5211 [hep-ph]].

\bibitem{CF2}
I.~Caprini and J.~Fischer,
``Expansion functions in perturbative QCD and the determination of $\alpha_s(M_\tau^2)$,''
Phys. Rev. D \textbf{84} (2011), 054019
[arXiv:1106.5336 [hep-ph]].


\bibitem{AACF}
G.~Abbas, B.~Ananthanarayan, I.~Caprini and J.~Fischer,
``Perturbative expansion of the QCD Adler function improved by renormalization-group summation and analytic continuation in the Borel plane,''
Phys. Rev. D \textbf{87} (2013) no.1, 014008
[arXiv:1211.4316 [hep-ph]];
``Expansions of $\tau$ hadronic spectral function moments in a nonpower QCD perturbation theory with tamed large order behavior,''
Phys. Rev. D \textbf{88} (2013) no.3, 034026
[arXiv:1307.6323 [hep-ph]].

\bibitem{Caprini2020}
I.~Caprini,
``Conformal mapping of the Borel plane: going beyond perturbative QCD,''
Phys. Rev. D \textbf{102} (2020) no.5, 054017
[arXiv:2006.16605 [hep-ph]].

\bibitem{ren}
  M.~Beneke,
  ``Renormalons,''
  Phys.\ Rept.\  {\bf 317} (1999), 1
  [hep-ph/9807443].  

\bibitem{Maiezza}
J.~Bersini, A.~Maiezza and J.~C.~Vasquez,
``Resurgence of the renormalization group equation,''
Annals Phys. \textbf{415} (2020), 168126
[arXiv:1910.14507 [hep-th]];
A.~Maiezza and J.~C.~Vasquez,
``Non-local Lagrangians from renormalons and analyzable functions,''
Annals Phys. \textbf{407} (2019), 78-91
[arXiv:1902.05847 [hep-th]];
``Resurgence of the QCD Adler function,''
[arXiv:2104.03095 [hep-ph]].

\bibitem{Cavalc}
E.~Cavalcanti,
``Renormalons beyond the Borel plane,''
Phys. Rev. D \textbf{103} (2021) no.2, 025019
[arXiv:2011.11175 [hep-th]];
``On the permanence of renormalons in compactified spaces,''
[arXiv:2011.04099 [hep-th]].

\bibitem{Pineda1}
C.~Ayala, X.~Lobregat and A.~Pineda,
``Superasymptotic and hyperasymptotic approximation to the operator product expansion,''
Phys. Rev. D \textbf{99} (2019) no.7, 074019
[arXiv:1902.07736 [hep-th]];
``Hyperasymptotic approximation to the top, bottom and charm pole mass,''
Phys. Rev. D \textbf{101} (2020) no.3, 034002
[arXiv:1909.01370 [hep-ph]].

\bibitem{Pineda2}
G.~S.~Bali, C.~Bauer and A.~Pineda,
``Perturbative expansion of the plaquette to ${\cal O}(\alpha^{35})$ in four-dimensional SU(3) gauge theory,''
Phys. Rev. D \textbf{89} (2014), 054505
[arXiv:1401.7999 [hep-ph]];
C.~Ayala, X.~Lobregat and A.~Pineda,
``Hyperasymptotic approximation to the plaquette and determination of the gluon condensate,''
JHEP \textbf{12} (2020), 093
[arXiv:2009.01285 [hep-ph]].

\bibitem{3dAQCD}
  C.~Ayala, G.~Cveti\v{c}, R.~K\"ogerler and I.~Kondrashuk,
  ``Nearly perturbative lattice-motivated QCD coupling with zero IR limit,''
 J. Phys. G \textbf{45} (2018) no.3, 035001
[arXiv:1703.01321 [hep-ph]].

\bibitem{IoffeBL}
B.~V.~Geshkenbein, B.~L.~Ioffe and K.~N.~Zyablyuk,
``The check of QCD based on the $\tau$-decay data analysis in the complex $q^2$-plane,''
Phys. Rev. D \textbf{64} (2001), 093009
[arXiv:hep-ph/0104048 [hep-ph]];
B.~L.~Ioffe,
``QCD at low energies,''
Prog. Part. Nucl. Phys. \textbf{56} (2006), 232-277
[arXiv:hep-ph/0502148 [hep-ph]].


\bibitem{LB1}
  D.~J.~Broadhurst,
  ``Large N expansion of QED: asymptotic photon propagator and contributions to the muon anomaly, for any number of loops,''
  Z.\ Phys.\ C {\bf 58} (1993), 339.
  
\bibitem{LB2}
  D.~J.~Broadhurst and A.~L.~Kataev,
  ``Connections between deep inelastic and annihilation processes at next to next-to-leading order and beyond,''
  Phys.\ Lett.\ B {\bf 315} (1993), 179
  [hep-ph/9308274].
  
\bibitem{Btes}
  M.~Beneke, ``Die Struktur der St\"orungsreihe in hohen Ordnungen,'' Ph.D. Thesis, Technische Universit\"at M\"unchen (1993).


\bibitem{Bo2017}    
D.~Boito, M.~Golterman, K.~Maltman and S.~Peris,
``Strong coupling from hadronic $\tau$ decays: A critical appraisal,''
Phys. Rev. D \textbf{95} (2017) no.3, 034024
[arXiv:1611.03457 [hep-ph]].

\bibitem{BoRee2019}
D.~Boito, M.~Golterman, K.~Maltman and S.~Peris,
``Evidence against naive truncations of the OPE from $e^+e^- \to$ hadrons below charm,''
Phys. Rev. D \textbf{100} (2019) no.7, 074009
[arXiv:1907.03360 [hep-ph]].

\bibitem{Neubert}  
  M.~Neubert,
  ``Scale setting in QCD and the momentum flow in Feynman diagrams,''
  Phys.\ Rev.\ D {\bf 51} (1995), 5924
  [hep-ph/9412265].
  
\bibitem{LandauC}
C.~Contreras, G.~Cveti\v{c} and O.~Orellana,
``pQCD running couplings finite and monotonic in the infrared: when do they reflect the holomorphic properties of spacelike observables?,''
J. Phys. Comm. \textbf{5} (2021) no.1, 015019
[arXiv:2008.03818 [hep-ph]].


\bibitem{HoangR}
A.~H.~Hoang and C.~Regner,
``Borel representation of $\tau$ hadronic spectral function moments in Contour-improved perturbation theory,''
[arXiv:2008.00578 [hep-ph]];
``On the difference between FOPT and CIPT for hadronic tau decays,''
[arXiv:2105.11222 [hep-ph]].
  
\bibitem{4lquarkthresh}
  Y.~Schr\"oder and M.~Steinhauser,
  ``Four-loop decoupling relations for the strong coupling,''
  JHEP {\bf 0601} (2006), 051
  doi:10.1088/1126-6708/2006/01/051
  [hep-ph/0512058];
  B.~A.~Kniehl, A.~V.~Kotikov, A.~I.~Onishchenko and O.~L.~Veretin,
  ``Strong-coupling constant with flavor thresholds at five loops in the anti-MS scheme,''
  Phys.\ Rev.\ Lett.\  {\bf 97} (2006), 042001
  [hep-ph/0607202].

\bibitem{Pich3}
A.~Pich and A.~Rodr\'\i{}guez-S\'anchez,
``Updated determination of $\alpha_s(m_\tau^2)$ from tau decays,''
Mod. Phys. Lett. A \textbf{31} (2016) no.30, 1630032
[arXiv:1606.07764 [hep-ph]].

\bibitem{Pich4}
A.~Pich,
``Precision physics with inclusive QCD processes,''
Prog. Part. Nucl. Phys. \textbf{117} (2021), 103846
[arXiv:2012.04716 [hep-ph]].

  
\bibitem{Bo2011}
D.~Boito, O.~Cat\`a, M.~Golterman, M.~Jamin, K.~Maltman, J.~Osborne and S.~Peris,
``A new determination of $\alpha_s$ from hadronic $\tau$ decays,''
Phys. Rev. D \textbf{84} (2011), 113006
[arXiv:1110.1127 [hep-ph]].




\bibitem{Bo2021}
D.~Boito, M.~Golterman, K.~Maltman, S.~Peris, M.~V.~Rodrigues and W.~Schaaf,
``Strong coupling from an improved $\tau$ vector isovector spectral function,''
Phys. Rev. D \textbf{103} (2021) no.3, 034028
[arXiv:2012.10440 [hep-ph]].  


\bibitem{prgs}
Mathematica programs (compatible with the version 11.1):
SumRPMSbALEPHM2m090.m (when $\langle O_{10} \rangle_{V+A}=0$);
SumRPMSbALEPHM2m090O10.m (when $\langle O_{10} \rangle_{V+A}=0$ is varied).
These programs call the subroutines: AdlerFunction4lMiniMOM.m; MSbarRenMod5A.save; aMSQ2complS\_almtauinput.m; expdataALEPH.m; SumRthMSbar.save; the program and the subroutines are available on www page http://www.gcvetic.usm.cl/ 

\bibitem{BKM}
D.~J.~Broadhurst, A.~L.~Kataev and C.~J.~Maxwell,
``Renormalons and multiloop estimates in scalar correlators: Higgs decay and quark mass sum rules,''
Nucl. Phys. B \textbf{592} (2001), 247-293
[arXiv:hep-ph/0007152 [hep-ph]].
  
\end{thebibliography}
\end{document}